\documentclass[aps,twocolumn,prx,
preprintnumbers,nofootinbib]{revtex4-2}
\usepackage{xpatch}
\makeatletter
\patchcmd{\@ssect@ltx}
    {\addcontentsline{toc}{#1}{\protect\numberline{}#8}}
    {}
    {}
    {}
\makeatother
\pdfoutput=1
\usepackage{color}
\usepackage{enumitem}
\usepackage{amsmath}
\usepackage[dvipsnames]{xcolor}
\usepackage{hyperref}
\usepackage{amsthm}

\usepackage{float}

\usepackage{enumitem} 
\usepackage{color}
\usepackage{enumitem}
\usepackage[dvipsnames]{xcolor}
\usepackage{hyperref}
\definecolor{amaranth}{rgb}{0.9, 0.17, 0.31}
\definecolor{forestForestGreen(web)}{rgb}{0.13, 0.55, 0.13}
\definecolor{blue(munsell)}{HTML}{005567}
\definecolor{bblue}{rgb}{0.0, 0.58, 0.71}
\hypersetup{
pdfstartview={FitH}, 
pdftitle={}, 
colorlinks=true, 
linkcolor=blue(munsell), 
citecolor=blue(munsell), 
filecolor=magenta, 
urlcolor=blue(munsell)
}

\setlist[itemize]{leftmargin=1.25em}
\setlist[enumerate]{leftmargin=1.25em}

\theoremstyle{definition}

\usepackage{pgfplots}
\usepackage{amssymb}
\usepackage{amsfonts}
\usepackage{graphicx}
\usepackage{epstopdf}
\usepackage{dcolumn}
\usepackage{amsmath}
\usepackage{latexsym,bm}
\usepackage{amsthm}
\usepackage{slashed}
\usepackage{float}
\usepackage{color}
\usepackage{url}
\usepackage{longtable}
\usepackage{rotating}
\usepackage[normalem]{ulem}
\usepackage{faktor}
\usepackage{tikz-cd}
\usepackage{tensor}
\usepackage{array}
\usepackage{tabularx}
\usepackage{makecell}
\usepackage{changepage}

\makeatletter
\def\l@subsubsection#1#2{}
\makeatother

\usepackage{tikz}
\usepackage{diagbox}
\usetikzlibrary{positioning}
\usetikzlibrary{calc}
\usetikzlibrary{decorations.pathreplacing,calligraphy}

\usepackage{xstring}
\usetikzlibrary{decorations.pathmorphing} 
\usetikzlibrary{decorations.markings} 
\usetikzlibrary{arrows} 
\usetikzlibrary{shapes} 
\usetikzlibrary{matrix} 
\usetikzlibrary{positioning} 
\usepackage[english]{babel} 
\usepackage[autostyle]{csquotes}
\usepackage{pifont}
\usetikzlibrary{shapes.multipart}
\usepackage{adjustbox}
\tikzset{->-/.style={decoration={
  markings,
  mark=at position .5 with {\arrow{>}}},postaction={decorate}}}

\newcommand{\fZ}{\mathfrak{Z}}
\newcommand{\bea}{\begin{eqnarray}}
\newcommand{\eea}{\end{eqnarray}}
\newcommand{\be}{\begin{equation}}
\newcommand{\ee}{\end{equation}}
\newcommand{\ba}{\begin{aligned}}
\newcommand{\ea}{\end{aligned}}
\newcommand{\bit}{\begin{itemize}}
\newcommand{\eit}{\end{itemize}}
\newcommand{\ben}{\begin{enumerate}}
\newcommand{\een}{\end{enumerate}}
\newcommand{\nn}{\nonumber}

\newcommand{\id}{\text{id}}

\newcommand{\UV}{\text{UV}}
\newcommand{\IR}{\text{IR}}

\newcommand{\Fun}{F}
\newcommand{\DFun}{P}
\newcommand{\IFun}{I}
\newcommand{\FF}{f}
\newcommand{\image}{\text{im}}
\renewcommand{\ol}{\overline}
\newcommand{\foli}{\varphi}

\newcommand{\SymTFT}{\text{SymTFT}}
\newcommand{\Bsym}{\mathfrak{B}^{\text{sym}}}
\newcommand{\Bphys}{\mathfrak{B}^{\text{phys}}}

\newcommand{\wt}{\widetilde}

\newcommand{\ot}{\otimes}

\newcommand{\Z}{{\mathbb Z}}

\newcommand{\Q}{{\mathbb Q}}

\newcommand{\cT}{\mathbf{T}}
\newcommand{\cI}{\mathcal{I}}

\newcommand{\cA}{\mathcal{A}}

\newcommand{\cC}{\mathcal{C}}
\newcommand{\cD}{\mathcal{D}}

\newcommand{\cF}{\mathcal{F}}
\newcommand{\cG}{\mathcal{G}}
\newcommand{\cH}{\mathcal{H}}
\renewcommand{\cI}{\mathcal{I}}

\newcommand{\cK}{\mathcal{K}}
\newcommand{\cL}{\mathcal{L}}
\newcommand{\cM}{\mathcal{M}}
\newcommand{\cN}{\mathcal{N}}
\newcommand{\cO}{\mathcal{O}}

\newcommand{\cQ}{\mathcal{Q}}

\newcommand{\cS}{\mathcal{S}}
\renewcommand{\cT}{\mathcal{T}}
\newcommand{\cU}{\mathcal{U}}

\newcommand{\cZ}{\mathcal{Z}}

\newcommand{\D}{\mathsf{D}}
\newcommand{\U}{\mathsf{U}}
\renewcommand{\L}{\mathsf{L}}

\def\UVcolor{MidnightBlue!60}
\def\IRcolor{Maroon!60}
\def\Scolor{ForestGreen}
\def\Ccolor{cyan}
\def\Ncolor{black!20}

\numberwithin{equation}{section}

\newcommand{\DW}{\text{DW}}

\newcommand{\bN}{\mathbb{N}}

\newcommand{\bZ}{\mathbb{Z}}

\newcommand{\fA}{\mathfrak{A}}
\newcommand{\fB}{\mathfrak{B}}

\newcommand{\cIB}{\mathfrak{I}}

\def\UV{\text{UV}}
\def\IR{\text{IR}}

\def\repa{\raise4pt\hbox{$\square$}\mkern-14mu\raise-4pt\hbox{$\square$}}
\def\repab{\overline{\raise4pt\hbox{$\square$}\mkern-14mu\raise-4pt\hbox{$\square$}\mkern-1mu}}

\DeclareMathOperator{\Aut}{Aut}
\DeclareMathOperator{\Hom}{Hom}

\newcommand{\RG}{\text{RG}}

\renewcommand{\Vec}{\mathsf{Vec}}
\newcommand{\Rep}{\mathsf{Rep}}
\newcommand{\Mod}{\mathsf{Mod}}

\renewcommand{\dim}{\text{dim}}

\newcommand{\TwoVec}{2\mathsf{Vec}}

\newcommand{\Ising}{\mathsf{Ising}}
\newcommand{\TY}{\mathsf{TY}}
\renewcommand{\Q}{\bm{Q}}
\newcommand{\sym}{\text{sym}}

\newcommand{\Fib}{\mathsf{Fib}}
\newcommand{\PD}{\text{PD}}

\tikzset{
  -latex/.style={->, >={Latex[length=1mm, width=0.75mm]}}
}
\tikzset{
  latex-latex/.style={<->, >={Latex[length=1mm, width=0.75mm]}}
}

\makeatother

\begin{document}

\title{
\vspace{2mm}
Categorical Anomaly Matching
}

\author{Andrea Antinucci}
\author{Christian Copetti}
\author{Yuhan Gai}
\author{Sakura Sch\"afer-Nameki}

\affiliation{Mathematical Institute,
University of Oxford, Woodstock Road, 
Oxford, OX2 6GG,  
United Kingdom}

\begin{abstract} 
\noindent 
Matching 't Hooft anomalies is a powerful tool for constraining the low-energy dynamics of quantum systems and their allowed renormalization group (RG) flows. For non-invertible (or categorical) symmetries, however, a key challenge has been the lack of a precise framework to characterize and quantify anomalies. We address this by identifying tensor functors between UV and IR symmetry categories as central to capturing these constraints.
To this end, we introduce Anomalous Simple Categories (ASCies) as fundamental building blocks of categorical anomalies. A given symmetry category may support multiple ASCies, each encoding distinct anomalous features.
These structures naturally arise in the context of the Symmetry Topological Field Theory (SymTFT), where tensor functors correspond to RG-interfaces between UV and IR SymTFTs, and ASCies are realized as particular such interfaces satisfying simple, universal criteria. We demonstrate the utility of this framework through examples involving anomalous 0-form, higher-form, and crucially, non-invertible symmetries in various spacetime dimensions. 
\end{abstract}


\maketitle

\tableofcontents

\section{Introduction}

What is the maximal amount of dynamical information that can be extracted from symmetries and anomalies? In studying RG-flows, one typically states that microscopic symmetries must be realized at long distances, and their anomalies must be reproduced by the infrared effective theory. This principle strongly constrains the possible IR scenarios and 
can even provide a criterion for whether two UV models flow to the same theory in the IR, i.e. are IR-dual. 

Anomaly matching is particularly important, and its full implications are yet to be uncovered, when applied to the vast generalization to higher-form symmetries \cite{Gaiotto:2014kfa} and most recently to  non-invertible,  or equivalently categorical, symmetries  \cite{Chang:2018iay,Kaidi:2021xfk,Choi:2021kmx,Bhardwaj:2022yxj} (for reviews see \cite{Schafer-Nameki:2023jdn,Shao:2023gho}).

In practice, implementing anomaly constraints can be challenging. For instance, some symmetries of the UV theory may become trivial at low energies, while new, emergent symmetries can arise that match UV anomalies via fractionalization \cite{Senthil:1999czm, Essin:2013rca, Barkeshli:2014cna, Chen:2014wse}, or more recently, transmutation \cite{Seiberg:2025bqy}. 
As a result, the space of possible IR theories is much richer than naive expectations suggest, and it becomes a nontrivial task to systematically \textbf{organize} and \textbf{classify} the full set of anomaly constraints.

At the same time, the very notion of an {\bf anomaly for a non-invertible symmetry} is considerably more subtle. Progress so far has largely focused on determining whether a given symmetry is anomalous or not~\cite{Thorngren:2019iar, Inamura:2021wuo, Kaidi:2023maf, Zhang:2023wlu, Cordova:2023bja, Antinucci:2023ezl}. However, an explicit {\bf quantification of categorical anomalies} -- a necessary ingredient for anomaly matching -- has until now remained out of reach.

It is this challenge that we will address in the current work. 
Consider a theory in $d$ spacetime dimensions.   
Our starting point is the observation that a complete formulation of the anomaly matching problem requires viewing symmetries as tensor $(d{-}1)$-categories $\cC$.
The categorical structure is not merely mathematical embellishment (even in the case of invertible symmetries), as it encodes not only the symmetry generators and their composition, but also, through the (higher) associativity constraints \cite{DouglasReutter,Copetti:2023mcq,Carqueville:2023jhb}, the 't~Hooft anomalies.

Let us denote the UV and IR symmetry categories by $\cC_\UV$ and $\cC_\IR$, respectively. Our goal is to formulate a criterion that determines when a given pair of such symmetries is compatible with 't~Hooft anomaly matching. More precisely, we seek to characterize the class of maps that preserve both the symmetry structure and the anomaly data along an RG-flow.
The appropriate notion describing such a map is that of a {\bf tensor functor}
\begin{equation}\label{funck}
    \Fun : \quad \cC_\UV \rightarrow \cC_\IR \,.
\end{equation}
The composition of symmetry generators is encoded in the tensor structure $\otimes$ of the symmetry category. A tensor functor is, by definition, a functor between categories that is compatible with this structure. Specifically, it satisfies
\begin{equation}\label{FunComp}
    \Fun(\D_1 \otimes \D_2) \cong \Fun(\D_1) \otimes \Fun(\D_2) \,,
\end{equation}
as well as compatibility with (higher) associativity constraints.

In the simplest case of invertible 0-form symmetries, described by groups $G_\UV$ and $G_\IR$ with anomalies $\omega_i \in H^{d+1}(BG_i, U(1))$, a tensor functor corresponds to a group homomorphism $\varphi : G_\UV \rightarrow G_\IR$, together with an identification
\be
[\varphi^* \omega_\IR] = [\omega_\UV] \,.
\ee
As we will see in several examples, even when only invertible symmetries are present -- but of different degrees -- a much more subtle structure can arise. In such cases, tensor functors may encode anomaly matching via symmetry fractionalization. 

When non-invertible symmetries are also involved, the situation becomes dramatically more intricate -- and more interesting.
The existence or absence of a tensor functor between two symmetry categories encodes a remarkable amount of physical information. It is therefore essential to develop this concept in detail.
Suppose a microscopic model has symmetry $\cC_\UV$. 
Then, we will show that compatibility with a low-energy effective field theory possessing symmetry $\cC_\IR$ requires the existence of a tensor functor (\ref{funck}). 
That is, a low-energy theory is consistent with the UV symmetry only if such a functor exists. This provides the most general and modern formulation of anomaly matching: it not only recovers the textbook criteria in standard cases, but also encompasses the categorical statement that {\em a symmetry is anomalous if it does not admit an {\bf SPT phase}}, meaning it lacks a fiber functor -- i.e., a tensor functor to the trivial category $(d{-}1)\Vec$.

What has so far been missing is a {\bf quantitative measure of anomalies for non-invertible symmetries}. One of the central insights of the present work is that anomalous building blocks of a categorical symmetry $\cC$ can be identified by considering  short exact sequences of tensor functors injecting into and projecting from $\cC$. 
This highlights the fundamental role of tensor functors in capturing the anomaly structure and constraining the possible IR physics.

Although tensor functors mathematically capture the full anomaly structure of categorical symmetries, it is often far more elegant -- and physically transparent -- to reformulate the problem in terms of the associated Symmetry Topological Field Theory (SymTFT). In fact, our main results and examples will be developed within this framework, which we will review below.

This reformulation becomes essential in higher-dimensions. For  fusion higher-categories, the notion of a tensor functor is not always well-defined. For example, in (2+1)d theories with fusion 2-category symmetries, a notion of monoidal (or tensor) 2-functor has been defined in~\cite{baez1996higher}, but a general framework remains elusive. In contrast, the SymTFT approach admits natural extensions to higher categories. We will adopt this perspective to generalize our measure of anomalies to spacetime dimensions $d >2$.

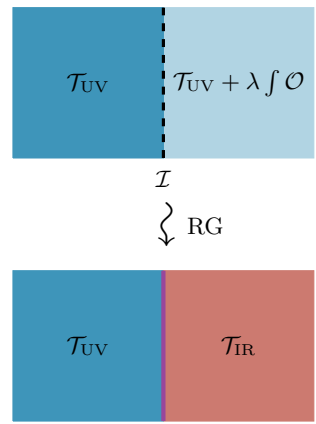
\begin{figure}[t!]
$
\begin{tikzpicture}
\begin{scope}[shift= {(0,0)}]
\draw [\UVcolor,  fill=\UVcolor] 
(-2,0) -- (-2,2) --(0,2) -- (0,0) -- (-2,0) ; 
\draw [\UVcolor,  fill=\UVcolor, opacity = 0.4] 
(0,0) -- (0,2) --(2,2) -- (2,0) -- (0,0) ; 
\draw [very thick, dashed] (0,0) -- (0,2);
\node at (-1, 1) {$\cT_{\UV}$};
\node at (1, 1) {$\cT_{\UV} + \lambda \int \cO$};
 \node[rotate=-90] at (0,-0.9) {\LARGE $\leadsto$};
  \node[right] at (0.2,-0.9) {$\RG$};
\end{scope}
\begin{scope}[shift= {(0,-3.5)}]
\draw [\UVcolor,  fill=\UVcolor] 
(-2,0) -- (-2,2) --(0,2) -- (0,0) -- (-2,0) ; 
\draw [\IRcolor,  fill=\IRcolor] 
(0,0) -- (0,2) --(2,2) -- (2,0) -- (0,0) ; 
\draw [ultra thick,  Purple] (0,0) -- (0,2);
\node at (-1, 1) {$\cT_{\UV}$};
\node at (1, 1) {$\cT_{\IR}$};
\node[above] at (0,3) {$\cI$};
\end{scope}
\end{tikzpicture}
$
\caption{
RG-interface: At the top we start with the UV:
the left half-space is the \textcolor{\UVcolor}{\bf UV theory $\cT_{\UV}$}, including the interface to the right half-space, which is $\cT_{\UV}$ with the $\cC_{\UV}$-symmetric relevant deformation (shown in light blue). Along the RG-flow the deformed theory flows to the \textcolor{\IRcolor}{\bf IR theory $\cT_\IR$}, separated by the RG-interface \textcolor{Purple}{$\cI$}. 
Such an interface  defines a  tensor functor  between the symmetries of the UV and IR theories. 
\label{fig:RGinterface}}
\end{figure}

\vspace{2mm}
\noindent 
{\bf RG-interfaces and Tensor Functors.}
{To physically motivate the relevance of tensor functors, we connect them to the concept of RG-interfaces.}
Consider a UV theory $\cT_\UV$ with symmetry $\cC_\UV$, which flows under a symmetric relevant deformation $\cO$ to a low-energy effective theory $\cT_\IR$ with faithfully acting symmetry $\cC_\IR$. A natural construction in this setting is the {\bf RG-interface}~\cite{Brunner:2007ur, Gaiotto:2012np}, obtained by turning on the deformation $\cO$ only in half of spacetime and flowing to the IR. See figure \ref{fig:RGinterface}.
The interface $\cI$ is $\cC_\UV$-symmetric, meaning that the $\cC_\UV$ symmetry defects can pass through it topologically. In particular, it encodes a tensor functor (\ref{funck}), which captures the symmetry aspects of the RG-interface. The existence of such a functor imposes strong constraints on the IR theory. A key goal of this work is to isolate this kinematical problem -- encoded in $\Fun$ -- from the more dynamical aspects of RG-flows.

\vspace{2mm}
\noindent 
{\bf Anomalous Simple Categories.}
While the 't Hooft anomaly matching conditions are elegantly implemented by the mathematics of tensor functors, a refinement of this concept is needed in order to better quantify 't Hooft anomalies, especially for non-invertible symmetries.
Intuitively, from a symmetry $\cC$ we would like to quotient out the largest possible anomaly-free subsymmetry, leaving behind a simpler generalized symmetry $\cS$ to express the 't Hooft anomaly of the original symmetry  $\cC$.
Mathematically, this concept is encoded in the notion of a short exact sequence of tensor categories \cite{bruguieres2011exact}:
\be\label{popcorn}
\cN \overset{\IFun}{\longrightarrow} \cC \overset{\DFun}{\longrightarrow} \cS \,,
\ee
where $\IFun$ is an embedding, and $\DFun$ is surjective. The interpretation of the sequence is the following:
\begin{itemize}
\item $\cN$ is chosen to be maximal (i.e. such that there is no other category that contains it and fits into this sequences) and corresponds to an anomaly-free sub-symmetry. The composition $\FF=\IFun \circ \DFun$ provides a map which ``forgets" the symmetry $\cN$, i.e. a fiber functor.

\item $\cS$ is the ``quotient" symmetry and represents the anomalous part of the symmetry $\cC$.
\end{itemize}
We will refer to a symmetry that has no non-anomalous sub-categories $\cN$ satisfying (\ref{popcorn}) as {\bf anomalous simple category (ASCy)} -- these are the simple building blocks of the anomalous parts of symmetries. 

In turn any sequence (\ref{popcorn}) with $\cN$ maximal implies that $\cS$ is an ASCy and thereby captures the anomalous part of the symmetry $\cC$. 
Important to note is that a given symmetry $\cC$ may have several inequivalent pairs $(\cN_i, \cS_i)$ associated to it that fit into a sequence such as (\ref{popcorn}). Each $\cN_i$ is a (maximal) non-anomalous sub-symmetry, whereas each ASCy $\cS_i$ captures different aspects of the anomaly of $\cC$. 

When applied to group-like symmetries the ASCies become precisely the anomalous quotient groups as required, however, the framework is far more general and applicable to any non-invertible/categorical symmetry.

\vspace{2mm}
\noindent 
{\bf SymTFT Realization.} 
While the characterization of anomalies via tensor functors -- and the notion of ASCies through short exact sequences of such functors -- offers a conceptual framework for quantifying anomalies of categorical symmetries, determining the relevant tensor functors in practice can be subtle, particularly in the case of general (i.e., non-invertible) categorical symmetries.
We show that recasting the problem in terms of the Symmetry Topological Field Theory (SymTFT) not only provides a precise reformulation of the above concepts, but also renders the entire framework more amenable to explicit computations.

The SymTFT \cite{Ji:2019jhk,  Gaiotto:2020iye, Apruzzi:2021nmk, Freed:2022qnc}, in a nutshell, separates the symmetry aspects from the dynamics of a theory. For a symmetry $\cC$ of a $d$-dimensional theory $\cT$ the SymTFT $\fZ(\cC)$ is a $(d+1)$-dimensional TQFT that lives on an interval with two boundary conditions: one that encodes the symmetries -- $\Bsym$ -- which is a topological boundary condition. The other encodes the dynamics: $\Bphys$. 
As the SymTFT is topological the size of the interval is immaterial and can be collapsed at no cost to regain the original theory. 

We will show that a tensor functor $\Fun: \cC_\UV \to \cC_\IR$ is equivalent to a configuration, where two SymTFTs {$\fZ(\cC_\UV)$ and $\fZ(\cC_\IR)$} are separated by a particular {topological} interface $\cI_\Fun$, which satisfies 
\be\label{BIB}
\Bsym_\UV \times \cI_{\Fun} =\Bsym_\IR \,.
\ee
This means when we fuse the interface {$\cI_\Fun$} onto the symmetry boundary of the UV theory, we obtain the IR-symmetry boundary. 
The situation is depicted in figure \ref{fig:ClubSandoVanilla3dUVIR}. 
The condition (\ref{BIB}) can be formulated entirely in terms of the topological defects of the SymTFT and is computationally much easier to check than the conditions on tensor functors. 

When applied to short exact sequences of tensor functors and ASCies in the context of higher dimensions and non-invertible symmetries, the SymTFT approach proves significantly easier to implement.

\vspace{2mm}
\noindent
{\bf Applications.}
We present numerous applications of this general framework. 
For 1+1d theories it can be applied to any fusion category symmetry. 
In this spacetime dimension, the analysis is completely general and systematic, thanks to the by now well-known results on condensable algebras. In turn, this connects to many of the mathematical results on tensor functors. 
In terms of examples, we will recover first some of the known results on anomalous invertible (group) symmetries, corroborating the validity of our general approach. Then we extend the analysis to non-invertible symmetries, including the Tambara-Yamagami categories. 

Although to our knowledge, tensor functors still need to be rigorously defined in higher dimensions, our SymTFT approach is easily extendable and we will apply these to examples in 2+1d and 3+1d with invertible and non-invertible symmetries.
For fusion 2-categories, that are symmetries of 2+1d theories, we can show many properties of tensor 2-functors and their realization in the SymTFT. Emboldened by this, in higher dimensions, we simply use the SymTFT approach as the defining framework for anomaly matching. 
In terms of examples, we  find  e.g.  generalizations of the Wang-Wen-Witten results \cite{Wang:2017loc} to non-invertible symmetries.

\newpage 
\clearpage

\onecolumngrid

\begin{figure}[t]
$$
\begin{tikzpicture} 
\begin{scope}[shift= {(-3,5)}]
\draw[thick,->] (0,0,0) -- (1,0,0) node[anchor=north east]{$y$};
    \draw[thick,->] (0,0,0) -- (0,1,0) node[anchor=north west]{$z$};
    \draw[thick,->] (0,0,0) -- (0,0,1) node[anchor=south]{$x$};
    \end{scope}
\begin{scope}[scale=1]
\draw [thick, <->] (-3.3, 0) -- (-3.3, 3); 
\draw [very thick,\UVcolor,  fill=\UVcolor]
(-3,0) -- (-1,3) --(2,3) -- (0,0) -- (-3,0) ; 
 \draw [\IRcolor,  fill=\IRcolor]
(0,0) -- (2,3) --(5,3) -- (3,0) -- (0,0) ; 
\draw [] (-3,0) -- (3,0)  ;
\draw [] (-1,3) -- (5,3)  ;
\draw [] (0,0) -- (2,3)  ;
\draw [fill=\UVcolor, opacity =0.8] (-3,0) -- (-3,3) --(0,3) -- (0,0) --(-3,0); 
\draw [fill=\UVcolor, opacity =0.5] (-1,3) -- (-1,6) --(2,6) -- (2,3) ; 
\draw [fill=\UVcolor, opacity =0.5]
(-3,3) -- (-1,6) --(2,6) -- (0,3) -- (-3,3) ; 
\draw [fill=\IRcolor, opacity =0.5] (2,3) -- (2,6) --(5,6) -- (5,3) ; 
\draw [fill=\IRcolor, opacity =0.5]
(0,3) -- (2,6) --(5,6) -- (3,3) -- (0,3) ; 
\draw [ fill=\IRcolor, opacity =0.8] (3,0) -- (3,3) --(5,6) -- (5,3) ; 
\draw [fill=Purple, opacity =0.8] (0,0) -- (0,3) --(2,6) -- (2,3) ; 
\draw [ fill=\IRcolor, opacity =0.5] (0,0) -- (0,3) --(3,3) -- (3,0) -- (0,0) ; 
\draw [Purple, ultra thick] (0,0) -- (0,3);
\draw [Purple, ultra thick] (2,3) -- (2,6);
 \node at (-1.5,1.5) {$\Bsym_\UV$}; 
\node at (1.5, 1.5) {$\Bsym_{\IR}$}; 
 \node at (0.5,4.5) {$\Bphys_{\UV}$}; 
 \node  at (3.5,4.5) {$\Bphys_{\IR}$}; 
\node at (-0.5,3.5) {$\fZ(\cC_{\UV})$}; 
\node at (2.8,3.5) {$\fZ(\cC_\IR)$}; 
\node[above] at (1.2,3) {$\cI_{F}$};
\draw [thick, <->] (3.4,1.5) -- (5.4,3+1.5);
\end{scope}
\begin{scope}[shift= {(9.5,1.5)}]
\draw [thick, fill=\UVcolor, opacity =1] (-3,0) -- (-3,3) --(0,3) -- (0,0) -- (-3,0) ; 
\draw [thick, fill=\IRcolor, opacity =1] (0,0) -- (0,3) --(3,3) -- (3,0) -- (0,0) ; 
\draw[ultra thick, Purple] (0,0) -- (0,3) ;
\node[above] at (0,3) {$\cI_{\RG}$};
 \node at (-1.5,1.5) {$\cT_\UV$}; 
\node at (1.5, 1.5) {$\cT_{\IR}$}; 
\end{scope}
\begin{scope}[shift= {(1,-5)}]
\draw [\UVcolor,  fill=\UVcolor]
(-3,0) -- (-3,3) --(0,3) -- (0,0) -- (-3,0) ; 
 \draw [\IRcolor,  fill=\IRcolor]
(0,0) -- (0,3) --(3,3) -- (3,0) -- (0,0) ; 
\draw [very thick] (-3,0) -- (3,0)  ;
\draw [very thick] (-3,3) -- (3,3)  ;
\draw [Purple, ultra thick] (0,0) -- (0,3)  ;
 \draw [Purple,fill=Purple] (0,0) ellipse (0.05 and 0.05);
  \draw [Purple,fill=Purple] (0,3) ellipse (0.05 and 0.05);
 \node[below] at (-1.5,0) {$\Bsym_\UV$}; 
 \node[above] at (-1.5,3) {$\Bphys_{\UV}$}; 
\node[below] at (1.5,0) {$\Bsym_{\IR}$}; 
 \node[above] at (1.5,3) {$\Bphys_{\IR}$}; 
\node at (-1.5,1.5) {$\fZ(\cC_\UV)$}; 
\node at (1.5,1.5) {$\fZ(\cC_\IR)$}; 
\node[above] at (0,3.1) {$\cI_{F}$};
\draw [thick, <->] (3.3, 0.3) -- (3.3, 2.7); 
\end{scope}
\begin{scope}[shift= {(9.5,-3.5)}]
\draw [very thick] (-3,0) -- (3,0) ;
  \draw [Purple,fill=Purple] (0,0) ellipse (0.07 and 0.07);
  \node[above] at (0,0) {$\cI_{\RG}$};
 \node[above] at (-1.5,0) {$\cT_\UV$}; 
\node[above] at (1.5, 0) {$\cT_{\IR}$}; 
\end{scope}
\draw [dashed]    (5,3) -- (9.5-3,1.5) --  (3,0)  ;
\draw [dashed]    (5,3+3) -- (9.5-3,1.5 +3) --  (3,3)  ;
\draw [dashed] (-3,0) -- (-2,-5); 
\draw [dashed] (3,0) -- (3+1,0-5); 
\draw [dashed] (5,3) -- (3 +1,3-5);
\draw [dashed] (-1,3) -- (-3 +1,3-5);
\draw [dashed] (1+3,-5) -- (9.5-3,-3.5);
\draw [dashed] (1 +3,-5+3) -- (9.5 -3,-3.5);
\draw [dashed, ->-] (9.5-3,1.5) -- (9.5-3,-3.5);
\draw [dashed, ->-] (9.5+3,1.5) -- (9.5+3,-3.5);
\end{tikzpicture}
$$
\caption{
The UV/IR SymTFTs with RG-Interface: in the 3d figure on the top left, we show the SymTFT interface $\cI_F$ between the two topological orders $\fZ (\cC_\UV)$ and $\fZ (\cC_\IR)$. The gapped symmetry boundary conditions $\Bsym_\UV$ and $\Bsym_\IR$ are shown in the front, the physical boundary conditions $\Bphys_\UV$ and $\Bphys_{\IR}$ in the back (both extending along the $y$-$z$-plane). After collapsing the interval (along the $x$ direction) separating the physical and symmetry boundaries, we obtain the figure to the right: This shows the RG-interface  $\cI_{\RG}$ separating the UV theory $\cT_\UV$  and the IR theory $\cT_\IR$. This type of figure will be shown in  section \ref{sec:TF} when discussing  the  RG-interfaces and tensor functors. 
Distinct from this is the projections of the SymTFT onto the $x$-$y$-plane, that we show frequently in the following as a cartoon of the 3d picture. This is shown in the lower part of the figure. Here $\cI_\Fun$ is the SymTFT interface, and only after collapsing further the $x$-interval do we get $\cI_\RG$. Finally in the lower right hand corner, we show the projection onto a 1d picture. \label{fig:ClubSandoVanilla3dUVIR}}
\end{figure}

\twocolumngrid

\vspace{2mm}
\noindent
{\bf LSM Anomalies for (Non-)Invertible Symmetries.}
Finally, discrete spacetime symmetries can be incorporated into the framework, by extending the SymTFT to a symmetry enriched (SE) TQFT: the SESymTFT. We discuss Lieb-Schultz-Mattis (LSM) anomalies and discuss the UV-IR symmetry matching, using such an SESymTFT. The SE is by translations and an automorphism of the Drinfeld center of $\cC$, i.e. the topological defects of the SymTFT. The existence of an LSM-anomaly is a simple criterion in the SymTFT -- namely the absence of an automorphism invariant SPT phase. The anomaly matching uses the exact sequence of categories applied to the SESymTFT. We reproduce known invertible examples and give a non-invertible example, where the UV symmetry is $\Rep (D_8)$ and we consider the $\Z_3$ triality symmetry as automorphism.

\vspace{2mm}
\noindent 
{\bf Plan of the Paper.}
This paper is organized as follows. In section~\ref{sec:TF}, we explain how RG-interfaces between QFTs naturally lead to the mathematical notion of tensor functors. We introduce various structural properties of tensor functors, and use them to define the concept of Anomalous Simple Categories (ASCies), from short exact sequences of symmetry categories. These serve as concrete invariants to detect and quantify anomalies of categorical symmetries. 
Section~\ref{sec: SymTFTTF} begins with a review of key aspects of SymTFT needed for our construction. We then reformulate the existence of a tensor functor between symmetry categories within the SymTFT framework, leading to the Matching Equation. We further reinterpret the previously introduced properties of tensor functors in terms of physical constraints on interfaces. 
In section~\ref{sec:ASC}, we study ASCies in more detail as tools to quantify anomalies for non-invertible symmetries, both from the tensor functor perspective and using the SymTFT formulation. 
We substantiate and illustrate the general framework with numerous examples.  
Section~\ref{sec:WWW} 
studies higher dimensional systems. After a brief review of symmetry fractionalization and transmutation, we apply our framework to study anomaly matching in spacetime dimension $d>2$, and connect to symmetry-preserving gapped phases, relating to results of \cite{Wang:2017loc}. 
In section~\ref{sec:LSM}, we apply our discussion to discrete spacetime symmetries and study LSM-type anomaly matching.
We conclude in section~\ref{sec:conclusion} with a summary and outlook on future directions.

\section{RG-Interfaces and Tensor Functors}
\label{sec:TF}
To motivate the role of tensor functors, we begin by examining their physical significance through a close analogy with RG-interfaces between QFTs. However, it is important to emphasize at the outset that tensor functors are a more general concept and need not arise solely from RG-interfaces.
A tensor functor encodes necessary conditions for UV and IR symmetries to be compatible, serving as a generalized version of the 't Hooft anomaly matching condition.
We will also introduce the notion of Anomalous Simple Categories (ASCies), which provide a natural measure of 't Hooft anomalies for non-invertible symmetries; this discussion will be expanded further in section~\ref{sec:ASC}.

An important point to keep in mind when generalizing to higher dimensions is that many of the concepts we discuss should extend to arbitrary spacetime dimension. However, mathematically precise statements about tensor functors are currently limited to 1+1d, i.e., to the setting of fusion categories. As mentioned in the introduction, we will employ the SymTFT reformulation presented in section \ref{sec: SymTFTTF} to extend our ideas to higher dimensions.

\subsection{From RG-Interfaces to Tensor Functors}

\begin{figure}
$$
\begin{tikzpicture}
\begin{scope}[shift= {(0,0)}]
\draw [\UVcolor,  fill=\UVcolor] 
(-2,0) -- (-2,2) --(0,2) -- (0,0) -- (-2,0) ; 
\draw [\UVcolor,  fill=\UVcolor, opacity = 0.4] 
(0,0) -- (0,2) --(2,2) -- (2,0) -- (0,0) ; 
\draw [very thick,  dashed] (0,0) -- (0,2);
\draw [thick] (-2,1) -- (2, 1);
\node at (-1, 1.5) {$\cT_{\UV}$};
\node at (1, 1.5) {$\cT_{\UV} + \lambda \int \cO$};
 \node[rotate=-90] at (0,-0.7) {{\LARGE $\leadsto$}};
 \node[right] at (0.2,-0.7) {$\RG$};
\node[below] at (-1, 1) {$\D$};
\node[below] at (1, 1) {$\D$};
\end{scope}
\begin{scope}[shift= {(0,-3.7)}]
\draw [\UVcolor,  fill=\UVcolor] 
(-2,0) -- (-2,2) --(0,2) -- (0,0) -- (-2,0) ; 
\draw [\IRcolor,  fill=\IRcolor] 
(0,0) -- (0,2) --(2,2) -- (2,0) -- (0,0) ; 
\draw [ultra thick, Purple] (0,0) -- (0,2);
\node at (-1, 1.5) {$\cT_{\UV}$};
\node at (1, 1.5) {$\cT_{\IR}$};
\node[above] at (0,2) {$\cI_{RG}$};
\draw [thick] (-2,1) -- (2, 1);
\node[below] at (-1, 1) {$\D$};
\node[below] at (1, 1) {$\Fun(\D)$};
\end{scope}
\end{tikzpicture}
$$
\caption{The RG-interface with symmetry defects:  in addition to figure \ref{fig:RGinterface} the topological defects $\D$ that generate the symmetry in the UV are shown. After the RG-flow, the defects in the IR theory are the image $\Fun (\D)$ under a tensor functor $\Fun: \cC_\UV\to \cC_\IR$  between the UV and IR symmetries. 
\label{fig:UVDefoRG}}
\end{figure}
Let us start by providing a clear, physical argument for the mathematical structures we will encounter throughout this section. 
The idea is that a UV-IR map on symmetries 
\be \Fun: \  \cC_{\UV} \to \cC_{\IR}
\ee
can be encoded in an {\bf RG-interface} \cite{Gaiotto:2012np}.
This can be described by turning on a $\cC_{\UV}$-symmetric relevant deformation in half space and flowing to its IR fixed point,
where the symmetry acting faithfully can generically be quite different from the one in the UV. For example, a subsector of the UV symmetry generators might act {trivially}, meaning that all of its corresponding (generalized) charges are screened. On the other end, new symmetries can emerge in the IR, giving rise to further selection rules which are absent in the UV theory. This means that certain IR symmetry charges must be identified under the UV symmetry action.

The complete UV/IR mapping process is far richer, but its key properties may be derived by physical consistency alone.
Indeed, as the process is continuous, topological defects $\D$ of $\cC_{\UV}$ remain topological throughout the half space RG-flow. This gives rise to a map $\Fun$  between UV and IR symmetry defects, which is compatible with the fusion and the structure of topological junctions of defects -- i.e. a tensor functor. This is shown in figure \ref{fig:UVDefoRG}.

A few obvious identifications are in place: 
{\bf UV symmetries can be screened} in the IR, i.e. the associated defects become trivial after the RG-flow. This means the 
{\bf $\Fun$ can have a kernel}, $\ker(\Fun)\not =1$. On the other hand the IR can have {\bf emergent symmetries}, i.e. not all symmetries are images of UV symmetry generators. This means that {\bf$\Fun$ is  not surjective}.

A dual perspective can be gained by discussing the fate of charged operators. In this discussion, we will only consider genuine, possibly extended, operators.
Operators at the IR fixed point always have a description in terms of their UV counterparts \cite{Zamolodchikov:1987ti} and their charge $q$ corresponds to a well-defined representation $\Fun^*(q)$ under the UV symmetry:
\be
\begin{tikzpicture}[scale=0.75,baseline={(0,0.75)}]
  \filldraw[color= \IRcolor, fill=\IRcolor] 
  (0,0) -- (2,0) -- (2,2) -- (0,2) -- cycle;
   \filldraw[color= \UVcolor, fill=\UVcolor] 
  (-2,0) -- (0,0) -- (0,2) -- (-2,2) -- cycle;
   \draw[Purple, ultra thick] (0,0) node[below,black] 
        {$\cI_{\RG}$} -- (0,2) ;
    \draw[fill=black] (1,1) node[above] {$q$} circle (0.05);
    \node at (2.75,1) {\LARGE$\leadsto$};
    \end{tikzpicture} 
      \begin{tikzpicture}[scale=0.75,baseline={(0,0.75)}]
         \filldraw[color= \UVcolor, fill=\UVcolor] 
  (-2,0) -- (0,0) -- (0,2) -- (-2,2) -- cycle;
           \filldraw[color= \IRcolor] (0,0) -- (2,0) -- (2,2) -- (0,2) -- cycle;
        \draw[Purple, ultra thick] (0,0) node[below,black] 
        {$\cI_\RG$} -- (0,2) ;
            \draw[fill=black] (-1,1) node[above] {$\Fun^*(q)$} circle (0.05);
\end{tikzpicture}
\ee
On the other hand, UV operators can decouple along the RG and can hence be absent from the low energy Hilbert space. Their charges can still exist as long as they are confined on the RG-interface though. This makes clear that, while the map $\Fun$ between symmetries goes $\cC_{\UV} \to \cC_{\IR}$, the representations of genuine operators are instead pulled back in the opposite direction. The significance of the dual map $\Fun^*$ will be apparent once we switch gears to the SymTFT.

\begin{figure}[t!]
 $$
\begin{tikzpicture}
\begin{scope}[shift= {(0,0)}]
  \filldraw[color= \UVcolor, fill=\UVcolor, opacity=0.4] 
  (0,0) -- (2,0) -- (2,2) -- (0,2) -- cycle;
   \filldraw[color= \UVcolor, fill=\UVcolor] 
  (-2,0) -- (0,0) -- (0,2) -- (-2,2) -- cycle;
    \draw[very thick,  dashed] (0,0) -- (0,2);
    \draw[thick] (-1,1) -- (2,1); 
    \draw[thick] (-2,1.5) -- (-1,1); 
    \node at (-1.7, 1.7)  {$\D_1$}; 
    \draw[thick] (-2,0.5)  -- (-1,1);
    \node at (-1.7, 0.3)  {$\D_2$}; 
    \draw[fill=black] (-1,1) circle (0.05);
    \node[above] at (-0.5,1) {$\D_{3}$};
     \node[above] at (1,1) {$\D_3$};
     \node[rotate=-90] at (0,-1) {\LARGE $\leadsto$};
          \node at (2.4,1) {$=$};
\end{scope}
\begin{scope}[shift= {(4.7,0)}]
 \filldraw[color= \UVcolor, fill=\UVcolor, opacity=0.4] 
  (0,0) -- (2,0) -- (2,2) -- (0,2) -- cycle;
   \filldraw[color= \UVcolor, fill=\UVcolor] 
  (-2,0) -- (0,0) -- (0,2) -- (-2,2) -- cycle;
    \draw[very thick, dashed] (0,0) -- (0,2);
    \draw[thick] (1,1) -- (2,1); 
    \draw[thick] (-2,1.5)  -- (0,1.5) -- (1,1); 
     \node at (-1.5, 1.7)  {$\D_1$}; 
    \draw[thick] (-2,0.5)  -- (0,0.5) -- (1,1); 
     \node at (-1.5, 0.2)  {$\D_2$}; 
    \draw[fill=black] (1,1) circle (0.05);
     \node[above] at (1.5,1) {$\D_3$};
       \node[rotate=-90] at (0,-1) {\LARGE $\leadsto$};
\end{scope}
\begin{scope}[shift= {(0,-4)}]
  \filldraw[color= \IRcolor, fill=\IRcolor] 
  (0,0) -- (2,0) -- (2,2) -- (0,2) -- cycle;
   \filldraw[color= \UVcolor, fill=\UVcolor] 
  (-2,0) -- (0,0) -- (0,2) -- (-2,2) -- cycle;
    \draw[ultra thick,  Purple] (0,0) -- (0,2);
    \draw[thick] (-1,1) -- (2,1); 
    \draw[thick] (-2,1.5) -- (-1,1); 
    \node at (-1.7, 1.7)  {$\D_1$}; 
    \draw[thick] (-2,0.5)  -- (-1,1);
    \node at (-1.7, 0.3)  {$\D_2$}; 
    \draw[fill=black] (-1,1) circle (0.05);
    \node[above] at (-0.5,1) {$\D_{3}$};
     \node[above] at (1,1) {$\Fun(\D_3)$};
     \draw[fill=\IRcolor] (0,1) circle (0.05);
      \node at (2.4,1) {$=$};
\end{scope}
\begin{scope}[shift= {(4.7,-4)}]
 \filldraw[color=\IRcolor, fill=\IRcolor] 
  (0,0) -- (2,0) -- (2,2) -- (0,2) -- cycle;
   \filldraw[color= \UVcolor, fill=\UVcolor] 
  (-2,0) -- (0,0) -- (0,2) -- (-2,2) -- cycle;
    \draw[ultra thick,  Purple] (0,0) -- (0,2);
    \draw[thick] (1,1) -- (2,1); 
    \draw[thick] (-2,1.5)  -- (0,1.5) -- (1,1); 
     \node at (-1.5, 1.7)  {$\D_1$}; 
    \draw[thick] (-2,0.5)  -- (0,0.5) -- (1,1); 
     \node at (-1.5, 0.2)  {$\D_2$}; 
    \draw[fill=black] (1,1) circle (0.05);
    \draw[fill=\IRcolor] (0,1.5) circle (0.05);
    \draw[fill=\IRcolor] (0,0.5) circle (0.05);
    \node[right] at (0,1.6) {$\Fun (\D_1)$};
    \node[right] at (0,0.4) {$\Fun (\D_2)$};
     \node[above] at (1.5,1) {$\Fun(\D_3)$};
     \node at (-0.75, -0.4) {$\times J(\D_1, \D_2, \D_3)$};  
\end{scope}
\end{tikzpicture}
$$
\caption{Passing a UV topological junction through the RG-interface $\cI$ gives rise to a linear map $J(\D_1,\D_2,\D_3)$ between the spaces of UV and IR topological junctions. For fusion categories these are vector spaces, and the map $J(\D_1,\D_2,\D_3)$ is implemented by matrix multiplication.\label{RGinterface_tensorproduct}}
\end{figure}
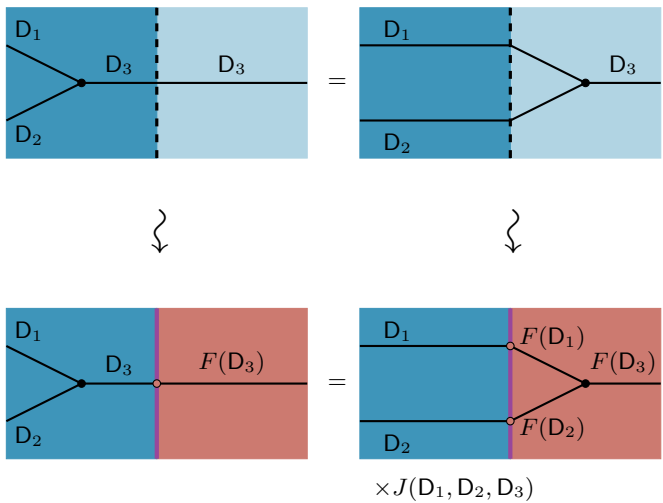

The observation that UV symmetry defects $\D$ remain topological throughout the RG-flow furthermore allows us to conclude that the RG-interface $\cI$ induces a map between the fusion structures in the UV and IR, see figure  \ref{RGinterface_tensorproduct}. 
It is by now established that the correct mathematical structure that describes symmetries in QFTs is that of the (higher) {\bf tensor  categories}. 
The fusion of topological defects is modeled by the tensor product, while higher-data associated fusion of the topological junctions are encoded in the (higher) morphisms of the category. 
For instance, for 1-categories the associativity condition on point junctions give rise to the {\bf F-symbols} that are analogs of 't Hooft anomalies. In that context $J(\D_1,\D_2,\D_3)$ are unitary matrices implementing a linear map between vector spaces 
$V_{\D_1 , \D_2}^{\D_3}=\Hom(\D_1\otimes \D_2, \D_3)$ and $V_{\Fun(\D_1), \Fun(D_2)}^{\Fun(\D_3)}=\text{Hom}(\Fun(\D_1) \otimes \Fun(\D_2), \Fun(D_3))$ 
\be
J (\D_1, \D_2, \D_3): \qquad V_{\D_1 , \D_2}^{\D_3} \to V_{\Fun(\D_1), \Fun(D_2)}^{\Fun(\D_3)}\,,
\ee
that satisfies certain compatibility conditions (see equation~\eqref{eqn:monoidal_structure_compatibility}). Different $J$s give rise to IR symmetries with the same fusion algebras, but with different F-symbols $\cF$:
\be \ba 
&{{\cF_{\IR}}^{\Fun(\D_1),\Fun(\D_2),\Fun(\D_3)}_{ \ \Fun(\D_4)}}_{\Fun(\D_5),\Fun(\D_6)} \cr
&={{\cF_{\UV}}^{\D_1,\D_2,\D_3}_{\ \D_4}}_{\D_5,\D_6} \, 
\left[J(\D_1,\D_2,\D_5) \right. \\ 
&\quad \left. J(\D_3,\D_5,\D_4) J^{-1}(\D_1,\D_6,\D_4) J^{-1}(\D_2,\D_3,\D_6) \right] \, , \ea \ee
where $\D_1' = \Fun(D_1)$ etc. 
Physically, this often coincides with the emergence of a 't Hooft anomaly for the IR symmetry. 
On the other hand, from the point of view of the UV symmetry, the linear maps $J(\D_1,\D_2,\D_3)$ correspond to a gauge transformation. Thus, the UV 't Hooft anomalies are preserved by the tensor functor, and the IR $F$-symbols can always be pulled back into consistent UV $F$-symbols by pre-composing then with $\Fun$. This is the essence of the 't Hooft anomaly-matching condition.

The above properties of $\cI$ (together with the appropriate extension to higher-codimensional junctions for higher categories) define the mathematical structure behind a {\bf tensor functor} \cite{EGNO}. Mathematically these are the natural maps between tensor categories that preserve their structure. Physically, this represents the intuitive idea that any topological operation (fusing, braiding) on UV symmetry defects must be mapped into an analogous operation in the IR. {Tensor functors represent the most basic such map, that is compatible with the fusion.}

Clearly, not every interface $\cI$ between $\cT_{\UV}$ and $\cT_{\IR}$ provides us with a tensor functor. As we have noticed, these interfaces must be $\cC_{\UV}$-symmetric. Symmetric interfaces and boundary conditions, and how they are described, have been studied in several recent works \cite{Bhardwaj:2024igy,Copetti:2024onh,Antinucci:2024izg, Antinucci:2025uvj, Choi:2023xjw,Choi:2024tri}. In such a formulation, symmetry breaking on an interface can be understood in terms of topological order parameters, much in the same way as the SSB breaking of a global symmetry is encoded in the IR through topological bulk order parameters \cite{Bhardwaj:2023fca,Bhardwaj:2023idu}. With this intuition in mind, a tensor functor is implemented by an interface which hosts no nontrivial topological charges under $\cC_{\UV}$ on its worldvolume.

\subsection{Tensor Functors} 
\label{sec:tensorfunctors}

Having motivated the relevance of tensor functors from a physics point of view, we will now turn to a more thorough {mathematical} investigation of {their} properties, and their interpretations in the context of generalized symmetries. 
Mathematically, this is standard material in the theory of fusion categories. We invite the interested reader to consult either the textbook \cite{EGNO}, or the growing set of physics-oriented lecture notes \cite{Schafer-Nameki:2023jdn,Shao:2023gho} for an introduction to categorical symmetries. 

The present section is entirely formulated in terms of symmetry categories and tensor functors between them. A physically more intuitive perspective on the same concepts will be presented in  sections \ref{sec: SymTFTTF} and \ref{sec:ASC} in terms of a SymTFT realization. This will aid in developing physical intuition, but more importantly, will provide a natural extension from fusion categories to fusion higher-categories.

A {\bf tensor functor} between two tensor categories
\be
\Fun: \ \cC_1 \to \cC_2 
\ee
is a functor, i.e. a map on objects and morphisms, satisfying certain conditions (the full mathematical definition is given in appendix \ref{app:Math}). 
The most important property is that it is compatible with the tensor structure, i.e. the fusion product: there is an isomorphism 
\be \label{eq: tensoriso}
 J_{\D_1,\D_2}:\quad  \Fun(\D_1) \otimes \Fun(\D_2) \cong 
 \Fun(\D_1 \otimes \D_2)
\ee
for all $\D_1,\D_2\in \cC_1$, satisfying compatibility condition~\eqref{eqn:monoidal_structure_compatibility}. 
Intuitively this is a map on symmetry generators, that is compatible with the fusion product.
A tensor functor can have various additional properties, which all will have a physical relevance in the following.

\vspace{2mm} \noindent \textbf{Injective Functors.} A 
 tensor functor 
$\IFun:  \ \cC_1 \to \cC_2$ is injective if 
\be\label{Finj}
\D_1 \not\cong \D_2 \Rightarrow \IFun (\D_1) \not\cong \IFun (\D_2)\,.
\ee
If furthermore, $\IFun$ is {faithful}, i.e., injective on morphisms, then it is called an \textbf{embedding}. 
This means the symmetry $\cC_1$ can be embedded into the symmetry $\cC_2$. 
For example, if we think of $\cC_1 = \cC_{\UV}$ and $\cC_2 = \cC_{\IR}$, then an embedding functor that is not an isomorphism or an equivalence of categories, indicates emergent symmetries in the IR. 

 \vspace{2mm} 
 \noindent 
 \textbf{Surjective Functors.} On the other hand we can have surjective functors: a tensor functor 
\be \label{Fsurj}
\DFun: \  \cC_1 \to \cC_2 
\ee
is surjective if any object $\D_2$ in $\cC_2$ appears in the image, i.e.  $\D_2 \subset \DFun(\D_1)$ for some $\D_1 \in \cC_1$. 
The notation $\D_2 \subset  \DFun(\D_1)$ means that $\D_2$ can be written in terms of the simple objects of $P(\D_1)$. 
{Surjective tensor functors describe a situation in which the UV symmetry may act non-faithfully in the IR.}

\vspace{1mm} \noindent \textbf{Normal Functors.} Another concept often appearing in the literature is that of a {normal} functor.
A functor 
\begin{equation}
    \Fun: \cC_1 \rightarrow \cC_2
\end{equation}
is said to be normal if for every object $\D_1$ of $\cC_1$, there is a sub-object $\D_1' \subset \D_1$ such that $\Fun(\D_1')$ is the largest trivial sub-object of $\Fun(\D_1)$, i.e. 
\be \label{eq:normal fun}
\Fun (\D_1) = n \, 1 \oplus \cdots \,,\quad 
\Fun (\D_1') = n\, 1  \,,
\ee
for some $n$, where $1$ is the identity object (i.e. the identity symmetry generator) and $\cdots$ does not contain further copies of $1$. If $\D_1$ is simple, {either} $\Fun (\D_1)= n \, 1$ or $1\not\subset \Fun(\D_1)$.
For normal functors between fusion categories, injectivity on objects is equivalent to having trivial kernels, thus such functors naturally appear in defining the analogue of a short exact sequence of groups for a fusion category \cite{bruguieres2011exact}. 
In our applications we will not find any relevant non-normal functors. Thus {we will} safely assume that all functors are normal throughout the paper.
More details on normal functors and examples of non-normal functors can be found in appendix \ref{app:AllThingsShouldBeNormal}.

\vspace{2mm}
\noindent
{\bf Examples.} 
A simple class of examples of tensor functors arise from group homomorphisms $G\to H$ which lift to tensor functors of $\Vec_G \to \Vec_H$.\footnote{When dealing with discrete groups, we will use $G$ to indicate the group, while $\Vec_G^\omega$ stands for the fusion category based on the group $G$, with 't Hooft anomaly $\omega \in H^3(G,U(1))$. This will be important as group homomorphisms will not always extend to interesting tensor functors on the category $\Vec_G^\omega$.} This becomes more interesting in the presence of anomalies. 
For example, the surjective functor
\be
\Fun :\quad \Vec_{\Z_4} \to \Vec_{\Z_2}^{\omega=1}  \,,
\ee
maps a $\bZ_4$ symmetry to a $\Z_2$ symmetry with nontrivial anomaly $\omega \in H^3 (\Z_2, U(1)) \cong \Z_2$. In this case the map on the objects is simply the projection $p:\bZ_4 \rightarrow \bZ_2$ taking $\mod 2$, however, the action on junctions is nontrivial: $J(2,1,3)= - J(1,2,3)= i$.
Notice that the anomaly also maps correctly, as $p^* (\omega) =0$, as required. This functor has been used extensively in the description of intrinsically gapless SPTs in the condensed matter literature \cite{Verresen:2019igf,Thorngren:2020wet, Wen:2022tkg, Bhardwaj:2024qrf}.
On the other hand the functor $\Vec_{\Z_4}^{\omega\not =0} \to \Vec_{\Z_2}^{\omega=1}$ is not a tensor functor, as it does not correctly pull-back the anomaly.

More generally, for any group homomorphism $\varphi: G \rightarrow H$, and anomalies $[\alpha]\in H^3(G,U(1))$ $[\omega]\in H^3(H, U(1))$, there is a tensor functor
\begin{equation}
    \Vec_G^{\alpha} \rightarrow \Vec_H^{\omega}\,,
\end{equation}
as long as $[\varphi^*\omega]=[\alpha]$ in $H^3(G,U(1))$.

On the other hand, for any normal subgroup $N \triangleleft G$ of a finite group $G$, there is also a normal tensor functor \cite{bruguieres2011exact}
\begin{equation}
    \Rep(G) \rightarrow \Rep(N)\,,
\end{equation}
given by restriction.

\subsection{Anomalous Simple Categories (ASCies)} \label{sec:ask def}

Tensor functors provide a key step toward extending familiar concepts  -- such as 't Hooft anomalies --  from symmetry groups to fusion category symmetries. We will show that a special class of surjective tensor functors, which give rise to \textbf{short exact sequences} of tensor categories, can significantly deepen our understanding of 't Hooft anomalies in this broader setting. Intuitively, such short exact sequences describe how anomaly-free (normal) subcategories can be consistently ``forgotten" along an RG-flow. A short exact sequence associated with the maximal such subcategory defines an \textbf{Anomalous Simple Category} (ASCy), which provides a distilled representation of the original 't Hooft anomaly.

To properly define ASCies, we will need to make a brief detour. Consider a surjective (normal) functor $\DFun$ and denote by $\ker (\DFun)$ its kernel. This is a fusion category $\cN$, generated by objects $\D$ such that $\Fun(\D)= n \, 1$.
This allows us to define a {\bf short exact sequence} of tensor categories \cite{bruguieres2011exact}:
\be\label{SES}
\cN \stackrel{\IFun}{\longrightarrow} \cC 
\stackrel{\DFun}{\longrightarrow} \cS \,,
\ee
where $\IFun$ is injective and $\DFun$ is surjective (and normal), and the exactness meaning 
\be
\image (\IFun) = \ker (\DFun)\,.
\ee
Furthermore 
\be\label{exactcond}
\DFun \circ \IFun = \FF  
\ee
where $\FF$ is a fiber functor, i.e. a functor to the trivial category $\Vec$ 
\be
\FF: \quad \cN \to \Vec  \,.
\ee
Whenever such an exact sequence exists, we call $\cN$ a {\bf normal subcategory} of $\cC$. Notice that for fusion categories, the dimensions of categories involved in the short exact sequence must  satisfy \cite{bruguieres2011exact}:
\be \label{eq: dimensionrel}
\dim(\cC) = \dim(\cN) \, \dim(\cS) \, ,
\ee
and $\dim(\cN)$ must be an integer.

The physical interpretation is the following: any normal subcategory $\cN$ is an {\bf anomaly free sub-category}, as it has a fiber functor by (\ref{exactcond}), and moreover it can be forgotten consistently in the IR, {leaving behind a self-consistent symmetry structure $\cS$}. 
We emphasize that not all anomaly free subcategories are normal, as we will see in examples. We will {amply} discuss the physical relevance of this. 
Note that \textbf{not} every normal subgroup $N \triangleleft G$ gives rise to a normal subcategory $\Vec_N$ of $\Vec_G^\omega$ even if $\omega|_N = 0$. 
An example of this is 
\be
\Vec_{\Z_2} \to \Vec_{\Z_4}^{\omega=2}  \,,
\ee
where $\Z_2 \triangleleft \bZ_4$ is non-anomalous but there is no surjective tensor functor from $\Vec_{\Z_4}^{\omega=2}$ to any category whose kernel is $\Vec_{\Z_2}$.

If $\cN$ is the {\bf maximal normal subcategory} of $\cC$, the image of $\DFun$, $\cS$, {is the simplest fusion category satisfying the 't Hooft anomaly matching conditions of $\cC$ and it} captures the anomaly of the {original} symmetry $\cC$.
We will discuss this in more detail in section \ref{sec:ASC}. Here maximal means: there does not exist $\cM$, another normal subcategory of $\cC$, with $\cN \subset \cM$ and $\cM \to\cC \to \cS'$.   

We define a category that does not have a non-trivial normal subcategory to be an {\bf anomalous simple  category (ASCy)}. Any category $\cS$ that fits into an exact sequence (\ref{SES}) with $\cN$ maximal is an ASCy, and captures aspects of the anomaly of $\cC$.
Notably, a given category $\cC$ may admit multiple exact sequences of the form (\ref{SES}) with the same normal subcategories $\cN_i$, but different ACSies $\cS_i$.

Note that maximality is crucial. E.g. relaxing maximality we have the exact sequence of categories
\be
\Vec_{\Z_2} \to \Vec_{\Z_4} \to \Vec_{\Z_2}^{\omega=1}\,,
\ee
where $\Vec_{\Z_2}$ is normal but not maximal. Indeed, {in this case} $\Vec_{\Z_4}$ is itself its maximal normal subcategory.

As we have already mentioned, the pullback operation implements 't Hooft anomaly matching at the level of fusion categories. In the case of ASCies, the pullback $\DFun^*$ can be used to describe the UV anomaly in terms of the anomalies of the ASCies $\cS_i$. Furthermore, as the pullback operation also naturally acts on symmetric gapped phases, described as module categories $\Mod_\cC$ over the UV symmetry:
\be
\DFun^*_i : \quad \Mod_{\cS_i} \longrightarrow \Mod_{\cC} \,.
\ee
ASCies describe the minimal ways in which the UV symmetry \textbf{must} be broken due to its 't Hooft anomaly. We will see detailed realizations of this mechanism in section \ref{sec:ASC}.

\vspace{2mm}
\noindent
{\bf Some low rank ASCies.}
Some examples of low rank ASCies are: 
\begin{itemize}
\item $\Vec$, 
\item $\Vec_{\bZ_2}^{\omega=1}$, 
\item the Fibonacci category $\Fib$,
\item the $\bZ_2$ Tambara-Yamagami categories $\TY (\Z_2, \pm)= \Ising_\pm$,
\item Haagerup category $\cH_3$.
\end{itemize}
Notice that in several of these cases, while an invertible, anomaly-free subcategory exists -- for example $\bZ_2$ in $\Ising$ or $\bZ_3$ in $\cH_3$ -- it is in fact not normal.

\section{Tensor Functors are SymTFT Interfaces}
\label{sec: SymTFTTF}

We have seen that (normal) tensor functors have an important physical relevance. 
However a comprehensive characterization can be challenging. This is when the SymTFT comes to the rescue. We can map the problem of finding tensor functors, normal categories, and ASCies in terms of the SymTFT, where they become compatibility conditions on certain boundary conditions and interfaces, which are relatively straightforward to check and can be systematically explored. 

In this section, after a brief review of the necessary SymTFT concepts, we expand RG-interfaces into a one-dimensional higher system with topological bulk (RG-sandwiches) and then isolate the kinematical part, that is encoded in the so-called SymTFT club quiche \cite{Bhardwaj:2023bbf}. This is comprised of two SymTFTs, separated by a topological interface and a gapped symmetry boundary for each. The condition on the existence of a tensor functor is recast as a ``Matching Equation", between the interface and the symmetry boundaries. This equivalence provides a dictionary between properties of  tensor functors and of SymTFT-quiches. 

The use of SymTFTs is not just a change of language—it provides a framework that readily extends to higher dimensions, where a direct categorical approach via tensor $(d{-}1)$-functors is still lacking.
The SymTFT approach is therefore a vital tool to quantify anomalies in higher-dimensional theories.

\subsection{SymTFT Sandwiches and Quiches}

In this section we will give a very brief summary of the SymTFT. 
The basic premise is to separate the kinematical symmetry aspects, from dynamics. This is done by extending a $d$-dimensional theory with symmetry $\cC$ to $d+1$ spacetime dimensions. The theory in $(d+1)$-dimensions is topological, and is obtained as a flat gauging of $\cC$. For abelian group-like symmetries the resulting theory is simply a Dijkgraaf-Witten theory for the abelian symmetry (with twist if the group was anomalous), which for any  fusion category symmetries becomes the Turaev-Viro TQFT. 
More generally, we will refer to it as  the SymTFT and denote it by $\fZ (\cC)$. 

The most salient features of the SymTFT are as follows: 
A theory $\cT$ with symmetry category $\cC$ is equivalent to the interval compactification of $\fZ (\cC)$ with 
\begin{itemize}
\item Symmetry boundary: $\Bsym$ that is gapped and on which the symmetry category is realized.
\item  Physical boundary: $\Bphys$, which contains all the dynamics.
\end{itemize}

This is usually depicted in terms of the SymTFT ``Sandwich":
\be\label{SymTFTSketch}
\begin{split}
\begin{tikzpicture}
\begin{scope}[shift={(0,0)}]
\draw [\UVcolor,  fill=\UVcolor] 
(0,0) --  (3,0) -- (3,3) --(0,3) -- (0,0); 
\draw [very thick] (0,0) -- (3,0)  ;
\draw [very thick] (0,3) -- (3,3) ;
\node at (0.8,1.7) {$\fZ(\cC)$} ;
\node[left] at (0,0) {$\Bsym$}; 
\node[left] at (0,3) {$\Bphys$};
\draw [very thick, ->-] (1.5,0)  -- (1.5,3) ;
\node[below] at (2, 1.5) {$\Q_a$};
 \draw [black,fill=\Ncolor] (1.5,0) ellipse (0.05 and 0.05);
 \draw [black,fill=\Ncolor] (1.5, 3) ellipse (0.05 and 0.05);
\end{scope}
\begin{scope}[shift={(4,1.5)}]
\node at (-0.5,0) {$=$} ; 
\draw [very thick] (0,0) -- (3,0) ;
\node[above] at (1.5,0) {$\cO_{a-1}$};
 \draw [black,fill=\Ncolor] (1.5,0) ellipse (0.05 and 0.05);
\end{scope}
\end{tikzpicture}
\end{split}
\ee
Every gapped boundary condition of $\fZ(\cC)$ can act as a symmetry boundary $\Bsym$. Different choices of $\Bsym$ give rise to symmetry categories that are related by generalized gauging, i.e. stacking with symmetric TQFTs before gauging. Two symmetries that are related by such a generalized gauging on the other hand have the same SymTFT and the same center. For any fixed $\cC$, $\fZ(\cC)$ has a canonical Dirichlet gapped boundary condition that realizes  the symmetry $\cC$. 

 For our considerations, the genuine topological defects of the SymTFT play a central role. Mathematically they form the Drinfeld center:
\be
\cZ (\cC) = \text{genuine topological defects of} \ \fZ(\cC)  \,.
\ee
Its topological defects of (spacetime) dimension $p$ will be denoted by $\Q_p$. Those that end on both symmetry and physical boundaries, as shown in (\ref{SymTFTSketch}) correspond to (generalized) charges and give rise to operators $\cO_{p-1}$ in the theory $\cT$. 

Gapped boundary conditions have a characterization in terms of Lagrangian algebras of the Drinfeld center $\cZ (\cC)$. For 1+1d, i.e. fusion category symmetries, these have a simple description in terms of non-negative linear combinations of anyons 
\be\label{Langis}
\cL = \bigoplus_{i} n_i a_i  \,,
\ee
summing over simple anyons $a_i \in \cZ (\cC)$, 
satisfying a set of compatibility conditions
\cite{davydov2013witt,Kong:2013aya,EGNO,cong2017hamiltonian}.\footnote{The algebra structure is important as well, and can be different or the same set of anyons. The subsequent analysis can easily be extended to incorporate this situation. }
Notably, a Lagrangian algebra has maximal dimension, i.e. 
\be
\dim(\cZ(\cC))=\dim(\cL)\,.
\ee
Here $d_a$ are the quantum dimensions of the anyons, $\dim(\cZ(\cC)) =\sqrt{\sum d_{a_i}^2}$  and $\dim(\cL):=  \sum_{i}n_i d_{a_i}$.
Lagrangian algebras classify the interfaces from the $\fZ(\cC)$ topological order to the trivial topological order. 

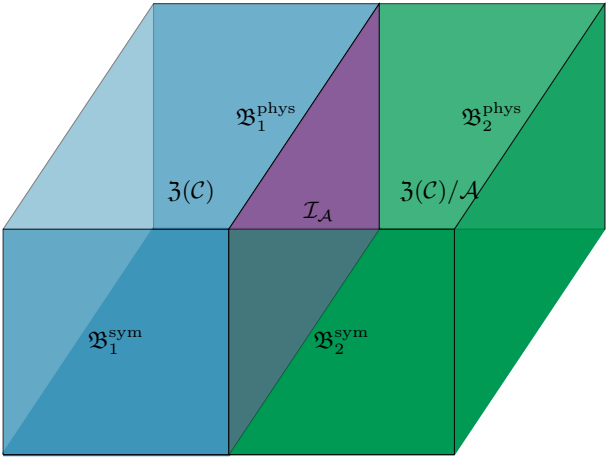
\begin{figure}
$$
\begin{tikzpicture} 
\begin{scope}
\draw [very thick,\UVcolor,  fill=\UVcolor]
(-3,0) -- (-1,3) --(2,3) -- (0,0) -- (-3,0) ; 
 \draw [\Scolor,  fill=\Scolor]
(0,0) -- (2,3) --(5,3) -- (3,0) -- (0,0) ; 
\draw [] (-3,0) -- (3,0)  ;
\draw [] (-1,3) -- (5,3)  ;
\draw [] (0,0) -- (2,3)  ;
\draw [fill=\UVcolor, opacity =0.8] (-3,0) -- (-3,3) --(0,3) -- (0,0) --(-3,0); 
\draw [fill=\UVcolor, opacity =0.5] (-1,3) -- (-1,6) --(2,6) -- (2,3) ; 
\draw [fill=\UVcolor, opacity =0.5]
(-3,3) -- (-1,6) --(2,6) -- (0,3) -- (-3,3) ; 
\draw [fill=\Scolor, opacity =0.5] (2,3) -- (2,6) --(5,6) -- (5,3) ; 
\draw [fill=\Scolor, opacity =0.5]
(0,3) -- (2,6) --(5,6) -- (3,3) -- (0,3) ; 
\draw [ fill=\Scolor, opacity =0.8] (3,0) -- (3,3) --(5,6) -- (5,3) ; 
\draw [fill=Purple, opacity =0.8] (0,0) -- (0,3) --(2,6) -- (2,3) ; 
\draw [ fill=\Scolor, opacity =0.5] (0,0) -- (0,3) --(3,3) -- (3,0) -- (0,0) ; 
 \node at (-1.5,1.5) {$\Bsym_1$}; 
\node at (1.5, 1.5) {$\Bsym_{2}$}; 
 \node at (0.5,4.5) {$\Bphys_{1}$}; 
 \node  at (3.5,4.5) {$\Bphys_{2}$}; 
\node at (-0.5,3.5) {$\fZ(\cC)$}; 
\node at (2.8,3.5) {$\fZ(\cC)/\cA$}; 
\node[above] at (1.2,3) {$\cI_{\cA}$};
\end{scope}
\end{tikzpicture}
$$
\caption{The club sandwich: the 3d figure shows the interface $\cI_\cA$ between the two topological orders $\fZ (\cC)$ and $\fZ (\cC)/\cA$ that are related by condensing an algebra $\cA$ in $\cZ (\cC)$. The gapped symmetry boundary conditions $\Bsym_i$ are shown in the front, the physical boundary conditions $\Bphys_i$ in the back. The projections will be denoted as in figure \ref{fig:ClubSandoVanilla3dUVIR}.
\label{fig:ClubSandoVanilla3d}}
\end{figure}

More generally one can define condensable algebras $\cA$, by relaxing the maximality condition. A condensable algebra means the anyons in that algebra are condensed, those that braid non-trivially with the anyons in the algebra confine, and the remaining ones pass through to a reduced topological order, denoted by $\cZ (\cC)/\cA$ . Again we can write $\cA= \oplus_i n_i a_i$. The dimension of the reduced topological order is then 
\be
\dim(\cZ (\cC)/\cA ) = \dim(\cZ (\cC)) / \dim (\cA) \,.
\ee
The resulting SymTFT takes the form of a {\bf club-sandwich} can be depicted as in figure \ref{fig:ClubSandoVanilla3d}. We will often project this to a 2d picture  from the top.

Note this is a slight modification of the standard SymTFT club sandwich in \cite{Bhardwaj:2023bbf}, used to study gapless phases, in that as for the SymTFT we rotated the diagram by 90 degrees but also we introduced interfaces that separate two choices of gapped boundary conditions $\Bsym_1$ and $\Bsym_2$, as well as two physical boundary conditions. The mathematics of the club sandwich is the same, but the physical application is rather distinct. 

The interface $\cI_\cA$ between the two topological orders is entirely fixed in terms of the condensable algebra. An equivalent, very useful way to think about this is in terms of a gapped boundary condition of the folded theory
\be
\cL_{\cI_\cA} \subset \cZ (\cC) \boxtimes \ol{\cZ (\cC)/\cA} \,.
\ee
We will use both formulations interchangeably in the following. 

Much of the information we require can be obtained from the SymTFT without specifying necessarily a physical boundary $\Bphys$.
Such a SymTFT sandwich without a $\Bphys$ is a {\bf quiche} or for the club-sandwich, a club quiche. 

In 1+1d for fusion categories, this can be systematically and comprehensively explored. 
For higher dimensions, a similarly comprehensive analysis exists in 2+1d for gapped boundary conditions have recently been classified in \cite{Bhardwaj:2024qiv, Wen:2024qsg, Bhardwaj:2025piv} and non-maximal condensable algebras in \cite{Bhardwaj:2025jtf, Wen:2025thg}. In higher dimensions, there is more circumstantial results for condensable algebras, though many results are known from e.g. (generalized) gauging.

\vspace{2mm}
\noindent 
\textbf{Generalized Charges and the Dual Tensor Functor.} 
Before introducing the SymTFT construction of UV/IR maps, we can already use the mathematical properties of the Drinfeld center to gain further insight on its structure. Recall that the Drinfeld center $\cZ(\cC)$ can be presented as pairs $(\D, b^\alpha_\D)$, where $\D \in \cC$ while $b^\alpha_\D$ is the collection of {\bf half-braiding} ($\alpha$ labels the various ones): for any other object $\D' \in \cC$ there is an isomorphisms
\begin{equation}
    b_{\D,\D'}^\alpha : \quad \D \otimes \D' \longrightarrow \D' \otimes \D
\end{equation}
satisfying certain compatibility conditions (see \cite[Definition 7.13.1]{EGNO}). The projection map 
\begin{equation}
    \pi :\quad  \cZ(\cC) \rightarrow \cC  
\end{equation}
is a tensor functor that forgets the half-braiding, namely $\pi(\D,b_\D^\alpha)=\D$. The charges $q\in \cQ_\cC$ are the simple objects in the kernel of $\pi$:
\begin{equation}
    \cQ_\cC = \ker(\pi)\subset \cZ(\cC) \ .
\end{equation}
Simple objects in $\cQ_\cC$ are of the form $(n\, 1,b_{1}^\alpha)\equiv(n \, 1, b^\alpha)$, and are the simple objects in the Lagrangian algebra $\cL_\cC$. Mathematically a charge $q=(n \, 1,b^\alpha)$ can be evaluated on symmetry generators $\D \in \cC$: $q(\D)=b^\alpha_{n \, 1,\D}$ and encodes the quantum numbers of the objects charged under the symmetry.

Any tensor functor $\Fun : \cC_1\rightarrow \cC_2$, whose tensor structure is compatible with the charge (i.e., satisfying equation~\eqref{eqn:compatibility_tensor_charge}) gives a dual map of charges $\Fun ^* :\cQ_{\cC_2} \rightarrow \cQ_{\cC_1}$, defined by precomposition\footnote{Under identifications of vector spaces $\Hom_{\cC_1}(n \, 1_{\cC_1} \otimes \D_1, X_1 \otimes n \, 1_{\cC_1}) \cong \Hom_{\cC_2}(n \, 1_{\cC_2} \otimes F(X_1), F(X_1) \otimes n \, 1_{\cC_2}).$}
\begin{equation}
    \Fun^*(q_2) (\D_1)=q_2(\Fun (\D_1)) \ .
\end{equation}
The physical meaning of this map is clear: while the tensor functor $\Fun: \cC_\UV \rightarrow \cC_\IR$ says that the UV symmetry still acts in the IR, $\Fun^*$ explains how the operators and states of the IR, naturally classified in representations of $\cC_\IR$, transform under the UV symmetry.

\subsection{RG-Quiches}

The problem of determining the existence and properties of a tensor functor $\Fun$ between two given categories $\cC_{\UV}$ and $\cC_{\IR}$ can be dramatically simplified using the SymTFT. A first natural guess would be to expand the RG-interface figure~\ref{fig:RGinterface} into a $(d+1)$-dimensional system: see (\ref{SymTFTSketch}).

The $\UV$ and $\IR$ SymTFTs  are now separated  by a topological interface $\cI_\Fun$ that terminates on the $d$-dimensional RG-interface. The interface $\cI_\Fun$ intersects along $\cIB_\Fun$ with the symmetry boundary, and separates the two boundary conditions $\Bsym_\UV$ and $\Bsym_\IR$. 
We assume that all defects can end on the physical boundary condition. Subject to this, all symmetry properties are independent of the physical boundaries, and are completely captured by the {\bf RG-quiche}: 
\be 
\label{RGQuiche}
\begin{split}
\begin{tikzpicture} 
 \draw [\UVcolor,  fill=\UVcolor]
(0,0) -- (0,3) --(3,3) -- (3,0) -- (0,0) ; 
 \draw [\IRcolor,  fill=\IRcolor]
(3,0) -- (3,3) --(6,3) -- (6,0) -- (3,0) ; 
\draw [very thick] (3,0) -- (3,3)  ;
\draw [very thick] (0,0) -- (6,0)  ;
\node[below] at (1.5,0) {$\Bsym_{\UV}$}; 
\node[below] at (4.5,0) {$\Bsym_{\IR}$}; 
\node at (1.5,1.5) {$\fZ(\cC_\text{UV})$}; 
\node at (4.5,1.5) {$\fZ(\cC_{\IR})$}; 
\node[above] at  (3,3) {$\cI_{\Fun}$};
\draw[fill= black] (3,0)  circle (0.05);
\node[below] at (3,0) {$\cIB_\Fun$};
\end{tikzpicture}
\end{split}
\ee
The SymTFTs with the symmetry boundary, but no physical boundary is the quiche, which is what will be key to our analysis. 

Notice that a generic interface will not in general describe a tensor functor.  The reason is simple: while tensor functors are maps with a precise direction -- which follows the physical RG-flow -- a topological interface $\cI$ in the SymTFT can be equally well interpreted as an interface between $\cZ(\cC_{\UV})$ and $\cZ(\cC_{\IR})$ or the other way around. Furthermore, notice that, while two symmetries may not be connected via any tensor functor, their centers always have interfaces between them, which are described by factorized products of gapped boundary conditions on the two sides. This tension resonates with a simple observation we have made earlier: the RG-interface should {not} break $\cC_\UV$.

\vspace{2mm} \noindent {\bf The SymTFT Matching Equation.}  These two observations come together in a simple equation relating the topological interface $\cI_{\Fun}$ between $\cZ(\cC_{\UV})$ and $\cZ(\cC_{\IR})$ to the existence of a tensor functor
\be
\Fun: \ \cC_{\UV} \to \cC_{\IR} \, .
\ee
We will dub this the {\bf SymTFT Matching Equation (ME)}. It takes the form:
\be
    \label{eq:masterequation}
  \text{(ME)}:\quad   \Bsym_\UV \times \cI_{\Fun} =\Bsym_\IR \,.
\ee
This simple looking equation is highly constraining, and at the same time straightforward to check in concrete examples. 
For fusion categories, 
\eqref{eq:masterequation} can be recast as a simple matrix equation, which makes it computationally straightforward to handle. Consider first the interface $\cI_{\Fun}$ alone (without gapped boundary conditions). This is equivalent, by folding along the interface to the topological order $\fZ(\cC_\UV)\boxtimes \ol{\fZ (\cC_{\IR})}$, with a gapped boundary condition:
\be 
\label{RGQuiche}
\begin{split}
\begin{tikzpicture} 
 \draw [\UVcolor,  fill=\UVcolor]
(0,0) -- (0,2) --(2,2) -- (2,0) -- (0,0) ; 
 \draw [\IRcolor,  fill=\IRcolor]
(2,0) -- (2,2) --(4,2) -- (4,0) -- (2,0) ; 
\draw [very thick] (2,0) -- (2,2)  ;
\node at (1,1) {$\fZ(\cC_\UV)$}; 
\node at (3,1) {$\fZ(\cC_{\IR})$}; 
\node[above] at  (2,2) {$\cI_F$};
 \begin{scope}[shift={(5,0)}]
 \node at (-0.5,1) {$=$}; 
 \draw [Violet,  fill=Violet, opacity =0.6]
(0,0) -- (0,2) --(3,2) -- (3,0) -- (0,0) ; 
\draw [very thick] (3,0) -- (3,2)  ;
\node at (1.5,1) {$\fZ(\cC_\UV)\boxtimes \ol{\fZ (\cC_{\IR})}$}; 
\node[above] at  (3,2) {$\fB_\Fun$};
\end{scope}
\end{tikzpicture}
\end{split}
\ee
The boundary condition $\fB_\Fun$ is defined by a folded Lagrangian algebra
\be 
\cL_{\Fun} \subset \cZ(\cC_\UV)\boxtimes \ol{\cZ (\cC_{\IR})} \,.
\ee
As the 2+1d bulk SymTFTs is a theory of anyons, we can 
expand $\cL_{\Fun}$ in terms of the simple anyons $a_\UV^i$ and $b_\IR^k$ of the two factors:
\be
\label{eqn:folded Lag}
\cL_{\Fun} = \bigoplus_{i, k} n_{i,k}\, a_\UV^i \ol{b^k_{\IR}} \, , \quad n_{i, k} \in \bN_0 \, .
\ee
For instance, if $\cI_{\Fun}$ is obtained by condensing an algebra $\cA_{\Fun} \subset \cZ (\cC_{\UV})$, then 
\be
\cL_{\Fun} = \cA_{\Fun} \otimes 1  \oplus \cdots \,.
\ee
However general RG-interfaces may not be given by condensation of algebras.

The folded Lagrangian defines a map 
\be
\phi_{\Fun}: \ \cZ (\cC_{\UV}) \to \cZ (\cC_{\IR})  \,,
\ee
given by 
\be\label{phiRG}
\phi_{\Fun}:\quad a^i_{\UV} \mapsto \bigoplus_{k} \,n_{i,k} \, b^k_{\IR} \,.
\ee
The anyons that do not appear in $\cL_\Fun$ are confined by the interface, and are mapped to the zero object. This map implements the action of passing an anyon across the interface $\cI_\Fun$.
Denoting by 
\be\ba
\cL_{\UV} & =  \bigoplus_i \,  v_i \, a_{\UV}^i \subset \cZ (\cC_{\UV}) \cr  \cL_{\IR} &= \bigoplus_{k} \,  w_k \, b_{\IR}^k \subset \cZ (\cC_{\IR})
\ea
\ee 
the Lagrangian algebras corresponding to the $\UV$ and $\IR$ symmetry boundaries, respectively, the Matching Equation (\ref{eq:masterequation}) becomes simply
\be\label{missy}
 \text{(ME)}:\quad \phi_{\Fun} (\cL_{\UV}) = \cL_{\IR} \, ,
\ee
which, at the level of objects, becomes the matrix relation:
\be \label{eq: mastermatrix}
\text{(ME)}: \quad \sum_{i} n_{i,k} v_i = w_k \,. 
\ee
The equivalence between the Matching Equation \eqref{eq:masterequation} and the existence of a tensor functor $\Fun : \cC_\UV \rightarrow \cC_\IR$ is proven in appendix~\ref{sec:proofOfME} for fusion 1-categories {and fusion 2-categories} (i.e., finite symmetries in two dimensions  and in three dimensions). In a broader context (e.g. higher dimensions and/or continuous symmetries), already giving a precise definition of a tensor functor is extremely complicated, and while our proof can probably be extended in a rigorous way once all the notions will be rigorously defined, we do not pursue this direction and instead we take \eqref{eq:masterequation} as a definition.

Let us comment now on higher dimensions. Gapped boundary conditions for $d+1$ dimensional TOs are well defined concepts. A topological interface $\cI_{\Fun}$ defines a  Lagrangian algebra of the folded TO. This should then also provide a map between the objects and (higher) morphisms of the centers $\cZ (\cC_{\UV})$ and $\cZ (\cC_{\IR})$. For 2+1d theories with fusion 2-categories such interfaces were in fact classified in \cite{Bhardwaj:2024qiv, Wen:2025thg, Bhardwaj:2025piv, Bhardwaj:2025jtf}. We will use this perspective to define a tensor functor for higher categories, using (\ref{missy}) in appendix \ref{app:F2C}.

\begin{figure}
$$
\begin{tikzpicture} 
\begin{scope}
\draw [\UVcolor,  fill=\UVcolor]
(-3,0) -- (-1,3) --(2,3) -- (0,0) -- (-3,0) ; 
 \draw [\IRcolor,  fill=\IRcolor]
(0,0) -- (2,3) --(5,3) -- (3,0) -- (0,0) ; 
\draw [very thick] (-3,0) -- (3,0)  ;
\draw [very thick] (-1,3) -- (5,3)  ;
\draw[very thick, purple] (0,0) -- (0,3);
\draw (0,0) -- (2,3)  ;
\draw [fill=\UVcolor, opacity =0.5] (-3,0) -- (-3,3) --(0,3) -- (0,0) ; 
\draw [fill=\UVcolor, opacity =0.5] (-1,3) -- (-1,6) --(2,6) -- (2,3) ; 
\draw [fill=\UVcolor, opacity =0.5]
(-3,3) -- (-1,6) --(2,6) -- (0,3) -- (-3,3) ; 
\draw [fill=\IRcolor, opacity =0.5] (3,0) -- (3,3) --(5,6) -- (5,3) ; 
\draw [fill=\IRcolor, opacity =0.5] (2,3) -- (2,6) --(5,6) -- (5,3) ; 
\draw [fill=\IRcolor, opacity =0.5]
(0,3) -- (2,6) --(5,6) -- (3,3) -- (0,3) ; 
\draw [fill=Purple, opacity =0.5] (0,0) -- (0,3) --(2,6) -- (2,3) ; 
 \node[below] at (-1.5,0) {$\Bsym_\UV$}; 
 
\node[below] at (1.5,0) {$\Bsym_{\IR}$}; 

\node at (0.2,4.5) {$\fZ(\cC_\UV)$}; 
\node at (3.2,4.5) {$\fZ(\cC_\IR)$}; 
\node[above] at (2,6) {$\cI_{\cA}$};
\draw[very thick] (-2,3) -- (4,3);
\draw[very thick] (-3,1.5) -- (3,1.5); 

\draw[dotted, ->-] (-2,3) -- (-3,1.5); \draw[dotted, ->-] (4,3) -- (3,1.5);
\node[left] at (-2.25,2.5) {$\pi_\UV$};
\node[right] at (3.75,2.5) {$\pi_\IR$};
\node[below] at (-1.5,1.5) {$\D$}; 
\node[below] at (2,1.5) {$\Fun(\D)$};
\draw[fill=black] (0,1.5) circle (0.05);
\draw[fill=purple] (1,3) circle (0.05);
\node[above] at (-0.5,3) {$a^\UV$}; 
\node[above] at (2.5,3) {$b^\IR$};
\node[right] at (0,2) {$\cIB_\Fun$};
\end{scope}
\end{tikzpicture}
$$
\caption{Extracting a tensor functor $\Fun$ from the folded Lagrangian algebra $\cL_\Fun$: unfolding it, results in lines $(a^\UV,b^\IR)$ inside the $\cL_\Fun$ algebra, which can be projected onto the boundary. If \eqref{eq:masterequation} is satisfied, these will give rise to a pair $(\D, \Fun(\D))$ related by the functor $\Fun$. \label{fig:ClubSandoUVIR}}
\end{figure}
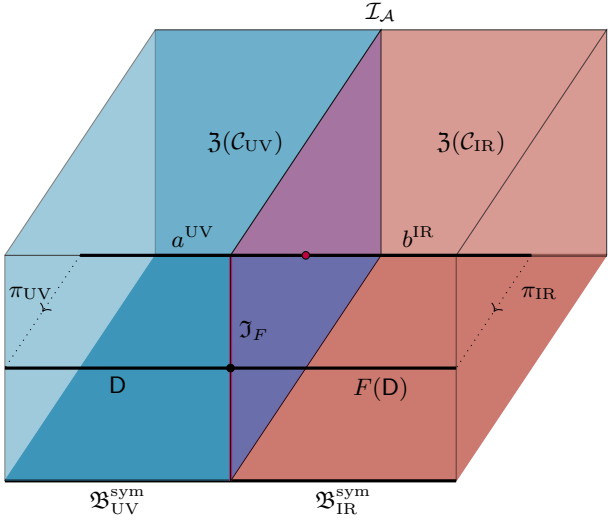

\vspace{2mm}\noindent {\bf Tensor Functor Dictionary and Dual Map.} We will now explain how the matching equation \eqref{eq:masterequation} implements the salient physical properties of RG-interfaces/tensor functors: the map between symmetry generators and the dual map on generalized charges.

First, let us spell out the dictionary between the bulk data, encoded in the algebra $\cL_{\Fun}$ and the structure of the functor $\Fun$. The interface algebra is a pair $(\cL_\Fun, m_{\Fun})$ of an object in the folded theory $\cZ(\cC_\UV)\boxtimes \ol{\cZ (\cC_{\IR})}$ and a multiplication map $m_\Fun : \cL_{\Fun} \otimes \cL_{\Fun} \to \cL_{\Fun}$ satisfying some well known properties \cite{davydov2013witt,Kong:2013aya,EGNO,cong2017hamiltonian}. The boundary projection:
\be
\pi(\cL_{\Fun}) \equiv \left( \pi_{\UV} \boxtimes \ol{\pi}_{\IR}\right) (\cL_\Fun) \, ,
\ee
defines an object in $\cC_{\UV} \boxtimes \ol{\cC_{\IR}}$. It decomposes into simple objects as:
\be
\pi(\cL_{\Fun}) = \bigoplus_{\D \in \cC_{\UV}} (\D , \, \ol{\Fun(\D)} ) \, ,
\ee
where $\Fun(\D)$ will define the tensor functor. Pictorially this is shown in figure \ref{fig:ClubSandoUVIR}.
The object $\pi(\cL_\Fun)$ has the structure of a Frobenius algebra \cite{Fuchs:2002cm}, and $\cIB_\Fun$ is a module over such an algebra. This means that there is a notion of product $J_\Fun$ on the objects $(\D,\ol{\Fun(\D)})$ composing $\pi(\cL_\Fun)$, which will eventually implement the isomorphism \eqref{eq: tensoriso}. The module structure implies that
pairs $(\D,\ol{\Fun(\D)})$ of boundary topological defects are allowed to end topologically on the interface $\cIB_\Fun$, in a way that is consistent with the product $J_\Fun$, see e.g. \cite{Choi:2023xjw}. 

Thus, the algebra structure will precisely give rise to the tensor functor $\Fun$.
To justify this claim consider a bulk anyon $(a^\UV, b^\IR) \in \cZ(\cC_\UV) \boxtimes \cZ(\cC_{\IR})$ belonging to the algebra $\cL_\Fun$ which remains nontrivial under $\pi$. This is mapped to a pair $(\D_{\UV},\D_{\IR})$ of boundary lines joining at $\cIB_\Fun$. We now bend the bulk interface $\cI$ onto $\Bsym_{\UV}$. From the Matching Equation, the topological line $\D_{\UV}$ is mapped into a topological line $\Fun(\D_\UV)$ -- this is a map between the gapped boundary conditions: 
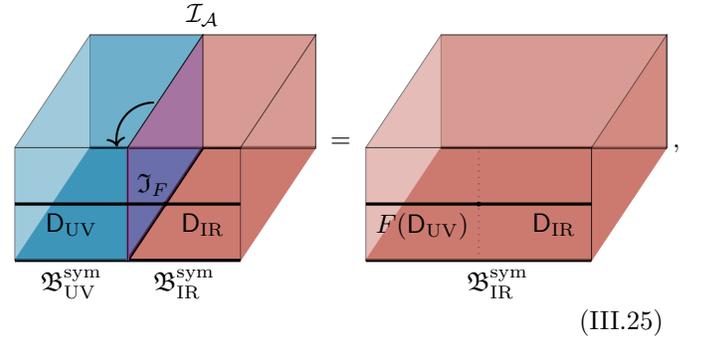
\be
\begin{tikzpicture}[baseline={(0,1.5)}]
\begin{scope}[scale=0.5,baseline={(0,1.5}]
\draw [\UVcolor,  fill=\UVcolor]
(-3,0) -- (-1,3) --(2,3) -- (0,0) -- (-3,0) ; 
 \draw [\IRcolor,  fill=\IRcolor]
(0,0) -- (2,3) --(5,3) -- (3,0) -- (0,0) ; 
\draw [very thick] (-3,0) -- (3,0)  ;
\draw [very thick] (-1,3) -- (5,3)  ;
\draw [very thick] (0,0) -- (2,3)  ;
\draw[very thick, Purple] (0,0)-- (0,3);
\draw [fill=\UVcolor, opacity =0.5] (-3,0) -- (-3,3) --(0,3) -- (0,0) ; 
\draw [fill=\UVcolor, opacity =0.5] (-1,3) -- (-1,6) --(2,6) -- (2,3) ; 
\draw [fill=\UVcolor, opacity =0.5]
(-3,3) -- (-1,6) --(2,6) -- (0,3) -- (-3,3) ; 
\draw [fill=\IRcolor, opacity =0.5] (3,0) -- (3,3) --(5,6) -- (5,3) ; 
\draw [fill=\IRcolor, opacity =0.5] (2,3) -- (2,6) --(5,6) -- (5,3) ; 
\draw [fill=\IRcolor, opacity =0.5]
(0,3) -- (2,6) --(5,6) -- (3,3) -- (0,3) ; 
\draw [fill=Purple, opacity =0.5] (0,0) -- (0,3) --(2,6) -- (2,3) ; 
 \node[below] at (-1.5,0) {$\Bsym_\UV$};  
\node[below] at (1.5,0) {$\Bsym_{\IR}$}; 

\node[above] at (2,6) {$\cI_{\cA}$};

\draw[very thick] (-3,1.5) -- (3,1.5); 
\node[below] at (-1.5,1.5) {$\D_\UV$}; 
\node[below] at (2,1.5) {$\D_\IR$};
\draw[fill=black] (1,1.5) circle (0.05);
\node[right] at (0,2) {$\cIB_\Fun$};
\draw [->,thick,domain=0:100, rotate=90] plot ({3.2+cos(\x)}, {-0.7+ sin(\x)});
\end{scope}
\end{tikzpicture}
\ = \ 
\begin{tikzpicture}[baseline={(0,1.5)}]
\begin{scope}[scale=0.5]
\draw [\IRcolor,  fill=\IRcolor]
(-3,0) -- (-1,3) --(5,3) -- (3,0) -- cycle ; 
\draw [very thick] (-3,0) -- (3,0)  ;
\draw [very thick] (-1,3) -- (5,3)  ;
\draw [dotted] (0,0) -- (0,3)  ;
\draw [fill=\IRcolor, opacity =0.5] (-3,0) -- (-3,3) --(3,3) -- (3,0) ; 
\draw [fill=\IRcolor, opacity =0.5] (-1,3) -- (-1,6) --(5,6) -- (5,3) ; 
\draw [fill=\IRcolor, opacity =0.5]
(-3,3) -- (-1,6) --(5,6) -- (5,3) -- (-3,3) ; 
 \draw [fill=\IRcolor, opacity =0.5] (3,0) -- (3,3) --(5,6) -- (5,3) ; 
\node[below] at (0.5,0) {$\Bsym_{\IR}$}; 
\draw[very thick] (-3,1.5) -- (3,1.5); 
\node[below] at (-1.5,1.5) {$\Fun(\D_{\UV})$}; 
\node[below] at (2,1.5) {$\D_\IR$};
\draw[fill=black] (0,1.5) circle (0.05);
\end{scope}
\end{tikzpicture} \, ,
\ee
we conclude that $\D_{\IR}$ are the simple objects belonging to $\Fun(\D_{\UV})$. 

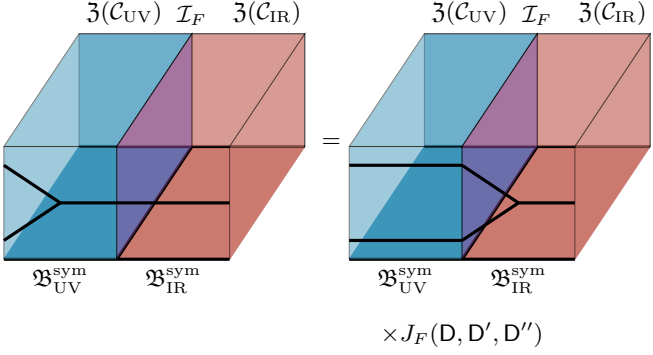
\begin{figure}[t!]
$$
\begin{tikzpicture}[baseline={(0,1.5)}]
\begin{scope}[scale=0.5]
\draw [\UVcolor,  fill=\UVcolor]
(-3,0) -- (-1,3) --(2,3) -- (0,0) -- (-3,0) ; 
 \draw [\IRcolor,  fill=\IRcolor]
(0,0) -- (2,3) --(5,3) -- (3,0) -- (0,0) ; 
\draw [very thick] (-3,0) -- (3,0)  ;
\draw [very thick] (-1,3) -- (5,3)  ;
\draw [very thick] (0,0) -- (2,3)  ;
\draw [fill=\UVcolor, opacity =0.5] (-3,0) -- (-3,3) --(0,3) -- (0,0) ; 
\draw [fill=\UVcolor, opacity =0.5] (-1,3) -- (-1,6) --(2,6) -- (2,3) ; 
\draw [fill=\UVcolor, opacity =0.5]
(-3,3) -- (-1,6) --(2,6) -- (0,3) -- (-3,3) ; 
\draw [fill=\IRcolor, opacity =0.5] (3,0) -- (3,3) --(5,6) -- (5,3) ; 
\draw [fill=\IRcolor, opacity =0.5] (2,3) -- (2,6) --(5,6) -- (5,3) ; 
\draw [fill=\IRcolor, opacity =0.5]
(0,3) -- (2,6) --(5,6) -- (3,3) -- (0,3) ; 
\draw [fill=Purple, opacity =0.5] (0,0) -- (0,3) --(2,6) -- (2,3) ; 
 \node[below] at (-1.5,0) {$\Bsym_\UV$};  
\node[below] at (1.5,0) {$\Bsym_{\IR}$}; 
\node[above] at (0.2,6) {$\fZ(\cC_\UV)$}; 
\node[above] at (4,6) {$\fZ(\cC_\IR)$}; 
\node[above] at (2,6) {$\cI_{\Fun}$};
\draw[very thick] (3,1.5) -- (0,1.5);
\draw[very thick] (-3,0.5) -- (-1.5,1.5); \draw[very thick] (-3,2.5) -- (-1.5,1.5) -- (0,1.5);
\end{scope}
\end{tikzpicture} 
\ = 
\begin{tikzpicture}[baseline={(0,1.5)}]
\begin{scope}[scale=0.5]
\draw [\UVcolor,  fill=\UVcolor]
(-3,0) -- (-1,3) --(2,3) -- (0,0) -- (-3,0) ; 
 \draw [\IRcolor,  fill=\IRcolor]
(0,0) -- (2,3) --(5,3) -- (3,0) -- (0,0) ; 
\draw [very thick] (-3,0) -- (3,0)  ;
\draw [very thick] (-1,3) -- (5,3)  ;
\draw [very thick] (0,0) -- (2,3)  ;
\draw [fill=\UVcolor, opacity =0.5] (-3,0) -- (-3,3) --(0,3) -- (0,0) ; 
\draw [fill=\UVcolor, opacity =0.5] (-1,3) -- (-1,6) --(2,6) -- (2,3) ; 
\draw [fill=\UVcolor, opacity =0.5]
(-3,3) -- (-1,6) --(2,6) -- (0,3) -- (-3,3) ; 
\draw [fill=\IRcolor, opacity =0.5] (3,0) -- (3,3) --(5,6) -- (5,3) ; 
\draw [fill=\IRcolor, opacity =0.5] (2,3) -- (2,6) --(5,6) -- (5,3) ; 
\draw [fill=\IRcolor, opacity =0.5]
(0,3) -- (2,6) --(5,6) -- (3,3) -- (0,3) ; 
\draw [fill=Purple, opacity =0.5] (0,0) -- (0,3) --(2,6) -- (2,3) ; 
 \node[below] at (-1.5,0) {$\Bsym_\UV$};  
\node[below] at (1.5,0) {$\Bsym_{\IR}$}; 
\node[above] at (0.2,6) {$\fZ(\cC_\UV)$}; 
\node[above] at (4,6) {$\fZ(\cC_\IR)$}; 
\node[above] at (2,6) {$\cI_{\Fun}$};
\draw[very thick] (0,0.5) -- (-3,0.5); \draw[very thick] (-3,2.5) -- (0,2.5);
\draw[very thick] (0,0.5) -- (1.5,1.5); \draw[very thick] (0,2.5) -- (1.5,1.5) -- (3,1.5);
\node at (0,-2) {$\times J_\Fun(\D,\D',\D'')$};
\end{scope}
\end{tikzpicture} 
$$
\caption{The map on topological junctions, shown in terms of the black lines, as they pass through the interface $\cI_\Fun$. The map is defined by  $J_\Fun$ in (\ref{JAJA}). To be precise, after unfolding we still need to compose the picture from the right with a 3-valent junction in $\cC_\IR$. \label{fig:smallassfigure}}
\end{figure}

Furthermore, all of the simple objects $\D_\UV$ must appear in $\pi(\cL_\Fun)$ as otherwise we would not have a consistent map between the UV and IR topological boundary conditions, as implied by the matching equation \eqref{eq:masterequation}.
As the maps $\pi_\UV$ and $\pi_\IR$ are themselves tensor functors, the multiplication $m_\Fun$ on $\cL_\Fun$ gives rise to a multiplication map $J_\Fun$ on $\pi(\cL_{\Fun})$. This is part of the general correspondence between Frobenius algebras in $\cC$ and Lagrangian algebras in the center $\cZ(\cC)$ \cite{Fuchs:2002cm,davydov2013witt}. 
Graphically:
\be 
\begin{tikzpicture}[baseline={(-1,1.5)}]
\begin{scope}[scale=0.5]
\draw [Violet,  fill=Violet, opacity=0.6]
(-3,0) -- (-1,3) --(2,3) -- (0,0) -- (-3,0) ; 
\draw [very thick] (-3,0) -- (0,0)  ;
\draw [very thick] (0,0) -- (2,3)  ;
\draw [fill=Violet, opacity =0.5] (-3,0) -- (-3,3) --(0,3) -- (0,0) ; 
\draw [fill=Violet, opacity =0.5] (-1,3) -- (-1,6) --(2,6) -- (2,3) ; 
\draw [fill=Violet, opacity =0.5]
(-3,3) -- (-1,6) --(2,6) -- (0,3) -- (-3,3) ; 
\draw[very thick] (-2,2) -- (-0.5,3); \draw[very thick] (-2,4) -- (-0.5,3) -- (1,3);
\draw[very thick] (-3,0.5) -- (-1.5,1.5); \draw[very thick] (-3,2.5) -- (-1.5,1.5) -- (0,1.5);
\node[below] at (-1.5,0) {$\Bsym_{\UV \boxtimes \ol{\IR}}$};  
\node at (2.75,4.5) {$\fB_\Fun$};
\draw[dotted, ->-] (-2,2) -- (-3,0.5);
\draw[dotted, ->-] (-2,4) -- (-3,2.5);
\draw[dotted, ->-] (1,3) -- (0,1.5);
 \node[right] at (0.65,2.25) {$\pi$};
\node[above] at (-1,6) {$\fZ(\cC_\UV)\boxtimes\ol{\fZ(\cC_{\IR})}$};
\end{scope}
\end{tikzpicture}
= \
\begin{tikzpicture}[baseline={(-1,1.5)}]
\begin{scope}[scale=0.5]
\draw [Violet,  fill=Violet, opacity=0.6]
(-3,0) -- (-1,3) --(2,3) -- (0,0) -- (-3,0) ; 
\draw [very thick] (-3,0) -- (0,0)  ;
\draw [very thick] (0,0) -- (2,3)  ;
\draw [fill=Violet, opacity =0.5] (-3,0) -- (-3,3) --(0,3) -- (0,0) ; 
\draw [fill=Violet, opacity =0.5] (-1,3) -- (-1,6) --(2,6) -- (2,3) ; 
\draw [fill=Violet, opacity =0.5]
(-3,3) -- (-1,6) --(2,6) -- (0,3) -- (-3,3) ; 
\draw[very thick] (-2,2) -- (1,2); \draw[very thick] (-2,4) -- (1,4);
\draw[very thick] (-3,0.5) -- (0,0.5); \draw[very thick] (-3,2.5) -- (0,2.5);

\node[below] at (-1.5,0) {$\Bsym_{\UV \boxtimes \ol{\IR}}$};  
\node at (2.75,4.5) {$\fB_\Fun$};
\draw[dotted, ->-] (1,2) -- (0,0.5);
\draw[dotted, ->-] (1,4) -- (0,2.5);
\draw[dotted, ->-] (-2,2) -- (-3,0.5);
\draw[dotted, ->-] (-2,4) -- (-3,2.5);
\node[right] at (0.75,3.25) {$\pi$};
\node[above] at (-1,6) {$\fZ(\cC_\UV)\boxtimes\ol{\fZ(\cC_{\IR})}$};
\end{scope}
\end{tikzpicture}
\ee
This defines a map
\be \label{JAJA}\ba
J_{\Fun} :  &(\D , \, \Fun(\D) ) \otimes (\D',  \, \Fun(\D')) \\
 &\ \ \ \ \ \ \ \   \longrightarrow (\D \otimes \D', \Fun(\D) \otimes \Fun(\D')) \, . 
\ea \ee
Upon unfolding the picture, $J_{\Fun}$ provides the map in figure \ref{RGinterface_tensorproduct} on topological junctions, see figure \ref{fig:smallassfigure}. 

\begin{figure}[t!]
$
\begin{tikzpicture}[baseline={(0,1.5)}]
\begin{scope}[scale=0.5]
\draw [\UVcolor,  fill=\UVcolor]
(-3,0) -- (-1,3) --(2,3) -- (0,0) -- (-3,0) ; 
 \draw [\IRcolor,  fill=\IRcolor]
(0,0) -- (2,3) --(5,3) -- (3,0) -- (0,0) ; 
\draw [very thick] (-3,0) -- (3,0)  ;
\draw [very thick] (-1,3) -- (5,3)  ;
\draw [very thick] (0,0) -- (2,3)  ;
\draw [fill=\UVcolor, opacity =0.5] (-3,0) -- (-3,3) --(0,3) -- (0,0) ; 
\draw [fill=\UVcolor, opacity =0.5] (-1,3) -- (-1,6) --(2,6) -- (2,3) ; 
\draw [fill=\UVcolor, opacity =0.5]
(-3,3) -- (-1,6) --(2,6) -- (0,3) -- (-3,3) ; 
\draw [fill=\IRcolor, opacity =0.5] (3,0) -- (3,3) --(5,6) -- (5,3) ; 
\draw [fill=\IRcolor, opacity =0.5] (2,3) -- (2,6) --(5,6) -- (5,3) ; 
\draw [fill=\IRcolor, opacity =0.5]
(0,3) -- (2,6) --(5,6) -- (3,3) -- (0,3) ; 
\draw[very thick, rounded corners] (1,3.5) -- (-0.3, 3.5) -- (-0.3,1.5);
\draw [black,fill=black](-0.3,1.5) ellipse (0.05 and 0.05);
\draw [black,fill=black] (1,3.5) ellipse (0.05 and 0.05);
\draw [fill=Purple, opacity =0.5] (0,0) -- (0,3) --(2,6) -- (2,3) ; 
 \node[below] at (-1.5,0) {$\Bsym_\UV$};  
\node[below] at (1.5,0) {$\Bsym_{\IR}$}; 
\node[above] at (0.2,6) {$\fZ(\cC_\UV)$}; 
\node[above] at (4,6) {$\fZ(\cC_\IR)$}; 
\node[above] at (2,6) {$\cI_{\Fun}$};
\draw[very thick] (-2.5,0) -- (-0.5,3);
\node at (-1.25,1) {$\D$};
\node[above] at (-0.3,3.5) {$q$};
\node at (5.5,3) {$=$};
\end{scope}
\begin{scope}[scale=0.5, shift={(9,0)}]
\draw [\UVcolor,  fill=\UVcolor]
(-3,0) -- (-1,3) --(2,3) -- (0,0) -- (-3,0) ; 
 \draw [\IRcolor,  fill=\IRcolor]
(0,0) -- (2,3) --(5,3) -- (3,0) -- (0,0) ; 
\draw [very thick] (-3,0) -- (3,0)  ;
\draw [very thick] (-1,3) -- (5,3)  ;
\draw [very thick] (0,0) -- (2,3)  ;
\draw [fill=\UVcolor, opacity =0.5] (-3,0) -- (-3,3) --(0,3) -- (0,0) ; 
\draw [fill=\UVcolor, opacity =0.5] (-1,3) -- (-1,6) --(2,6) -- (2,3) ; 
\draw [fill=\UVcolor, opacity =0.5]
(-3,3) -- (-1,6) --(2,6) -- (0,3) -- (-3,3) ; 
\draw [fill=\IRcolor, opacity =0.5] (3,0) -- (3,3) --(5,6) -- (5,3) ; 
\draw [fill=\IRcolor, opacity =0.5] (2,3) -- (2,6) --(5,6) -- (5,3) ; 
\draw [fill=\IRcolor, opacity =0.5]
(0,3) -- (2,6) --(5,6) -- (3,3) -- (0,3) ; 
\draw[very thick, rounded corners] (1,3.5) -- (-1.5, 3.5) -- (-1.5,1.5);
\draw [black,fill=black](-1.5,1.5) ellipse (0.05 and 0.05);
\draw [black,fill=black] (1,3.5) ellipse (0.05 and 0.05);
\draw [fill=Purple, opacity =0.5] (0,0) -- (0,3) --(2,6) -- (2,3) ; 
 \node[below] at (-1.5,0) {$\Bsym_\UV$};  
\node[below] at (1.5,0) {$\Bsym_{\IR}$}; 
\node[above] at (0.2,6) {$\fZ(\cC_\UV)$}; 
\node[above] at (4,6) {$\fZ(\cC_\IR)$}; 
\node[above] at (2,6) {$\cI_{\Fun}$};
\draw[very thick] (-1.5,0) -- (0.5,3);
\node at (-0.35,1) {$\D$};
\node[above] at (-0.75,3.75) {$q$};
\node at (0,-2) {$\times \, q(\D)$};
\end{scope}
\end{tikzpicture} 
$
\caption{If the matching equation \eqref{eq:masterequation} is not satisfied, some UV charges $q$ give rise to topological edge modes on the interface $\cIB_\Fun$. The half-brading ensures that the UV symmetry $\cC_\UV$ then is broken by the interface. \label{fig:bigassfigure}}
\end{figure}
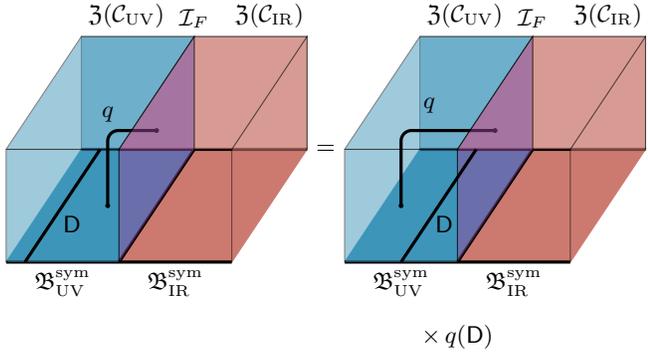
The matching equation \eqref{eq:masterequation}, furthermore, implements correctly the dual map $\Fun^*$ on charges. Consider a charge $q^\IR$ in $\cZ(\cC_{\IR})$. This is an anyon $b^\IR$ inside the Lagrangian algebra $\cL_{\IR}$. Then {$\Fun^*(b^\IR)$} must now be a -- possibly non-simple -- UV generalized charge, as a consequence of \eqref{eq: mastermatrix}.

The dictionary we have just proposed provides a convenient algorithm to extract the data of the tensor functor $\Fun$ from those of a SymTFT interface $\cI_\Fun$ satisfying the matching equation. We will make use of it in several instances throughout the rest of the paper.

\vspace{2mm} \noindent {\bf Symmetric Interfaces from SymTFT.} Let us now come to the interpretation of the tensor functor as a $\cC_{\UV}$-symmetric interface. Notice that the product on the LHS of \eqref{eq:masterequation} defines a map between gapped boundary conditions of $\cZ(\cC_{\UV})$ and $\cZ(\cC_{\IR})$. In general, the resulting gapped BC is not indecomposable, as noticed for example in \cite{Bhardwaj:2023bbf}. This translates into the existence of nontrivial charges $q \, \in \, \cZ(\cC_{\UV})$ which can terminate on both $\Bsym_{\UV}$ and $\cI_{\Fun}$. According to \cite{Bhardwaj:2024igy} these describe topological order parameters for the interface $\cIB_\Fun$, which are half-braidings on $\cC_{\UV}$, see figure \ref{fig:bigassfigure}.

Interfaces for which $\Bsym_{\UV} \times \cI$ is not indecomposable thus break the $\cC_{\UV}$ symmetry and cannot define a tensor functor between $\cC_{\UV}$ and $\cC_{\IR}$.

This is elegantly avoided in the case of a tensor functor through the dual map on charges. Consider an IR charge $b^\IR$ and its image $\Fun^*(b^\IR)$. Simple objects in $\Fun^*(b^\IR) \otimes b^\IR$ define the topological edge modes on the interface $\cIB_\Fun$.  
However, these edge modes are neutral with respect to the $\cC_{\UV}$ symmetry, since both $\Fun^*(b^\IR)$ and $b^\IR$ are contained within the same interface Lagrangian algebra $\cL_\Fun$, indicating that the $\cC_\UV$ symmetry remains unbroken in the presence of the interface.

\subsection{From Tensor Functors to RG-Interfaces}

The physical relevance of tensor functors becomes apparent when considering specific properties, such as injectivity, surjectivity and normal functors. 
We will identify these properties in the SymTFT RG-Interface picture, using the definition (\ref{missy}).
We establish a dictionary between basic properties of tensor functors that were discussed in section \ref{sec:tensorfunctors} and their counterparts in terms of interfaces. We furthermore provide the physical interpretation, which will be put to use in the next sections.

\subsubsection{Injective Functors}  
\label{sec:injective}

We defined an injective functor $\Fun$ (or embedding) in~(\ref{Finj}). Recall that this captures the case when the IR symmetry $\cC_\IR$ is larger than the UV symmetry $\cC_\UV$, allowing for {\bf emergent symmetries}, while the UV symmetry still acts faithfully in the IR.  
Dually, 
\be
\Fun^* :\  \cQ_{\IR} \rightarrow \cQ_{\UV}
\ee
is surjective, with kernel given by the set of IR charges that are neutral under the UV symmetry.

An injective functor $\IFun$ is realized by a {\bf condensation interface} $\cI_\IFun$, specified by a condensable algebra 
\be
\cA_\IFun \subset \cL_{\IR} \subset \cZ (\cC_{\IR}) \,,
\ee 
where $\cL_{\IR}$ denotes the symmetry boundary. Condensing the anyons in $\cA_\IFun$ within $\cZ(\cC_\IR)$ yields the new theory $\cZ(\cC_\UV) = \cZ(\cC_\IR)/\cA_\Fun$. We refer to a condensable algebra $\cA \subset \cL_{\IR}$ as an {\bf electric algebra} (with respect to $\cL_{\IR})$.

This algebra has a simple physical interpretation: 
$\IFun:\cC_\UV \rightarrow \cC_\IR$ is an embedding if and only if the dual functor on the charges $\IFun^*: \cQ_{\IR} \rightarrow \cQ_{\UV}$ is surjective.  The algebra {$\cA_\IFun \subset \cL_{\IR}$} is the kernel of this map, i.e. the set of IR charges that are neutral under the UV symmetry. The SymTFT depiction is as follows:
\be 
\label{RGQuicheInj}
\begin{split}
\begin{tikzpicture} 
 \begin{scope}[shift={(1,0)}]
 \draw [\UVcolor,  fill=\UVcolor]
(-0.5,0) -- (-0.5,3) --(3,3) -- (3,0) -- (-0.5,0) ; 
 \draw [\IRcolor,  fill=\IRcolor]
(3,0) -- (3,3) --(6,3) -- (6,0) -- (3,0) ; 
\draw [very thick] (3,0) -- (3,3)  ;
\draw [very thick] (-0.5,0) -- (6,0)  ;
\node[below] at (1.5,0) {$\Bsym_{\UV}: \, \cL_\UV$}; 
\node[below] at (4.5,0) {$\Bsym_{\IR}:\, \cL_\IR$}; 
\node at (1.3,1.5) {$\fZ(\cC_\text{UV})=\fZ(\cC_\IR)/\cA_{\IFun}$}; 
\node at (4.5,1.5) {$\fZ(\cC_{\IR})$}; 
\node[above] at  (3,3) {$\cI_{\IFun}$};
\draw[thick, rounded corners](3,1.2) -- (4.2,1.2) -- (4.2,0);
\draw [black,fill=black] (3,1.2) ellipse (0.05 and 0.05);
\draw [black,fill=black] (4.2, 0) ellipse (0.05 and 0.05);
\node[right] at  (4.2, 0.6) {$a\in \cA_{\IFun}$};
\end{scope}
\end{tikzpicture}
\end{split}
\ee
All defects in $\cA_\IFun$ become trivial in $\fZ(\cC_\UV)$, while those that braid non-trivially with {$\cA_\IFun$} are projected out (or confined) by the interface.

The left-hand side of~\eqref{missy} corresponds to a boundary condition for $\fZ(\cC_\UV)$, whose Lagrangian algebra is obtained by \textbf{sequentially gauging} $\cA_\IFun$ and then $\cL_{\UV}$. By the Matching Equation, this must coincide with the condensation of $\cL_\IR$. This is only possible if  $\cA_\IFun$ is a subalgebra of $\cL_{\IR}$. That is, $\cA_\IFun$ must be an electric condensable algebra (with respect to $\cL_{\IR}$).

We conclude by noticing that our discussion gives a simple strategy to derive several examples of injective tensor functors: given an IR SymTFT $\fZ(\cC_\IR)$, we condense an algebra {$\cA_\IFun$} of our choosing. This gives rise to a UV SymTFT $\fZ(\cC_\UV)$, for which we choose a symmetry boundary $\Bsym_\UV$. As the condensation happened on the right, {$\Bsym_\UV \times \cI_\IFun$} must be an indecomposable IR boundary condition $\fB_\IR$, corresponding to a symmetry {$\cC_\IR$}. Choosing $\Bsym_\IR = \fB_\IR$ we will always define an injective tensor functor {$\IFun: \cC_\UV \to \cC_\IR$}.  We will apply this logic e.g. in section \ref{sec:WWW}.

\subsubsection{Surjective Functors} 

In the absence of emergent symmetries, the tensor functor between UV and IR symmetries  
\be
\DFun: \cC_\UV \rightarrow \cC_\IR
\ee
is surjective, see~(\ref{Fsurj}). It may have a kernel $\ker(\DFun) \subset \cC_\UV$, encoding the UV symmetries that act trivially at long distances.
The corresponding interface is a condensation interface $\cI_\DFun$ from $\fZ(\cC_\UV)$ to $\fZ(\cC_\IR)$, specified by a condensable algebra 
\[
\cA_\DFun \subset \cZ(\cC_{\UV}).
\]
This setup contrasts with the previous (injective) case: here the algebra $\cA_\DFun$ lives in $\cZ(\cC_\UV)$, rather than in $\cZ(\cC_\IR)$. However, the situation is not symmetric. Equation~\eqref{eq:masterequation} implies that the composed boundary $\Bphys_\UV \times \cI_\DFun$ must define a simple boundary condition of $\fZ(\cC_\IR)$, which imposes the constraint
\be
\cL_{\UV} \cap \cA_\DFun = \{1\} \,,
\ee
i.e., the condensable algebra must intersect the UV Lagrangian algebra only in the trivial object. We refer to such $\cA_\DFun$ as a {\bf magnetic algebra} (with respect to $\cL_{\UV}$). The SymTFT setup describing surjective functors is depicted as follows:
\be 
\label{RGQuicheSurj}
\begin{split}
\begin{tikzpicture} 
 \begin{scope}[shift={(1,0)}]
 \draw [\UVcolor,  fill=\UVcolor]
(0,0) -- (0,3) --(3,3) -- (3,0) -- (0,0) ; 
 \draw [\IRcolor,  fill=\IRcolor]
(3,0) -- (3,3) --(6.5,3) -- (6.5,0) -- (3,0) ; 
\draw [very thick] (3,0) -- (3,3)  ;
\draw [very thick] (0,0) -- (6.5,0)  ;
\node[below] at (1.5,0) {$\Bsym_{\UV}: \, \cL_\UV$}; 
\node[below] at (4.5,0) {$\Bsym_{\IR}:\, \cL_\IR$}; 
\node at (1.3,1.5) {$\fZ(\cC_\text{UV})$}; 
\node at (4.7,1.5) {$\fZ(\cC_{\IR})= \fZ(\cC_\UV)/\cA_\DFun$}; 
\node[above] at  (3,3) {$\cI_{\DFun}$};
\draw[thick, rounded corners](3,1.2) -- (1.8,1.2) -- (1.8,0);
\draw [black,fill=black] (3,1.2) ellipse (0.05 and 0.05);
\draw [black,fill=black] (1.8, 0) ellipse (0.05 and 0.05);
\node[left] at  (1.8, 0.6) {$1$};
\end{scope}
\end{tikzpicture}
\end{split}
\ee
The intuition is that the magnetic algebra $\cA_\DFun$ represents the set of symmetries that become trivial in the IR. More precisely, once projected to the boundary $\Bsym_\UV$ using $\pi_\UV$, $\cA_\DFun$ becomes $\ker(\DFun)$.

A crucial fact is that, specifying what subsymmetry becomes trivial in the IR does {not} uniquely specify $\cC_\IR$. There can be different magnetic algebras $\cA$ with the same projection onto $\Bsym_\UV$, which give rise to inequivalent TQFTs after condensation. This has an important physical interpretation: the remaining, ``quotient'' symmetry, can have emergent anomalies\footnote{Emergent anomalies recently found many applications in characterizing {intrinsically gapless topological phases} \cite{Verresen:2019igf, Li:2023mmw, Bhardwaj:2024qrf, Antinucci:2024ltv}.}.

\vspace{0.2cm}
\noindent
{\bf Example: Emergent Anomalies in (1+1)d.}
Let us illustrate the above remark with the simplest example of a $(1+1)$d theory with 
\be
\cC_\UV = \Vec_{\bZ_4} \,,
\ee
where only the quotient $\bZ_4 / \bZ_2 \cong \bZ_2$ acts faithfully in the IR. 
The dynamical condition for this symmetry reduction is that all excitations created by operators with $\bZ_4$ charge $1$ are heavy, while those with charge $2$ are light. There are, however, two distinct ways in which this can occur: either with or without an emergent anomaly for the IR $\bZ_2$ symmetry.
This example has already been discussed in section \ref{sec:TF}, we now revisit it from the perspective of the SymTFT.

 The UV SymTFT $\fZ(\Vec_{\bZ_4})$ is the untwisted $\bZ_4$ Dijkgraaf-Witten (DW) theory in $2+1$ dimensions:
\begin{equation}
    S=\frac{2\pi i}{4}\int a \cup\, db  \ ,
\end{equation}
whose topological lines are 
\begin{equation}
    \Q_{(n_{e},n_{m})}=\exp\left(\frac{2\pi in_{e} }{4}\int a +\frac{2\pi in_{m}}{4}\int b\right) \,, 
\end{equation}
with $ \ n_{e} , n_{m}\in \bZ_4$. They have $\bZ_4\times \bZ_4$ fusion rules $\Q_{n_e,n_m}\otimes\Q_{n_e',n_m'} = \Q_{n_e + n_e', n_m + n_m'}$ and spins
\begin{equation}
    \theta_{n_{e},n_{m}}=\exp{\left(\frac{2\pi i n_{e} n_{m}}{4}\right)}\,.
\end{equation}
They are generated by the electric and magnetic lines
\begin{equation}
    e:= \Q_{(1,0)}\ ,\ \ \ \ \ m :=\Q_{(0,1)} \ .
\end{equation}
The canonical Dirichlet boundary $\Bsym_{\text{UV}}$ corresponds to the Lagrangian algebra $\cL_{\bZ_4}=1\oplus e\oplus e^2\oplus e^3$. 

Denoting by $\D^{n}$, $n=0,1,2,3$ the symmetry generators on $\Bsym_\UV$, we have $\D^{n}=\pi _\UV(m^{n}e^{n_{e}})$, for any $n_{e}$. The $\bZ_2$ subgroup is generated by 
\begin{equation}
    \D^2=\pi_\UV(m^2)=\pi_\UV(m^2e^{n_{e}}) \ .
\end{equation}
There are two choices for the RG-interface:
\begin{itemize}
    \item 
    $\cA_0=1\oplus m^2$. The lines in $\cZ(\cC_\IR)$ are generated by
    $1\sim m^2,e':=e^2\sim e^2m^2,m'=m\sim m^3, e'm':=e^2m\sim e^2m^3$ and form the center of a non-anomalous $\bZ_2$, namely $\cZ(\Vec_{\bZ_2})$.
    \item      $\cA_1=1\oplus m^2e^2$. 
Now $\cZ(\cC_\IR)$ is generated by 
$1\sim m^2e^2, e'=e^2\sim m^2, s=em\sim e^3m^3 ,\overline{s}=em^3\sim e^3m$. These form the so-called double semion TQFT -- or twisted $\bZ_2$ DW theory -- and describe the center of an anomalous $\bZ_2$, namely $\cZ(\Vec_{\bZ_2}^1)$.
\end{itemize}
Using \eqref{eq:masterequation} $\Bsym_\IR$ must be in both cases the canonical Dirichlet boundary corresponding to $1\oplus e'$. 

The upshot of our analysis is that there are two different surjective functors 
\be
\DFun_{\omega=0,1} : \Vec_{\bZ_4} \rightarrow \Vec_{\bZ_2} ^{\omega=0,1}
\ee
with same kernel $\ker(\DFun_\omega)=\Vec_{\bZ_2}$, but different IR symmetry: in the first case the IR SymTFT describes an anomaly free $\bZ_2$, while in the second case there is an emergent anomaly for $\bZ_2$. 
We will see in section \ref{sec:TY} that, if $\cC_\UV$ involves non-invertible symmetries, not only the anomaly, but even the symmetry group can be different in the IR, while the trivialized subsymmetry stays the same.

\subsubsection{Fiber Functors}

A fiber functor of a $(d-1)$-category $\cC_\UV$ is a tensor functor $\Fun: \cC_\UV\rightarrow (d-1)\, \Vec$ into the ``trivial" category $(d-1)\Vec$. Physically, it encodes the mapping of a UV symmetry along an RG that ends in a trivially gapped theory, that is, an SPT.

For this reason the absence of a fiber functor is often taken as a definition of an anomalous symmetry \cite{Thorngren:2019iar}. 
While our goal is to give a more fine-grained description of anomalies, \eqref{eq:masterequation} should certainly recover this coarse characterization.  

As the SymTFT for the trivial symmetry $(d-1)\Vec$ is a trivial invertible topological order, the interface $\cI_{\Fun}$ now becomes a topological boundary condition for $\fZ(\cC_\UV)$, given by a Lagrangian algebra $\cL_F$. 
Equation \eqref{eq:masterequation} then states that Lagrangian algebras $\cL_\UV$ and $\cL_F$ must intersect trivially:
\be
\begin{tikzpicture} 
 \begin{scope}[shift={(0,0)}]
\draw [\UVcolor,  fill=\UVcolor]
(-3,0) -- (-3,3) --(0,3) -- (0,0) -- (-3,0) ; 
 \draw [\Ncolor,  fill=\Ncolor]
(0,0) -- (0,3) --(3,3) -- (3,0) -- (0,0) ; 
\draw [very thick] (0,0) -- (0,3)  ;
\draw [very thick] (-3,0) -- (3,0);
\node[below] at (-1.5,0) {$\Bsym_\cS$}; 
\node[below] at (1.5,0) {$\Vec$}; 
\node at (-1.5,2.2) {$\fZ(\cC)$}; 
\node at (1.5,2.2) {$\fZ (\Vec)$}; 
\node[above] at  (0,3) {$\cI_\Fun$};
\draw [black,fill=black] (0,1) ellipse (0.05 and 0.05);
\draw [black,fill=black] (-1.2, 0) ellipse (0.05 and 0.05);
 \draw[thick, rounded corners](3-3,1) -- (1.8-3,1) -- (1.8-3,0);
\node[above] at  (2.5-3, 1) {$1$};
\end{scope}
\end{tikzpicture}
\ee

This implies that a fiber functor is described by a magnetic Lagrangian algebra in $\cZ(\cC_\UV)$. The absence of a magnetic Lagrangian algebra in the center has recently been used in several works \cite{Antinucci:2022vyk, Kaidi:2023maf, Zhang:2023wlu, Antinucci:2023ezl, Cordova:2023bja} as a characterization of anomalous categorical symmetries.

\subsubsection{Normal Subcategories} 
\label{sec:normal subcat}

We now formulate the SymTFT realization of  normal subcategories as defined in section \ref{sec:ask def}, i.e. $\cN$ that fits into the short exact sequence 
\be\label{SESAgain}
\cN \stackrel{\IFun}{\longrightarrow} \cC 
\stackrel{\DFun}{\longrightarrow} \cS \,.
\ee
A normal subcategory is the kernel of a surjective (normal) functor $\DFun$. To each functor $\IFun$ and $\DFun$ we can associate algebras $\cA_{\IFun}$ and $\cA_{\DFun}$ in $\cZ(\cC)$, which realize them in the SymTFT in terms of interfaces.
Following our previous remarks $\cA_\IFun$ is an electric algebra, $\cA_\IFun\subset \cL_\cC$, while $\cA_\DFun$ is magnetic, $\cA_\DFun \cap \cL_\cC=\left\{1\right\}$. The requirement that $\IFun$ and $\DFun$ correctly combine into a short exact sequence translates into the condition that the algebra
\be \label{eq:trivially braiding algebra}
\cA_\IFun \otimes \cA_\DFun  \in \cZ (\cC) 
\ee
is a Lagrangian algebra, see figure \ref{fig:greybluegreen}.
Maximality of $\cA_\IFun \otimes\cA_\DFun$ follows from the relation \eqref{eq: dimensionrel} between quantum dimensions in a short exact sequence.
 
\begin{figure}
$$
\begin{tikzpicture} 
\begin{scope}[scale=0.9]
 \begin{scope}[shift={(0,0)}]
\draw [\Ncolor,  fill=\Ncolor]
(-3,0) -- (-3,3) --(0,3) -- (0,0) -- (-3,0) ; 
 \draw [\Ccolor,  fill=\Ccolor]
(0,0) -- (0,3) --(3,3) -- (3,0) -- (0,0) ; 
 \draw [\Scolor,  fill=\Scolor]
(3,0) -- (3,3) --(6,3) -- (6,0) -- (3,0) ; 
\draw [very thick] (0,0) -- (0,3)  ;
\draw [very thick] (3,0) -- (3,3)  ;
\draw [very thick] (-3,0) -- (6,0)  ;
\node[below] at (-1.5,0) {$\Bsym_\cN$}; 
\node[below] at (1.5,0) {$\Bsym_\cC$}; 
\node[below] at (4.5,0) {$\Bsym_\cS$}; 
\node at (-1.5,2.2) {$\fZ(\cN)=\fZ(\cC)/\cA_\IFun$}; 
\node at (1.5,2.2) {$\fZ(\cC)$}; 
\node at (4.5,2.2) {$\fZ(\cS)=\fZ(\cC)/\cA_{\DFun}$}; 
\node[above] at (0,3) {$\cI_\IFun$};
\node[above] at  (3,3) {$\cI_\DFun$};
\draw[thick, rounded corners](0,1) -- (1.2,1) -- (1.2,0);
\draw [black,fill=black] (0,1) ellipse (0.05 and 0.05);
\draw [black,fill=black] (1.2, 0) ellipse (0.05 and 0.05);
\node[above] at  (0.6, 1) {$a\in \cA_{\IFun}$};
\draw[thick, rounded corners](3,1) -- (1.8,1) -- (1.8,0);
\draw [black,fill=black] (3,1) ellipse (0.05 and 0.05);
\draw [black,fill=black] (1.8, 0) ellipse (0.05 and 0.05);
\node[above] at  (2.5, 1) {$1$};
\draw[thick, rounded corners](-3,1.5) -- (4.5,1.5) -- (4.5,0);
\draw [black,fill=black] (4.5, 0) ellipse (0.05 and 0.05);
\end{scope}
 \begin{scope}[shift={(0,-5)}]
\draw [\Ncolor,  fill=\Ncolor]
(-3,0) -- (-3,3) --(1.4,3) -- (1.4,0) -- (-3,0) ; 
 \draw [\Scolor,  fill=\Scolor]
(1.6,0) -- (1.6,3) --(6,3) -- (6,0) -- (1.55,0) ; 
\draw [very thick] (1.4,0) -- (1.4,3)  ;
\draw [very thick] (1.6,0) -- (1.6,3)  ;
\draw [very thick] (-3,0) -- (1.4,0)  ;
\draw [very thick] (6,0) -- (1.6,0)  ;
\node[below] at (-0.5,0) {$\Bsym_\cN$}; 
\node[below] at (3.5,0) {$\Bsym_\cS$}; 
\node at (-0.5,1.5) {$\fZ(\cN)=\fZ(\cC)/\cA_\IFun$}; 
\node at (4,1.5) {$\fZ(\cS)=\fZ(\cC)/\cA_{\DFun}$}; 
\node[above] at (1.5,3) {$\cL_{\cN, \text{mag}} \otimes \cL_{\cS}$};
\draw[thick, rounded corners](1.4,1) -- (0.5,1) -- (0.5,0);
\draw [black,fill=black] (1.4,1) ellipse (0.05 and 0.05);
\draw [black,fill=black] (0.5, 0) ellipse (0.05 and 0.05);
\node[left] at  (0.5, 0.5) {$1$};
\node[above] at (1.5, 3.6) {$\Downarrow$};
\end{scope}
\end{scope}
\end{tikzpicture}
$$
\caption{SymTFT representation of the short exact sequence of tensor functors (\ref{SESAgain}). The topological order on the left describes the center of $\fZ (\cN)$, i.e. of the normal subcategory. The exactness is shown at the top: any topological defect in $\fZ (\cN)$ ends on the symmetry boundary of $\Bsym_\cS$. The other two sets of lines simply indicate injectivity and surjectivity. 
Due to the exactness, collapsing the middle section, $\fZ (\cC)$,  $\fZ (\cN)$  and $\fZ (\cS)$ are separated by the trivial topological order, i.e. the interface between them is a product of two gapped boundary conditions: $\cL_{\cN, \text{mag}}$ and $\cL_\cS$, respectively.  
\label{fig:greybluegreen}}
\end{figure}

On the other hand, fusing the two interfaces $\cI_\IFun$ and $\cI_\DFun$ together, we find a factorized interface between $\fZ(\cN)$ and $\fZ(\cS)$, described by condensation of the algebras  $\cL_{\cN, \text{mag}}\boxtimes \ol{\cL_{\cS}}$. 
Here $\cL_{\cN, \text{mag}}$ is a magnetic Lagrangian algebra -- $\cL_{\cN} \cap \cL_{\cN, \text{mag}} =\{1\}$ -- describing a fiber functor for $\cN$.

The corresponding functors fit into a short exact sequence (\ref{SESAgain}).
The exactness $\image(\IFun) = \ker(\DFun)$ has the following pictorial interpretation (see the above figure): 
each anyon that can end on $\Bsym_\cS$ is the image of some anyon in $\fZ(\cN)$, which cannot end on $\Bsym_\cN$.

Due to the existence of the fiber functor, {\bf normal subcategories are anomaly-free}. However it is important to emphasize that not all anomaly-free subcategories are normal, as we will see in an example shortly. The physical interpretation of a normal subcategory is as a part of the symmetry that we can gap-out while preserving the full symmetry $\cC$. On the other hand, to gap out an anomaly free but not normal subcategory one needs to break explicitly the remaining symmetry. This fact provides strong predictions: it forbids the existence of $\cC$-symmetric relevant deformations that gap a non-normal but anomaly-free subsymmetry.

\vspace{0.2cm}
\noindent \textbf{Non-Example: \boldmath ${\Vec_{\bZ_2}\subset \Vec_{\bZ_4}^{\omega=2}}$.}  
In section \ref{sec:TF} we have described an anomaly-free subcategory that is not a normal subcategory:
the $\bZ_2$ subgroup of a $\bZ_4$ symmetry in (1+1)d with anomaly $2\in H^3(B\bZ_4,U(1))=\bZ_4$: $\Vec_{\bZ_4}^{\omega=2}$.

It is useful to recall how to see that there is a non-anomalous $\Z_2$-subgroup. For this consider the inflow action
\begin{equation}\label{eq:arounf}
    S_{\text{inflow}}=\frac{4\pi i}{4}\int A \cup \beta (A) 
\end{equation}
with $A\in H^1(X,\bZ_4)$ the background field, and $\beta(A)=\frac{\delta A}{4}$ the Bockstein map. The subgroup $\bZ_2\subset \bZ_4$ is anomaly free: $S_{\text{inflow}}$ becomes trivial if $A=2A'$.  Let us recast it via the SymTFT.

$\cZ(\Vec_{\bZ_4}^{\omega=2})$ has lines $e^{n_{e}}m^{n_{m}}$, $n_{e},n_{m}\in \bZ_4$, with spin 
\begin{equation}
    \theta_{n_{e},n_{m}}=\exp{\left(-\frac{2\pi i}{4}\left(n_{e}n_{m}-\frac{1}{2}n_{m}^2\right)\right)}
\end{equation}
and the canonical Lagrangian algebra for the $\Vec_{\bZ_4}^{\omega=2}$ symmetry is 
\be
\cL_{\Vec_{\bZ_4}^{\omega=2}}= 1\oplus e \oplus e^2 \oplus e^3 \,.
\ee 
This contains as a subalgebra $\cA_\IFun =1 \oplus e^2$ and it is easy to check that
\begin{equation}
    \cZ(\Vec_{\bZ_4}^{\omega=2})/\cA_\IFun =\cZ(\Vec_{\bZ_2}) \ .
\end{equation}
Thus we have an injective functor $\IFun : \Vec_{\bZ_2}\rightarrow \Vec_{\bZ_4}^{\omega=2}$ that embeds an anomaly-free $\bZ_2$ inside $\bZ_4$ with anomaly $2$.
However there are no non-trivial condensable algebras of $\cZ(\Vec_{\bZ_4}^{\omega=2})$ that intersect trivially with $\cL_{\Vec_{\bZ_4}^{\omega=2}}$, hence $\Vec_{\bZ_4}^{\omega=2}$ does not have surjective functors {with kernel $\image(\IFun)=\Vec_{\bZ_2}$}.

This shows that while $\bZ_2$ is anomaly-free, it is not a normal subcategory.
While this category has a fiber functor, it is not of the form $\FF = \DFun \circ \IFun$. The physical meaning is that $\bZ_2$ can be gapped, but it cannot be gapped if it is embedded inside $\Vec_{\bZ_4}^{\omega=2}$. In other words, gapping out $\bZ_2$ requires breaking explicitly $\bZ_4$ down to $\bZ_2$.

\section{ASCies as a Measure of Anomalies}
\label{sec:ASC}

Defining and quantifying anomalies for non-invertible symmetries has been a challenge thus far. We will now make propose that there is a simple characterization in terms of ASCies, which are obtained from short exact sequences of tensor functors, which will provide a characterization of all anomalous aspects, in terms of categories $\cS_i$, of a given category $\cC$. 

We will first discuss this in the realm of tensor functors, and then use the reformulation in terms of the SymTFT, that allows exploring things systematically and concretely, for any categorical symmetry. 
We conclude with some interesting examples of anomaly matching through ASCies.

\subsection{ASCies and Anomalies of Fusion Category Symmetries}

Our mathematical description of tensor functors and categorical exact sequences via the SymTFT gives interesting insights into the classification of 't Hooft anomalies for non-invertible symmetries. 

Let us briefly recall the salient features of the problem. In the case of (higher) groups $G$, anomalies are captured by $G$-symmetric $(d+1)$-dimensional Symmetry Protected Topological (SPT) phases, which are invertible TQFT and can be composed by stacking.  Thus such anomalies form a group and they can be compared by making use of the group structure. Such an observation was implicitly used ever since the first instances of 't Hooft anomaly matching \cite{tHooft:1979rat}.

In the case of non-invertible symmetries, the closest analogue of an anomaly theory is the SymTFT—a {\bf non-invertible} $(d+1)$-dimensional topological order. However, the absence of a group structure on the space of such TQFTs prevents a direct extension of anomaly matching \cite{Kong:2020cie}.
Thus, while an obstruction theory to the existence of a trivially gapped phase has been developed in several instances \cite{Thorngren:2019iar,Apte:2022xtu,Damia:2023ses,Antinucci:2023ezl,Cordova:2023bja,Hsin:2025ria}, there is currently no obvious way to relate these obstructions, nor to understand the physical process leading to their cancellation. 

This problem is of great physical relevance: two anomalous symmetries are expected to have kinematical obstructions in being connected by an RG-flow if their anomalies do not match. Thus, defining precisely the meaning of this is of crucial importance to extract the maximal amount of constraints from symmetries.

Tensor functors, as discussed in the previous sections, will play a role in addressing this important issue.
On a mundane level, if we have a surjective tensor functor 
\be
\DFun:\quad \cC_{\UV} \to \cC_{\IR} \, ,
\ee
then the pullback $\DFun^*$ implements the 't Hooft anomaly matching conditions, see section \ref{sec:TF}.
More generally, we will see that short exact sequences of tensor functors 
\be
\cN\overset{\IFun}{\longrightarrow} \cC \overset{\DFun}{\longrightarrow} \cS
\ee
will allow us to quantify anomalies for any categorical symmetry $\cC$. 
The idea is that an anomaly-free sub-symmetry that is also normal, $\cN$ in the above sequence, can be consistently gapped out at low energy, and thus forgotten.

It is then natural to propose that an anomaly of the symmetry $\cC$ is encoded in the image of $P$ with ``maximal" kernel, denoted by $\cN_{\text{max}}$. 
For given $\cC$, there can be several choices of maps $\DFun_{i}$  with maximal kernel $\cN_{\text{max}, i}$ and $\cS_{i}$:
\be\label{Huskiemax}
\cN_{\text{max}, i} \overset{\IFun_i}{\longrightarrow} \cC \overset{\DFun_i}{\longrightarrow} \cS_i \,.
\ee
However each sequence results in a simple category $\cS_i$, which is fully anomalous, i.e. is an {\bf ASCy}.
Notice that any category which is not simple, will inevitably fit into one (or more) short exact sequences, and thus eventually lead to an ASCy.

The collection of ASCies $\{\cS_i\}$ should be thought of as ``building blocks" for anomalies of the category $\cC$:
\be
\fA(\cC)= \{\cS_i \text{ satisfying (\ref{Huskiemax})}\}\,.
\ee
 Their symmetric gapped phases -- described by $\cS_i$-module categories $\cM \in \Mod_{\cS_i}$ -- can be pulled back by pre-composition along the tensor functor:
\be
\Mod_{\cC}\  \overset{\DFun_i^*}{\longleftarrow}\  \Mod_{\cS_i} \, .
\ee
In practice, this provides a map between $\cS_i$-symmetric phases -- in which all the $\cS_i$ symmetry is broken -- and $\cC$-symmetric phases whose symmetry breaking pattern is enforced by the anomaly. 

For example, if a symmetry is anomaly-free, a fiber functor $\FF : \cC \to \Vec$ singles out a specific $\cC$-preserving gapped phase by pulling back the unique trivially gapped phase for $\Vec$ along $\FF^*$:
\be
\begin{tikzpicture}
\draw[fill = blue] (0,0) circle (0.05); 
\draw[thick, red,-latex,looseness=8,  out=60, in=120] (0.15,0.15) to (-0.15,0.15);
\draw[fill = blue] (2,0) circle (0.05) ;
\draw[<-] (0.5,0) -- (1.5,0);
\node[above] at (1,0) {$\FF^*$};
\node[red, above] at (0,0.85) {$\cC$};
\end{tikzpicture}
\ee
If a category $\cC$, admits multiple inequivalent ACSies $\cS_i$, this is the natural generalization of the fact that a symmetry category can admit multiple inequivalent fiber functors.
In practice, the pullbacks along $\DFun_i$ of the $\cS_i$-symmetric gapped phases will describe all the maximally symmetry-preserving gapped phases for $\cC$.

We will use this characterization of ASCies as measures of anomalies, and in practice the reformulation of short exact sequences using the SymTFT of section \ref{sec: SymTFTTF}, to determine for  given symmetry $\cC$ the set of ASCies $\fA(\cC)$.

\vspace{2mm} \noindent \textbf{ASCies from the SymTFT.} In section \ref{sec:ask def} we defined Anomalous Simple Categories (ASCies) as categories without any non-trivial normal subcategory. 
In the SymTFT we can characterize ASCies as being categories {\bf $\cC$ such that $\cZ(\cC)$ do not have {non-trivial} magnetic algebras $\cA\cap \cL_\cC=\left\{1 \right\}$}. 
This criterion makes it straightforward to determine if a symmetry category also is an ASCy.

The SymTFT framework provides a computationally powerful tool to determine the possible maximal short exact sequences, and we demonstrate the effectiveness of this approach in several examples in the next section.

\subsection{Example: Anomalous Groups}

The first examples are anomalous group symmetries $\Vec_{G}^\omega$. We will now state some general results and then specialize to cyclic groups to give concrete examples.

Given a short exact sequence of finite groups (here $N\triangleleft G$ is a normal subgroup)
\begin{equation}
    1 \rightarrow N \rightarrow G \xrightarrow{p} G/N \rightarrow 1\,,
\end{equation}
one could wonder whether it always induces an exact sequence of fusion categories based on $\Vec_G^\omega$ for any $\omega\in H^3(G, U(1))$. The answer is negative, as we show below.  
However, if $\omega$ is such that $\omega=p^*(\omega')$ for some $\omega'\in H^3(G/N,U(1))$, then there is (at least) one surjective functor $\DFun: \Vec_G^\omega \rightarrow \Vec_{G/N}^{\omega'}$ and a short exact sequence of fusion categories \cite{bruguieres2011exact} 
\begin{equation}
    \Vec_{N} \rightarrow \Vec_{G}^{p^*(\omega')} \xrightarrow{P} \Vec_{G/N}^{\omega'} \, , \ \ \ \omega = p^* (\omega')  \, .
\end{equation}
The latter equation encodes the UV/IR anomaly matching. Thus, an exact sequence of groups can be lifted to an exact sequence of the corresponding (anomalous) fusion categories if and only if the 't Hooft anomaly matching conditions are satisfied. Notice that while the anomaly must trivialize on $N$, $\omega|_N =0$, this condition is not sufficient to ensure the existence of the exact sequence.

\vspace{2mm}
\noindent
{\bf Cyclic Groups with Anomalies.}
Consider anomalous cyclic groups, $\cC= \Vec_{\bZ_N}^\omega$, with $\omega \in H^3(\bZ_N,U(1))$ its group co-homology anomaly, to wit:
\begin{equation}
    \omega(a,b,c) = \exp\left( \frac{2\pi i k}{N^2}a(b+c-[b+c]_N) \right) \,,
\end{equation}
where $k\in \{0,1,\cdots, N-1\}$, and $[ \quad]_N$ denotes reduction modulo $N$. Its SymTFT is described by a twisted $\bZ_N$ Dijkgraaf-Witten theory, with twist $\omega$. Its spectrum consists of $N^2$ invertible lines 
\be
\Q_{n_e,n_m} = e^{n_e} m^{n_m} \, ,
\ee with spins \cite{Barkeshli:2014cna}:
\be
\theta_{n_e,n_m} = \exp\left(- \frac{2\pi i}{N} \left( n_e n_m - \frac{k}{N} n_m^2\right) \right) \, ,
\ee
and identifications
\be
(n_e,n_m)\sim (n_e+N,n_m)\sim (n_e+2k, n_m+N) \ .
\ee
Any subgroup $\bZ_n$ of $\bZ_N$, $N=n \ell$, is a normal subgroup and sits inside a short exact sequence of groups:
\be
1 \longrightarrow \bZ_n \overset{\iota}{\longrightarrow} \bZ_N \overset{\pi}{\longrightarrow} \bZ_\ell \longrightarrow 1 \, . 
\ee
This does not, however, always lift into a short exact sequence of categories. 
A necessary condition is obviously that $\bZ_n$ is anomaly-free, that is:
\be
\iota^*(\omega) = 0 \, , \quad \text{or} \quad k = k' n \, .
\ee
Contrary to expectations, this is not sufficient to make $\bZ_n$ into a normal subcategory, as discussed in the example of $\cC=\Vec_{\bZ_4}^{\omega=2}$ in the previous section.
This example, as well as several ones we will present below, highlight a striking physical consequence of the concept of normal subcategory. An anomaly free subcategory $\cN$, if not normal, {\bf cannot} be gapped out while preserving the quotient symmetry $\cS$. This highlights the usefulness of the notion of short exact sequences of tensor functors, and the notion of normal subcategories in the context of symmetric RG-flows.

Another important property of ASCies, is that one can in general associate several of them to a given symmetry. This already happens in the case of cyclic groups, as we now show via an example.

\vspace{2mm}
\noindent 
{\bf Example: $\Z_8^{\omega=4}$.}
The simplest example is given by $\cC = \Vec_{\bZ_8}^\omega$ with anomaly $\omega=4$. The largest normal subcategory is now $\Vec_{\bZ_2}$, which fits in two different short exact sequences:
\be \ba
&\Vec_{\bZ_2} \overset{\IFun}{\longrightarrow} \Vec_{\bZ_8}^{\omega=4} \overset{\DFun_{1}}{\longrightarrow} \Vec_{\bZ_4}^{\omega=1} \, , \\
&\Vec_{\bZ_2} \overset{\IFun}{\longrightarrow} \Vec_{\bZ_8}^{\omega=4} \overset{\DFun_{-1}}{\longrightarrow} \Vec_{\bZ_4}^{\omega=-1} \, ,
\ea \ee
with the functors $P_{\pm 1}$ defined in equation~\eqref{eqn:Pk cyclic group}. 
Leading to
\be
\fA (\Vec_{\Z_8}^{\omega=4}) = \left\{\Vec_{\bZ_4}^{\omega=1},  \Vec_{\bZ_4}^{\omega=3} \right\} \,.
\ee
At the level of SymTFT, the embedding functor $\IFun: \Vec_{\bZ_2}\rightarrow \Vec_{\bZ_8}^{\omega=4}$ can be obtained from the condensation of the electric algebra 
$\cA_{\IFun} = 1 \oplus e^2 \oplus e^4 \oplus e^6$ in $\cZ(\Vec_{\bZ_8}^{\omega=4})$, which gives the reduced topological order $\cZ (\Vec_{\bZ_2})$. To work out the possible surjective normal functors and the ASCies, we look for magnetic algebras in $\cZ(\Vec_{\bZ_8}^{\omega=4})$. It turns out that there are two (maximal) magnetic algebras, which implement the functors $\DFun_{\pm1}$:
\be
\cA_{\DFun_{1}} = 1 \oplus m^4   \, , \qquad
\cA_{\DFun_{-1}} = 1 \oplus e^4 m^4 \, ,
\ee
and 
\be \ba
&\cZ(\bZ_8^{\omega=4})/\cA_{\DFun_{1}}  &&= \cZ(\bZ_4^{\omega=1})\, , \\
&\cZ(\bZ_8^{\omega=4})/\cA_{\DFun_{-1}} &&= \cZ(\bZ_4^{\omega=-1}) \, .
\ea \ee 
The situation can be summarized succinctly via the RG-quiche:
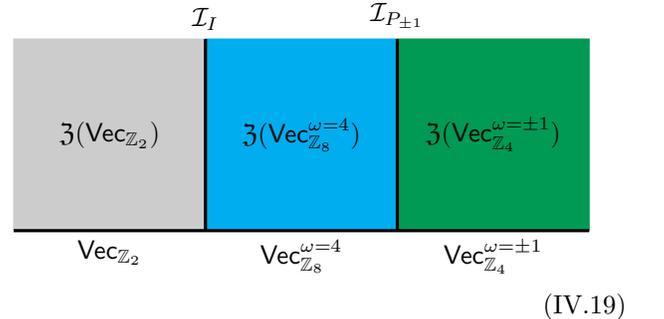
\be
\label{eq: RGQuicheTY1}
\begin{tikzpicture} 
\begin{scope}[scale=0.85]
\draw [\Ncolor,  fill=\Ncolor]
(-3,0) -- (-3,3) --(0,3) -- (0,0) -- (-3,0) ; 
 \draw [\Ccolor,  fill=\Ccolor]
(0,0) -- (0,3) --(3,3) -- (3,0) -- (0,0) ; 
 \draw [\Scolor,  fill=\Scolor]
(3,0) -- (3,3) --(6,3) -- (6,0) -- (3,0) ; 
\draw [very thick] (0,0) -- (0,3)  ;
\draw [very thick] (3,0) -- (3,3)  ;
\draw [very thick] (-3,0) -- (6,0)  ;
\node[below] at (-1.5,0) {$\Vec_{\bZ_2}$}; 
\node[below] at (1.5,0) {$\Vec_{\bZ_8}^{\omega=4}$}; 
\node[below] at (4.5,0) {$\Vec_{\bZ_4}^{\omega=\pm 1}$}; 
\node at (-1.5,1.5) {$\fZ(\Vec_{\bZ_2})$}; 
\node at (1.5,1.5) {$\fZ(\Vec_{\bZ_8}^{\omega=4})$}; 
\node at (4.5,1.5) {$\fZ(\Vec_{\bZ_4}^{\omega= \pm 1})$}; 
\node[above] at (0,3) {$\cI_{\IFun}$};
\node[above] at (3,3) {$\cI_{\DFun_{\pm 1}}$};
\end{scope}
\end{tikzpicture}
\ee
Notice that the difference between the two IR anomalies, which is $2 \mod 4$, pulls back to the trivial anomaly for $\bZ_8$: thus, some of the IR anomalies must be emergent. There is however no canonical way to assign which of the two ASCies has an emergent anomaly, contrary to what happens in the case of an anomaly-free UV symmetry. We believe that this observation deserves further study. 

Finally let us comment that, in this case, the two ASCies are exchanged by an automorphism of $\Vec_{\bZ_8}^{\omega=4}$. To see this, notice that the center of $\Vec_{\bZ_8}^{\omega=4}$ has a time-reversal symmetry \cite{Delmastro:2019vnj} -- which is also a symmetry of $\Vec_{\bZ_8}^{\omega=4}$ -- implemented by:
\be
\Theta \, \left( \begin{array}{c}
    n_e \\
    n_m 
\end{array} \right) = \left( \begin{array}{cc}
   1 & 3 \\
  0 & -1 
\end{array} \right) \, \left( \begin{array}{c}
    n_e \\
    n_m 
\end{array} \right) \, .
\ee
This is broken by the choice of algebra $\cA_{\DFun_{\pm 1}}$, as $\Theta \cA_{\DFun_{1}} = \cA_{\DFun_{- 1}}$ and leads to different ASCies $\Vec_{\bZ_4}^{\omega=\pm 1}$. It is important to stress that this is not the general case: as we will see in examples below, a given categorical symmetry can have several different ASCies not related by automorphisms.

\vspace{2mm} \noindent \textbf{A more general sequence.}
The previous example fits into a larger family of short exact sequence of abelian categories:
\be
\Vec_{\bZ_n} \ \overset{\IFun}{\longrightarrow}\  \Vec_{\bZ_N}^{\omega = k n^2 } 
\ \stackrel{\DFun_k}{\longrightarrow} \ 
\Vec_{\bZ_\ell}^{\omega = k} \,, \quad   N = n \ell \, .
\ee
The projection $\DFun_k$ acts by $\DFun_k(a) = [a]_\ell$, but has a nontrivial action on junctions: 
\be\ba\label{eqn:Pk cyclic group}
\DFun_k \left( \begin{tikzpicture}[baseline={(0,0)}]
 \draw (0,0) -- (90:0.5) node[above] {$c$};   \draw (0,0) -- (210:0.5) node[left] {$a$};    \draw (0,0) -- (-30:0.5) node[right] { $b$};  
 \draw[fill=black] (0,0) circle (0.05);
\end{tikzpicture} \right) 
&= J_k(a,b) \,  \begin{tikzpicture}[baseline={(0,0)}]
 \draw (0,0) -- (90:0.5) node[above] {$[c]_\ell$};   \draw (0,0) -- (210:0.5) node[left] {$[a]_\ell$};    \draw (0,0) -- (-30:0.5) node[right] { $[b]_\ell$};  
 \draw[fill=black] (0,0) circle (0.05);
\end{tikzpicture} \, ,  \\[1em]
J_k(a,b) &= \exp\left( \frac{2 \pi i k}{\ell^2} a (b-[b]_\ell) \right)  \,.
\ea\ee
The same anomaly $\omega = k n^2$ can be represented by different values of $k \in \bZ_\ell$, provided the short exact sequence of groups does not split. This is related to an automorphism of the $\bZ_N$ algebra. More precisely, $k n^2$ and $k'n^2$ give the same anomaly if $(k-k') = \ell/\gcd(n,\ell) \mod \ell$. Which gives $\gcd(n,\ell)$ different short exact sequences. This also shows that split sequences of cyclic groups give rise to unique short exact sequences of tensor categories.

We conclude that the set of ASCies for this category is 
\begin{equation}
    \Vec_{\bZ_\ell}^{\nu = k} \in \fA (\Vec_{\Z_N}^{\omega})\,,
\end{equation}
for $N= n \ell$, $\omega= k n^2$, and for all $m>1$ such that $m\mid \ell$, $m^2\nmid k$.
At the level of SymTFT, this is translated into the existence of several magnetic algebras of the same quantum dimension. This is simple to understand: the normal subcategory $\Vec_{\bZ_n}$ is described by condensation of the electric algebra:
\be
\cA_{\IFun} = \bigoplus_{a=0}^\ell \, e^{n a} \, .
\ee
We now want to find the possible magnetic algebras which are mutually local with respect to $\cA_\IFun$. These must be spanned by anyons of the form $e^a \, m^{\ell b}$, and are bosons provided that $a = n c$. Now fix a value $c$, the algebra $\cA_{\DFun_c}$ generated by $e^{n c} m^{\ell}$ is a magnetic algebra describing an exact sequence of fusion categories. Some of these algebras are identified, this happens if $(c-c') \equiv n q \mod \ell$, for some integer $q$, leaving only $\gcd(n,\ell)$ distinct magnetic algebras. This matches the counting of the short exact sequences. It is also possible to check that the IR topological orders are $\cZ(\Vec_{\bZ_\ell}^{k'})$ where $k' = k + r \ell/\gcd(n,\ell)$, for some integer $r$, and the magnetic generator descends from $m e^{- r n /\gcd(n,\ell)}$ in the UV.

\subsection{Example: Tambara-Yamagami Categories} \label{sec:TY}
Tambara-Yamagami categories $\TY(A,\chi,\epsilon)$ \cite{TambaYama} are a prime candidate to test our ideas in the realm of non-invertible symmetries. \
Recall that a $\TY(A,\chi,\epsilon)$ category is described by an abelian group $A$ and a duality defect $\cD$ satisfying:
\be
\cD \otimes \cD = \bigoplus_{a \, \in \, A} a \, .
\ee
The associator between two $a$ defects and $\cD$ is encoded in a symmetric bicharacter $\chi$ on $A$, while $\epsilon$ is the Frobenius-Schur indicator of the self-dual defect $\cD$.
The mathematical structure of their anomalies -- in the sense of the obstruction theory to the existence of a fiber functor -- is well understood \cite{meir2012module,Thorngren:2019iar,Zhang:2023wlu, Antinucci:2023ezl}. The obstruction to the existence of a fiber functor comes into two layers:
\begin{enumerate}
    \item A {\bf first obstruction} encoded in the existence of a duality-invariant $A$-SPT. This is described by a Lagrangian subgroup of $A$ with respect to the pairing $\chi$. A vanishing first obstruction means that the invertible part of $\TY(A,\chi,\epsilon)$ can be trivially gapped. 
    \item A {\bf second obstruction}, which is akin to a pure anomaly for $\cD$. This vanishes if and only if the Frobenius-Schur indicator can be trivialized in the $A$-invariant SPT. The precise mathematical formulation of such trivialization can be found in \cite{meir2012module,Thorngren:2019iar,Antinucci:2023ezl}.
\end{enumerate}

The structure of categorical exact sequences involving $\TY$ is known \cite{bruguieres2014central}: normal subcategories of $\TY$ are either subgroups of its invertible symmetry, or are the full $\TY$ itself. We now elucidate some interesting aspects of the interplay between the aforementioned obstruction theory and the structure of surjective tensor functors.

\vspace{2mm}
\noindent\textbf{Example: Ising Fusion Category.} The Ising fusion category, that is $\TY(\bZ_2,+)$ is by itself an ASCy. This can be immediately seen from the non-existence of magnetic condensable algebras in the Drinfeld center \cite{2009arXiv0905.3117G}. There is, however, an electric condensable algebra  
\be
\cA_\IFun = 1\oplus X_{0, -1} \subset \cL_{\TY(\bZ_2,+)}\,,
\ee
where $X_{0, -1}\in \cZ(\TY(\bZ_2,+))$ is a simple object of quantum dimension 1 (we follow the notation used in \cite{2009arXiv0905.3117G} to label the simple objects), that corresponds to the embedding $\IFun : \Vec_{\bZ_2}\rightarrow \TY(\bZ_2,+)$. Indeed while $\Vec_{\bZ_2}$ is an anomaly free subsymmetry, is not a normal one. This extremely simple fact alone predicts a universal fact: there cannot be $\TY(\bZ_2,+)$ symmetric relevant deformations that trivialize $\bZ_2$. This is obviously true for instance in the Ising CFT, where the $\varepsilon_{(1/2,1/2)}$ deformation, that gaps $\bZ_2$, breaks the duality symmetry.

\vspace{2mm}
\noindent\textbf{Example: \boldmath $\TY(\bZ_4,+)$.} We  examine an example of a $\TY$ symmetry where the duality defect can be gauged, but the full symmetry still has a 't Hooft anomaly. We will characterize its anomalies by analyzing its ASCies. 
To do so, consider the Drinfeld center $\cZ(\TY(\bZ_4,+))$ \cite{2009arXiv0905.3117G}. It is composed of the anyons as shown in Table~\ref{tab:spinsTYZ4}.
\begin{table}[t]
      \centering
      {
\renewcommand{\arraystretch}{1.5}
    \begin{tabular}{|c|c|c|c|c|c|c|c|}
    \hline
       & $X_{0,\pm1}$  & $X_{1,\pm \zeta_8}$ & $X_{2,\pm1}$ & $X_{3,\pm\zeta_8}$ & $Y_{1,0}$ & $Y_{2,0}$ & $Y_{3,0}$  \\
       
       \hline

       $\theta$ & $1$ & $-i$ & $1$ & $-i$ & $1$ & $1$ & $1$ \\
       
       \hline

       $d$ & $1$ & $1$ & $1$ & $1$ & $2$ & $2$ & $2$ \\
       \hline \hline
       & $Y_{1,2}$ & $Y_{1,3}$ & $Y_{3,2}$ &  $Z_{\rho_0,\pm\zeta_{16}}$ & $Z_{\rho_1,\pm 1}$ & $Z_{\rho_2,\smash{\raisebox{0.1ex}{\scalebox{0.75}{$\pm \zeta_{16}^{-3}$}}}}$ & $Z_{\rho_3,\pm 1}$          \\
       \hline
       $\theta$ & $-1$ & $i$ & $-1$ & $\pm \zeta_{16}$ & $\pm 1$ & $\pm \zeta_{16}^{-3}$ & $\pm 1$ 
         \\
         \hline
        $d$ & $2$ & $2$ & $2$ & $2$ & $2$ & $2$ & $2$  \\
        \hline
    \end{tabular}}
    \caption{Spins and quantum dimensions of the lines of $\cZ(\TY(\bZ_4))$, where $\zeta_8 = \exp(\frac{2\pi i}{8})$ and $\zeta_{16}= \exp(\frac{2\pi i}{16})$, and we name the anyons following notations in \cite{2009arXiv0905.3117G}. Note that $X_{0,+1} =1$.
    }
    \label{tab:spinsTYZ4}
\end{table}

We refer the reader to \cite{2009arXiv0905.3117G} for the relevant fusion rules.
The canonical Dirichlet boundary for $\TY(\bZ_4,+)$ has Lagrangian algebra:
\be
\cL_{\TY(\bZ_4,+)} = 1 \oplus X_{0,-1} \oplus Y_{1,0} \oplus Y_{2,0} \oplus Y_{3,0} \, .
\ee
Non-maximal algebras for this center have been classified in \cite{Bhardwaj:2023bbf}. 
There are only two magnetic algebras, both of which are two dimensional:
\be
\cA_{\DFun_+} = 1 \oplus X_{2,+1} \, , \quad \cA_{\DFun_-} = 1 \oplus X_{2,-1}\, .
\ee
We find, using standard techniques of anyon condensation, that:
\be \ba
\cZ(\TY(\bZ_4,+))/\cA_{\DFun_+} &=  \cZ(\Vec_{\bZ_4}^{\omega=1}) \, , \\
\cZ(\TY(\bZ_4,+))/\cA_{\DFun_-}  &= \cZ(\Vec_{\bZ_4}) \, .
\ea \ee
The boundaries $\Bsym_{\IR \pm}$ are determined using \eqref{eq:masterequation}:
\be\ba
\fB_{ \TY(\bZ_4,+)} \times \cI _{\DFun_+} &= \fB_{\Vec_{\bZ_4}^{\omega=1}}, \\
\fB_{ \TY(\bZ_4,+)} \times \cI_{\DFun_-} &= \fB_{\Vec_{\bZ_2 \times \bZ_2}^{\omega}} \, ,
\ea\ee 
They both have the same trivially-braiding electric algebra (see \eqref{eq:trivially braiding algebra}), that is:
\be
\cA_{\IFun} =1 \oplus X_{2,-1} \oplus Y_{2,0} \, .
\ee
Its condensation corresponds to the injective functor $\Vec_{\bZ_2} \overset{\IFun}{\longrightarrow} \TY(\bZ_4,+)$.
We are thus led to the RG-quiches:
\be
\label{eq: RGQuicheTY1}
\begin{tikzpicture} 
\begin{scope}[scale=0.85]
\draw [\Ncolor,  fill=\Ncolor]
(-3,0) -- (-3,3) --(0,3) -- (0,0) -- (-3,0) ; 
 \draw [\Ccolor,  fill=\Ccolor]
(0,0) -- (0,3) --(3,3) -- (3,0) -- (0,0) ; 
 \draw [\Scolor,  fill=\Scolor]
(3,0) -- (3,3) --(6,3) -- (6,0) -- (3,0) ; 
\draw [very thick] (0,0) -- (0,3)  ;
\draw [very thick] (3,0) -- (3,3)  ;
\draw [very thick] (-3,0) -- (6,0)  ;
\node[below] at (-1.5,0) {$\Vec_{\bZ_2}$}; 
\node[below] at (1.5,0) {$\TY(\bZ_4,+)$}; 
\node[below] at (4.5,0) {$\Vec_{\bZ_4}^{\omega=1}$}; 
\node at (-1.5,1.5) {$\fZ(\Vec_{\bZ_2})$}; 
\node at (1.5,1.5) {$\fZ(\TY(\bZ_4,+))$}; 
\node at (4.5,1.5) {$\fZ(\Vec_{\bZ_4}^{\omega=1})$}; 
\node[above] at (0,3) {$\cI_{\IFun}$};
\node[above] at (3,3) {$\cI_{\DFun_+}$};
\end{scope}
\end{tikzpicture}
\ee
\be
\label{eq: RGQuicheTY2}
\begin{tikzpicture} 
\begin{scope}[scale=0.85]
\draw [\Ncolor,  fill=\Ncolor]
(-3,0) -- (-3,3) --(0,3) -- (0,0) -- (-3,0) ; 
 \draw [\Ccolor,  fill=\Ccolor]
(0,0) -- (0,3) --(3,3) -- (3,0) -- (0,0) ; 
 \draw [\Scolor,  fill=\Scolor]
(3,0) -- (3,3) --(6,3) -- (6,0) -- (3,0) ; 
\draw [very thick] (0,0) -- (0,3)  ;
\draw [very thick] (3,0) -- (3,3)  ;
\draw [very thick] (-3,0) -- (6,0)  ;
\node[below] at (-1.5,0) {$\Vec_{\bZ_2}$}; 
\node[below] at (1.5,0) {$\TY(\bZ_4,+)$}; 
\node[below] at (4.5,0) {$\Vec_{\bZ_2 \times \bZ_2}^{\omega}$}; 
\node at (-1.5,1.5) {$\fZ(\Vec_{\bZ_2})$}; 
\node at (1.5,1.5) {$\fZ(\TY(\bZ_4,+))$}; 
\node at (4.5,1.5) {$\fZ(\Vec_{\bZ_4})$}; 
\node[above] at (0,3) {$\cI_{\IFun}$};
\node[above] at (3,3) {$\cI_{\DFun_-}$};
\end{scope}
\end{tikzpicture}
\ee
We conclude that $\TY(\bZ_4,+)$ has two ASCies associated to it, via the exact sequences:
\be
\begin{tikzcd}
       &  & \Vec_{\bZ_4}^{\omega=1} \\
   \Vec_{\bZ_2}  \arrow[r, "\IFun"]  & \TY(\bZ_4,+) \arrow[ur, outer sep=-1, pos=.65, "\DFun_+"]   \arrow[dr, pos=.35, "\DFun_-"]  & \\
       &  & \Vec_{\bZ_2 \times \bZ_2}^\omega
\end{tikzcd}.
\ee
From the folded Lagrangian algebras corresponding to the interfaces $\cI_{\DFun _\pm}$ we can also exhibit explicitly the map on objects:
\be \label{eq:functore Ppm} \ba
&\DFun_+(\cD) = \eta \oplus \eta^3 \, , \, &&\DFun_+(a) = \eta^2 \, , \, &&&\DFun_+(a^2) = 1 \, , \\
&\DFun_-(\cD) = \eta_1 \oplus \eta_2 \, , \, &&\DFun_-(a) = \eta_1 \eta_2 \, , \, &&&\DFun_-(a^2) = 1 \, ,
\ea\ee
where $\eta\in \Vec_{\bZ_4}^{\omega=1}$ and $\eta_1,\eta_2\in \Vec_{\bZ_2\times\bZ_2}^{\omega}$ denote the generators of each symmetry.
Let us comment on this result. 
\begin{enumerate}
\item As in previous examples, there is no unique ASCy. Furthermore,  $\DFun_+$ and $\DFun_-$ cannot be related by any automorphism of $\cZ(\TY(\bZ_4, +))$. Indeed any automorphism relating $X_{2,1}$ with $X_{2,-1}$ must also exchange even and odd $\Sigma_a^\pm$, but this is forbidden by their spins. 

\item Again, as in the $\Ising$ example, while the invertible symmetry $\Vec_{\bZ_4}$ is anomaly-free, it does not correspond to a normal subcategory: it can only be preserved at the price of breaking the duality symmetry $\cD$.

\item On the other hand, it is very instructive to consider the fate of the generator of the $\bZ_4$ symmetry after the normal subcategory $\Vec_{\bZ_2}$ has been quotiented out.
It can be explicitly checked the tensor functors $\DFun_\pm$ are only consistent if the corresponding object in the IR has a nontrivial $\bZ_2$ anomaly. 
While in other cases -- such as the $\Vec_{\bZ_4} \to \Vec_{\bZ_2}^\omega$ RG map -- we had interpreted this as an emergent anomaly, let us stress that in the present example this is not quite correct. 
As the $\bZ_2$ quotient must be broken in all of the gapped phases of $\TY(\bZ_4,+)$, it was anomalous already in the full UV $\TY$ symmetry, due to its nontrivial interplay with the duality symmetry. 
Its description in the ASCies is however much simpler, and it pulls back through $P^*_\pm$ to a -- significantly more complex -- mixed anomaly involving the non-invertible duality symmetry.

\item It is also instructive to study the map between gapped phases of  $\TY(\bZ_4,+)$ {and those of its ASCies}. The UV symmetry has three distinct phases \cite{Thorngren:2019iar,Bhardwaj:2023idu}: 
\be\ba
&\TY(\bZ_4,+):  \cr 
&\begin{tikzpicture}
\begin{scope}[scale=1.8, shift={(0,0)}] 
     \draw[-latex]  (-0.25,-0.15) -- (-0.25,+0.15);   \draw[-latex]  (0.25,0.15) -- (0.25,-0.15);
      \draw[-latex]  (-0.15,0.25) -- (0.15,0.25);   \draw[-latex]  (0.15,-0.25) -- (-0.15,-0.25);
    \draw[-latex,red] (1,0) -- (0.3,0.25);  \draw[-latex,red] (1,0) -- (0.3,-0.25);
     \draw[-latex,red] (1,0) to[bend right=30] (-0.25,0.3);  \draw[-latex,red] (1,0) to[bend left=30] (-0.25,-0.3);
       \draw[fill=\IRcolor] (1,0) circle (0.05);
          \foreach  \x in {-1,1}
     \foreach  \y in {-1,1} {
    \draw[fill=blue] (\x*0.25,\y*0.25) circle (0.05);
    }
    \end{scope}
\begin{scope}[scale=1.8, shift={(1.8,0)}] 
    \draw[red,latex-latex] (-0.2,0.3) -- (0.2,0.3);  \draw[red,latex-latex] (-0.2,0.3) -- (0.2, -0.3);
      \draw[red,latex-latex] (-0.2,-0.3) -- (0.2,-0.3);  \draw[red,latex-latex] (-0.2,-0.3) -- (0.2, 0.3);
      \draw[latex-latex]  (-0.3,-0.2) --  (-0.3,+0.2);    \draw[latex-latex]  (0.3,-0.2) --  (0.3,+0.2);
      \foreach  \x in {-1,1}
     \foreach  \y in {-1,1} {
    \draw[fill=blue] (\x*0.3,\y*0.3) circle (0.05);
    }
        \end{scope}
\begin{scope}[scale=1.8, shift={(3,0)}] 
    \draw[latex-latex]  (-0.3,0) --  (0.3,0);
    \draw[red,-latex] (-0.4,0) to[bend left=30] (0.4,0.1); 
    \draw[red,-latex] (0.4,0) to[bend left=30] (-0.4,-0.1); 
    \draw[red,-latex, looseness=8,  out=60, in=120] (-0.4,0) to (-0.45,0.1);
     \draw[red,-latex, looseness=8,  out=-120, in=-60] (0.4,0) to (0.45,-0.1);

     \draw[fill=blue] (-0.4,0) circle (0.05);    \draw[fill=blue] (0.4,0) circle (0.05);
     \end{scope}
\end{tikzpicture}
\ea
\ee 
where dots of the same color denote gapped vacua with the same Euler counterterm and we have indicated in black/red the action of the invertible and non-invertible symmetry generators, respectively. The first phase is present in all $\TY$ categories and corresponds to the maximal breaking of the $\TY$ symmetry -- i.e. the regular module category -- the other two instead preserve the $\bZ_2$ subgroup of $\TY(\bZ_4,+)$. The two ASCies associated to $\TY(\bZ_4,+)$ have, respectively, one and three gapped phases:
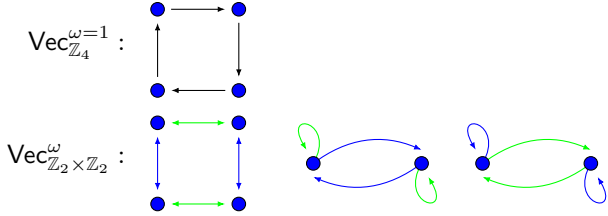
\be\ba
\Vec_{\bZ_4}^{\omega=1} &: \ \  
\begin{tikzpicture}[baseline={(0,0)}]
\begin{scope}[scale=1.8]
\draw[-latex]  (-0.3,-0.2) -- (-0.3,+0.2);   \draw[-latex]  (0.3,0.2) -- (0.3,-0.2);
      \draw[-latex]  (-0.2,0.3) -- (0.2,0.3);   \draw[-latex]  (0.2,-0.3) -- (-0.2,-0.3);
         \foreach  \x in {-1,1}
     \foreach  \y in {-1,1} {
    \draw[fill=blue] (\x*0.3,\y*0.3) circle (0.05);
    }
    \end{scope}
      \end{tikzpicture} \\
\Vec_{\bZ_2 \times \bZ_2}^\omega &: \ \     
\begin{tikzpicture}[baseline={(0,0)}] 
\begin{scope}[scale=1.8]
\draw[latex-latex,blue]  (-0.3,-0.2) -- (-0.3,+0.2);   \draw[latex-latex,blue]  (0.3,0.2) -- (0.3,-0.2);
      \draw[latex-latex,green]  (-0.2,0.3) -- (0.2,0.3);   \draw[latex-latex,green]  (0.2,-0.3) -- (-0.2,-0.3);
         \foreach  \x in {-1,1}
     \foreach  \y in {-1,1} {
    \draw[fill=blue] (\x*0.3,\y*0.3) circle (0.05);
    }
    \end{scope}
\begin{scope}[scale=1.8, shift={(1.25,0)}]
    \draw[blue,-latex] (-0.4,0) to[bend left=30] (0.4,0.1); 
    \draw[blue,-latex] (0.4,0) to[bend left=30] (-0.4,-0.1); 
    \draw[green,-latex, looseness=8,  out=60, in=120] (-0.4,0) to (-0.45,0.1);
     \draw[green,-latex, looseness=8,  out=-120, in=-60] (0.4,0) to (0.45,-0.1);
     \draw[fill=blue] (-0.4,0) circle (0.05);    \draw[fill=blue] (0.4,0) circle (0.05);   
    \end{scope}
\begin{scope}[scale=1.8, shift={(2.5,0)}]
    \draw[green,-latex] (-0.4,0) to[bend left=30] (0.4,0.1); 
    \draw[green,-latex] (0.4,0) to[bend left=30] (-0.4,-0.1); 
    \draw[blue,-latex, looseness=8,  out=60, in=120] (-0.4,0) to (-0.45,0.1);
     \draw[blue,-latex, looseness=8,  out=-120, in=-60] (0.4,0) to (0.45,-0.1);
     \draw[fill=blue] (-0.4,0) circle (0.05);    \draw[fill=blue] (0.4,0) circle (0.05);   
     \end{scope}
    \end{tikzpicture}
\ea\ee
It is clear that they describe (all) the non-maximal symmetry-breaking phases of $\TY(\bZ_4,+)$: we conclude that those {\bf symmetry-breaking patterns are anomaly-enforced}.
\end{enumerate}

\vspace{2mm}
\noindent \textbf{\boldmath$\TY(\bZ_2 \times \bZ_2,\chi_d,-)$.} Another interesting example is that of a $\TY$ category whose anomaly strictly lies in the ``second obstruction", that is in a nontrivial Frobenius-Schur indicator. To this end we consider the $A=\bZ_2 \times \bZ_2$ Tambara-Yamagami category that does not admit a fiber functor, i.e., the one with diagonal bicharacter $\chi_d$ and nontrivial FS indicator $\epsilon=-1$. We will henceforth drop $\chi_d$ for ease of notation.
The anyon composition of its center is shown in Table \ref{tab:spinsTYZ2Z2}. 
\begin{table}[t]
      \centering
      {
    \renewcommand{\arraystretch}{1.5}
    \begin{tabular}{|c|c|c|c|c|c|}
    \hline
       & $X_{(0,0),\pm 1}$  & $X_{(1,0),\pm i}$ & $X_{(0,1),\pm i}$ & $X_{(1,1),\pm 1}$ & $Y_{(0,0),(1,0)}$  \\
       
       \hline

       $\theta$ & $1$ & $-1$ & $-1$ & $1$ & $1$
         \\
       
       \hline

       $d$ & $1$ & $1$ & $1$ & $1$ & $2$  \\

        \hline \hline

       & $Y_{(0,0),(0,1)}$ & $Y_{(0,0),(1,1)}$ & $Y_{(1,0),(0,1)}$ & $Y_{(1,0),(1,1)}$ & $Y_{(0,1),(1,1)}$ \\

       \hline

       $\theta$ & $1$ & $1$ & $1$ & $-1$ & $-1$ \\
       \hline
       $d$ & $2$ & $2$ & $2$ & $2$ & $2$\\
       
       \hline 
       \end{tabular}

    \vspace{0.3em}
    
       \begin{tabular}{|c|c|c|c|c|}
    \hline 
       & $Z_{\rho_1,\smash{\raisebox{0.05ex}{\scalebox{0.75}{$\pm \zeta_8^{3}$}}}}$ & $Z_{\rho_2,\pm\zeta_8}$ & $Z_{\rho_3,\pm i}$ & $Z_{\rho_4,\pm i}$ \\

       \hline

       $\theta$ & $\pm e^{\frac{3\pi i}{4}}$ & $\pm e^{\frac{\pi i}{4}}$ & $\pm i$ & $\pm i$ \\
       \hline
       $d$ & $2$ & $2$ & $2$ & $2$ \\
       \hline
    \end{tabular}
    }
    \caption{Spins and quantum dimensions of the simple lines of $\cZ(\TY(\bZ_2\times \bZ_2))$. Here $\zeta_8 = \exp(\frac{2\pi i}{8})$ and $X_{(0,0),+1}=1$.}
    \label{tab:spinsTYZ2Z2}
\end{table}
The symmetry boundary condition is described by the Lagrangian algebra:
\begin{align}
    \cL_{\TY(\bZ_2 \times \bZ_2,-)} & =  1 \oplus X_{(0,0),-1} \oplus Y_{(0,0),(1,0)} \nn\\
& \quad \oplus Y_{(0,0),(0,1)} \oplus Y_{(0,0),(1,1)}\,.
\end{align}

There is only a single largest magnetic algebra in this case, which has dimension 4 \cite{GaiSchaferNamekiWarman}:
\be
 \cA_{\DFun} = 1 \oplus X_{(1,1),+1} \oplus Y_{(1,0), (0,1)} \, .
\ee
Anyon condensation leads to the double-semion theory as the reduced TO:
\be
\cZ(\TY(\bZ_2 \times \bZ_2,-))/\cA_{\DFun} = \cZ(\Vec_{\bZ_2}^{\omega=1}) \, ,
\ee
The map $\phi _\DFun$ that defines this functor is 
\be\phi_{\DFun}:\quad 
\left\{\ba
 1 \oplus X_{(1,1),+1} \oplus Y_{(1,0), (0,1)} &\mapsto  1 \cr 
X_{(0,0),-1} \oplus X_{(1,1),-1} \oplus Y_{(0,0),(1,0)} &\mapsto e \cr 
(Z_{\rho_3,i} \oplus Z_{\rho_4,i}) &\mapsto s \cr 
(Z_{\rho_3,-i} \oplus Z_{\rho_4, -i})& \mapsto \bar{s} \,.
\ea \right. \ee 
By using the projections on the boundary categories we find the action of $\DFun: \TY (\Z_2\times\Z_2, -) \to \Vec_{\Z_2}^{\omega=1}$ on objects:
\be
\DFun(\cD) = 1\oplus\eta \, , \ \DFun(a_1) = \DFun(a_2) = 1\, .
\ee
The electric algebra satisfying \eqref{eq:trivially braiding algebra} and braiding with $\cA_\DFun$ trivially is:
\be
\cA_{\IFun} = 1 \oplus X_{(0,0),-1} \, .
\ee
It leads to the embedding $\Vec_{\bZ_2 \times \bZ_2} \overset{\IFun}{\rightarrow} \TY(\bZ_2 \times \bZ_2, -)$ giving rise to the RG-quiche:
\be
\label{eq: RGQuicheTY3}
\begin{tikzpicture} 
\begin{scope}[scale=0.85] 
\draw [\Ncolor,  fill=\Ncolor]
(-3,0) -- (-3,3) --(0,3) -- (0,0) -- (-3,0) ; 
 \draw [\Ccolor,  fill=\Ccolor]
(0,0) -- (0,3) --(3,3) -- (3,0) -- (0,0) ; 
 \draw [\Scolor,  fill=\Scolor]
(3,0) -- (3,3) --(6,3) -- (6,0) -- (3,0) ; 
\draw [very thick] (0,0) -- (0,3)  ;
\draw [very thick] (3,0) -- (3,3)  ;
\draw [very thick] (-3,0) -- (6,0)  ;
\node[below] at (-1.5,0) {$\Vec_{\bZ_2\times\bZ_2}$}; 
\node[below] at (1.5,0) {$\TY(\bZ_2\times\bZ_2,-)$}; 
\node[below] at (4.5,0) {$\Vec_{\bZ_2}^{\omega=1}$}; 
\node at (-1.5,1.5) {$\fZ(\Vec_{\bZ_2\times \bZ_2})$}; 
\node at (1.5,1.5) {$\fZ(\TY(\bZ_2^2,-))$}; 
\node at (4.5,1.5) {$\fZ(\Vec_{\bZ_2}^{\omega=1})$}; 
\node[above] at (0,3) {$\cI_{\IFun}$};
\node[above] at (3,3) {$\cI_{\DFun}$};
\end{scope}
\end{tikzpicture}
\ee
The categorical exact sequence:
\be
\Vec_{\bZ_2 \times \bZ_2} \overset{\IFun}{\longrightarrow} \TY(\bZ_2 \times \bZ_2 , -) \overset{\DFun}{\longrightarrow} \Vec_{\bZ_2}^{\omega=1} \, ,
\ee
conforms to our intuition: the $\TY(\bZ_2 \times \bZ_2,-)$ symmetry has trivial first obstruction to a fiber functor (a $\bZ_2 \times \bZ_2$ SPT which is invariant under gauging exists), but the duality symmetry still suffers from a  ``group-cohomology" 't Hooft anomaly, reflected in the nontrivial Frobenius-Schur indicator \cite{Antinucci:2023ezl}. The $\Vec_{\bZ_2}^{\omega=1}$ SSB phase pulls back to the minimal SSB phase for our TY category: 
\be
\begin{tikzpicture}
  \begin{scope}[scale=2]
    \draw[ red,-latex] (-0.4,0) to[bend left=30] (0.4,0.1); 
    \draw[red,-latex] (0.4,0) to[bend left=30] (-0.4,-0.1); 
    \draw[ red,-latex, looseness=8,  out=60, in=120] (-0.4,0) to (-0.45,0.1);
     \draw[red,-latex, looseness=8,  out=-120, in=-60] (0.4,0) to (0.45,-0.1);
     \draw[fill=blue] (-0.4,0) circle (0.05);    
     \draw[fill=blue] (0.4,0) circle (0.05);
     \end{scope}
\end{tikzpicture}
\ee
while the SSB of the invertible symmetry is not anomaly-enforced. Similar lessons can be drawn whenever the first obstruction for the $\TY$ category vanishes, and resonate with our physical picture of surjective tensor functors as anomaly quantifiers. We will discuss a similar example in (3+1)d in section \ref{sec:noninv WWW}.

\section{Emergent Higher-Form Symmetries and Fractionalization}\label{sec:WWW}

In this section we study anomaly matching in space-time dimension $d>2$, where a key new ingredient compared to (1+1)d is the presence of higher-form symmetries \cite{Gaiotto:2014kfa}. 
{The higher-dimensional generalization of tensor functors and short exact sequences for categories remains largely unexplored mathematical territory. However, by employing the SymTFT approach, we are able to navigate around some of these challenges and nevertheless formulate a precise quantification of anomalies in higher dimensions. We believe that all of these concepts should ultimately possess a rigorous mathematical foundation, and we support this perspective with a detailed discussion for the 2+1d case in appendix \ref{app:F2C}.}

Although they often emerge only at long distances, higher-form symmetries can play a crucial role in matching the anomalies of microscopic 0-form symmetries. The main concept here  is {\bf symmetry fractionalization}, as developed originally in the condensed matter literature \cite{Senthil:1999czm, Essin:2013rca,  Barkeshli:2014cna, Chen:2014wse}, while its manifestation in field theoretic RG-flows is known as {\bf symmetry transmutation} \cite{Seiberg:2025bqy} (see \cite{Delmastro:2022pfo, Brennan:2022tyl, Antinucci:2024bcm, Brennan:2025acl} for previous studies).
At the level of tensor functors, this is implemented by the fact that 0-form symmetry defects can map in the IR to composite defects built from emergent higher-form symmetries. In section \ref{sec:AnoEme}, we show how to incorporate such functors into the SymTFT interface framework. Sections \ref{sec:SET} and \ref{sec:nonmin} address symmetry-preserving gapped realizations, connecting with \cite{Wang:2017loc}, and we comment on non-invertible examples in section \ref{sec:noninv WWW}.

\subsection{Fractionalization and Transmutation}

Before starting the SymTFT interface analysis, let us briefly review what are the type of phenomena that we want to describe. A distinct feature of RG-flow in space-time dimension $d>2$ is that, the background field $A_1^{(\UV)}$ for a $\UV$ 0-form symmetry can sometimes force a non-trivial background field 
\begin{equation} \label{eq:symtrasm}
    B_{p+1}^{(\IR)}=\cF(A_1^{(\UV)})
\end{equation}
for a $p$-form symmetry that emerges at long distance, in the sense that its selection rules are only valid below some energy scale.

To understand the physics behind relations like \eqref{eq:symtrasm}, recall that the typical way in which $p$-form symmetries arise at long distance is that some $p$-dimensional defects $W_p$, which are endable in the UV, become unbreakable at long distances.
Letting 
\be
\partial W_p = \cO_{p-1} \,,
\ee
with $\cO_{p-1}$ a non-genuine operator\footnote{Often non-genuine operators refer to those attached to topological operators of one dimension higher. Here we use it also in the context of not necessarily topological operators.}, $W_p$ becomes unbreakable at long distance if $\cO_p$ becomes extremely heavy and decouples from the theory in the IR. In this limit, $W_p$ can carry a nontrivial emergent $p$-form symmetry charge at low energies.

\begin{figure}
    \centering
    $
\begin{tikzpicture}
\draw[thick,black](-3,-0.25) -- (-3,1.5) -- (-2+0.2,0.75) -- (-2+0.2,-1.75) -- (-3,-1) -- (-3,-0.25);
\draw[thick,black](-1,-0.25) -- (-1,1.5) -- (0+0.2,0.75) -- (0+0.2,-1.75) -- (-1,-1) -- (-1,-0.25);
\draw [thick,blue] (-4,-0.25) -- (-2.5,-0.25);
\draw [thick,blue] (-2+0.25,-0.25) -- (-0.5,-0.25);
\draw [thick,blue,dashed](-2.5,-0.25) -- (-2,-0.25);
\draw [thick,blue,dashed](-0.5,-0.25) -- (0,-0.25);
\draw [thick,blue] (0+0.25,-0.25) -- (1,-0.25);
\draw [black,fill=black] 
(-2.5,-0.25) ellipse (0.05 and 0.05);
\draw [black,fill=black] 
(-0.5,-0.25) ellipse (0.05 and 0.05);
\node[blue] at (-4.5,-0.25) {$W_p$};
\node[black] at (-2.2,1.5) {$g_1$};
\node[black] at (-0.2,1.5) {$g_2$};
\draw[thick,->] (-1.5,-2) -- (-1.5, -3);
\begin{scope}[shift={(1,-5)}]
\draw[thick,black](-3.1,-0.25) -- (-3.1,1.5) -- (-2+0.25,0.75) -- (-2+0.25,-1.75) -- (-3.1,-1) -- (-3.1,-0.25);
\draw[thick,blue] (-4,-0.25) -- (-2.5,-0.25);
\draw [thick,blue] (-2+0.35,-0.25) -- (-0.2,-0.25);
\draw [thick,blue,dashed](-2.5,-0.25) -- (-2,-0.25);
\node[blue] at (-4.5,-0.25) {$W_p$};
\node[black] at (-2,1.5) {$g_1g_2$};
\draw[Purple] (-2.5,-0.25) ellipse (0.25 and 0.5);
\node[above,scale=0.75] at (-2.5,0.25) {$\cF(g_1,g_2)$};
\draw [fill=black] (-2.5,-0.25) ellipse (0.05 and 0.05);
\node[rotate = 90] at (-2.5,-2) {$=$};
\end{scope}
\begin{scope}[shift={(1,-9)}]
    \draw[thick,black](-3.1,-0.25) -- (-3.1,1.5) -- (-2+0.25,0.75) -- (-2+0.25,-1.75) -- (-3.1,-1) -- (-3.1,-0.25);
\draw[thick,blue] (-4,-0.25) -- (-2.5,-0.25);
\draw [thick,blue] (-2+0.35,-0.25) -- (-0.2,-0.25);
\draw [thick,blue,dashed](-2.5,-0.25) -- (-2,-0.25);
\node[blue] at (-4.5,-0.25) {$W_p$};
\node[black] at (-2,1.5) {$g_1g_2$};
\draw [fill=black] (-2.5,-0.25) ellipse (0.05 and 0.05);
\node at (-5.7,-0.25) {$  \gamma_W (g_1, g_2)\, \times$};
\end{scope}
\end{tikzpicture}
    $
\caption{The projective action of $G_\UV$ on $W_p$ can be understood in terms of symmetry fractionalization. In this case $\gamma_W(g_1,g_2) = \cF(g_1,g_2)[W_p]$ is matched by the topological charge of $W_p$ under the fractionalized higher form symmetry.}
    \label{fig:projective line}
\end{figure}
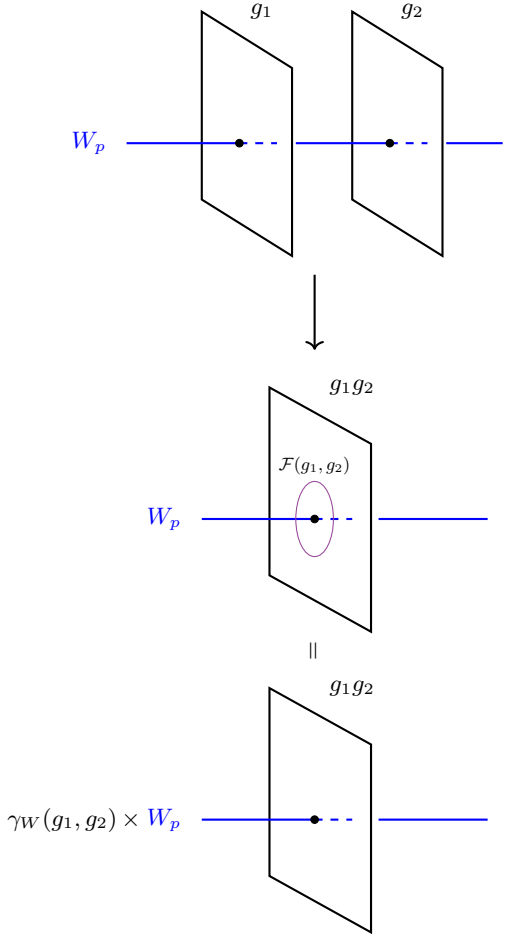

If the system has a UV 0-form symmetry $G_\UV$, all gauge invariant local operators transform in linear representations of $G_\UV$. On the other hand non-genuine operators $\cO_{p-1}$, which are not  gauge invariant, can transform in projective (higher) representations of $G_\UV$ \cite{Brennan:2022tyl,Delmastro:2022pfo,Bhardwaj:2023wzd}. At low energies, even if $W_p$ becomes unbreakable, it must transform in the same projective representation as $\cO_{p-1}$ see figure \ref{fig:projective line}. Such a projectivity can be understood as a manifestation of a defect anomaly \cite{Antinucci:2024izg}.

Put differently, in the IR, $W_p$ has to be charged under the 0-form symmetry $G_\UV$, i.e. a  network of $p$-form symmetry defects -- under which the $p$-dimensional operator is charged -- is induced at junctions of the $G_\UV$ symmetry defects, see figure \ref{fig:projective line}. The precise relation between the backgrounds is \eqref{eq:symtrasm}.

\vspace{2mm} 
\noindent {\bf Example.}
As a typical example consider $SU(N)$ QCD with massive quarks. The baryon symmetry $U(1)_B$ assigns unit charge to baryonic operators constructed out of $N$ quarks. Thus the (non gauge invariant) quark operator $q(x)$ has fractional $1/N$ charge under $U(1)_B$. Below the mass scale of the quarks, the Wilson lines $W_R$ become unbreakable and are charged under the emergent $\bZ_N^{(1)}$ 1-form center symmetry: the fundamental Wilson line is the world-line of $q(x)$, and carries $1/N$ charge under $U(1)_B$, implying that a background $A_1^{(\UV)}$ activates a background for $\bZ_N^{(1)}$:
\begin{equation}
    B_2^{(\IR)}\equiv c_1\left(A_1^{(\UV)}\right) \, (\text{mod }N) \in H^2(X, \bZ_N)\,,
\end{equation}
where $c_1\left(A_1^{(\UV)}\right)$ denotes the first Chern-class.

For our discussion, it will be crucial to understand the role played by this phenomenon for anomaly matching \cite{Delmastro:2022pfo, Brennan:2022tyl, Antinucci:2024bcm, Seiberg:2025bqy}: the inflow action of the UV symmetry is reproduced in the IR by an anomaly of the emergent higher-form symmetries through relations as \eqref{eq:symtrasm}. Indeed the UV anomaly for $G_\UV$ can be matched, in the IR, by a symmetry $G$ that either does not act on local operators, or it does but in a non-anomalous way.
This is due to the fact that the emergent higher-form symmetries can come with an anomaly, which may or may not involve $G$.
Plugging \eqref{eq:symtrasm} into the IR anomaly inflow action, we can match the UV one. This type of matching by higher-form symmetries is crucial in the discussion of symmetry preserving gapped phases for anomalous symmetries, such as the Wang-Wen-Witten construction \cite{Wang:2017loc}.
Let us illustrate this through a few elementary examples.

\vspace{2mm}\noindent {\boldmath{ $U(1)$ \bf in Even Dimensions.}}
An elementary example (see \cite{Antinucci:2024bcm} for more details) is a $U(1)$ symmetry in $d\in 2\bZ$ space-time dimensions with perturbative anomaly
\begin{equation}\label{eq:inflow U(1)}
    S_{\text{inflow}}^{(\UV)}=\frac{ik}{(2\pi)^{d/2}\left(d/2+1\right)!}\int _{X_{d+1}} A_1^{(\UV)}\wedge  \Big( dA_1^{(\UV)} \Big)^{d/2} \ .
\end{equation}
If the symmetry is spontaneously broken, then the IR is a compact Goldstone boson $\phi\sim \phi +2\pi$ where the $U(1)$ symmetry acts non-linearly, but it is (naively) non-anomalous, as there is no $U(1)$ WZW term. However the IR also has an emergent $U(1)^{(d-2)}$ $(d-2)$-form symmetry corresponding to the conservation of the winding number of $\phi$. The two symmetries have a mixed anomaly
\begin{equation}
    S_{\text{inflow}}^{(\IR)}=\frac{i}{2\pi}\int _{X_{d+1}} B^{(\IR)}_{d-1} \wedge dA^{(\IR)}_1 \ .
\end{equation}
This matches the UV anomaly in the IR only if $A_1^{(\IR)}=A_1^{(\UV)}$ and if $A_1^{(\UV)}$ induces a background for a $(d-2)$-form symmetry 
\begin{equation}
    B^{(\IR)}_{d-1}=\frac{k}{(2\pi)^{d/2-1} \left(d/2+1\right)!}  A_1^{(\UV)}\wedge   \left(dA_1^{(\UV)}\right)^{d/2-1} \,.
\end{equation}

\vspace{2mm}
\noindent {\boldmath{{\bf $U(1)_0$ in (2+1)d with Fermions.}}}
Another interesting example arises in $(2+1)$d $U(1)$ gauge theory with $N_f\in 2\bZ$ fermions $\psi_i$ and trivial total Chern-Simons level. The flavor symmetry is $G_\UV=U(N_f)/\bZ_{\frac{N_f}{2}}$, and if $N_f>2$ it has a self-anomaly \cite{Hsin:2024wki}
\begin{equation}
    S_{\text{inflow}}^{(\UV)}=\frac{2\pi i}{2N_f/2} \int _{X_4}w_2(A_1^{(\UV)}) \cup w_2(A_1^{(\UV)})\,,
\end{equation}
where $w_2(A_1^{(\UV)})\in H^2(X_4,\bZ_{N_f/2})$ represents the obstruction of lifting the $G_\UV$ bundle to a $U(N_f)$ bundle. Notice that the fermions $\psi_i$, that are not gauge invariant, transform linearly under $U(N_f)$ but projectively under $G_\UV$. Adding a mass $M>0$ for the fermions, the theory flows in the IR to a pure Chern-Simons theory $U(1)_{N_f/2}$. The Wilson line of charge $1$, being the world-line of $\psi_i$, becomes unbreakable and charged under and an emergent $\bZ_{N_f/2}^{(1)}$ 1-form symmetry. The fractional charge of $\psi_i$ then implies that $A_1 ^{(\UV)}$ forces a background 
\begin{equation}\label{eq:bckground CS}
    B_2^{(\IR)}=w_2(A_1^{(\UV)})
\end{equation}
for $\bZ_{N_f/2}^{(1)}$. While the flavor symmetry does not act on local operators in the IR, its anomaly is matched by the anomaly of the 1-form symmetry 
\begin{equation}
    S_{\text{inflow}}=\frac{2\pi i}{2N_f/2} \int _{X_4} B_2^{(\IR)}\cup B_2^{(\IR)}
\end{equation}
using \eqref{eq:bckground CS}.

While all relations like \eqref{eq:symtrasm} have an interpretation in terms of symmetry fractionalization \cite{Barkeshli:2014cna} of the UV 0-form symmetry with the emergent higher-form symmetry, we will see in the next subsections how to phrase them -- when the symmetries involved are finite -- as precise statements on the functor $\Fun : \cC_\UV \rightarrow\cC_\IR$ that we will derive from SymTFT interfaces.

\subsection{Anomaly Matching with Emergent Symmetries} 
\label{sec:AnoEme}

We now present two scenarios how an anomalous 0-form symmetry in the UV, can be matched with an emergent higher-form symmetry. In the current section we will construct a functor from 
\be
\cC_\UV = G_{\UV}^{(0), \omega}
\ee
to the IR symmetry, which involves a 0-form symmetry $G= G_\UV$ in terms of groups, but with trivial anomaly, and a $(d-2)$-form symmetry $\widehat{K}^{(d-2)}$, which have a mixed anomaly $\nu$: 
\be
\cC_\IR = G^{(0), \omega=0} \times^\nu \widehat{K}^{(d-2)} \,.
\ee
In section \ref{sec:SET} we furthermore construct a gapped phase, which is $G^{(0)}$-symmetric, which has an emergent $K^{(1)}$ 1-form symmetry, and no 0-form symmetry acting non-trivially on local operators. 

\subsubsection{General Analysis Starting with $G^{(0), \omega}_\UV$}

We present now a simple higher dimensional implementation of the RG-interfaces realized in the SymTFT construction. A related discussion also appeared recently in \cite{Perez-Lona:2025ncg}. The starting point is an anomalous invertible symmetry in $d$ spacetime dimensions given by a 0-form symmetry $G_\UV \cong G$ with anomaly $\omega \in H^{d+1}(BG_\UV,U(1))$. We denote this symmetry by 
\be
\cC_\UV= G^{(0),\omega}_\UV \equiv (d-1) \Vec_{G}^\omega\,,
\ee
and its generators by $\D_{d-1}^g$. We will denote $p$-form symmetries $A$ (whenever the degree is specified) by $A^{(p)}$. 
We want to use the SymTFT interfaces to produce a tensor functor $\Fun: \cC_\UV\rightarrow \cC_\IR$, where, 
in $\cC_\IR$ the UV symmetry $G^{(0),\omega}_\UV$ acts in a non-anomalous fashion. By this, we mean that the UV self-anomaly is matched by an emergent higher-form symmetry, and $G^{(0)}$ may act non-faithfully on local operators in the IR. Inspired by \cite{Wang:2017loc}, we construct $\cC_\IR$ in two steps: 
\begin{enumerate}
\item First, we find a group $\widetilde{G}$ which extends $G_\UV$ while trivializing the anomaly $\omega$, i.e. there is a short exact sequence
\begin{equation}\label{eq:group ext}
    1\longrightarrow K \overset{\iota}{\longrightarrow} \widetilde{G}\overset{p}{\longrightarrow} G_\UV \longrightarrow 1
\end{equation}
such that the pull-back $p^* \omega$ of the $G_\UV$ anomaly is trivial in $\wt{G}$: $p^*(\omega)=0$. We will assume that the extension is central, so that $K$ is abelian, and the extension data is determined by a class 
\be 
\nu \in H^2 (G_\UV, K) \,.
\ee
The SymTFT for this non-anomalous symmetry $\wt{G}$ will have an interface with  $\cZ (G^{(0),\omega}_\UV)$ obtained by condensing defects in $\cZ (\wt{G}^{(0)})$ that form an algebra $\cA$\footnote{When the bulk has dimension $\geq 4$ we use the terminology of condensable (and Lagrangian) algebras in a loose sense. For us they will be a collection of mutually local objects that can simultaneously terminate on an interface/boundary. Notice that they often involve defects of different dimensionality, tied together in a consistent manner. Lagrangian algebras are still maximal, in the sense that all topological operators outside the algebra braid nontrivially with some of its generators.}
\be
\fZ(G^{(0),\omega})=\fZ(\widetilde{G}^{(0)})/\cA \,,
\ee
which in the SymTFT takes the following form with the interface $\cI_\cA$ defined by the condensable algebra $\cA$: 
\be 
\label{RGQuiche WWW0}
\begin{split}
\begin{tikzpicture} 
 \draw [\UVcolor,  fill=\UVcolor]
(0,0) -- (0,3) --(3,3) -- (3,0) -- (0,0) ; 
 \draw [\IRcolor,  fill=\IRcolor]
(3,0) -- (3,3) --(6,3) -- (6,0) -- (3,0) ; 
\draw [very thick] (3,0) -- (3,3)  ;
\node at (1.5,1.5) {$\fZ(G^{(0),\omega}_\UV)$}; 
\node at (4.5,1.5) {$\fZ(\widetilde{G}^{(0)})$}; 
\node[above] at  (3,3) {$\cI_{\cA}$};
\end{tikzpicture}
\end{split}
\ee

\item Next, we construct an appropriate gapped BC $\Bsym_\IR$ for $\cZ (\wt{G}^{(0)})$, described by a Lagrangian algebra $\cL_{\IR}$. This will correspond to the IR symmetry $\cC_{\IR}$ to which $G_\UV^{(0),\omega}$ admits a tensor functor.
As we have pointed out in section \ref{sec:injective}, given a Lagrangian algebra $\cL_\IR$ that contains $\cA$ 
\be
\cA \subset \cL_{\IR}\,,
\ee
then the associated symmetries $\cC_\UV$ and $\cC_\IR$ automatically satisfy the (ME) \eqref{eq:masterequation}, defining an injective tensor functor:
\be
\IFun: \quad G^{(0),\omega}_\UV \to \cC_{\IR} \,.
\ee
From our construction of $\wt{G}$ \eqref{eq:group ext}, the gapped boundary condition $\Bsym_\IR$ can be reached from the canonical Dirichlet boundary condition in $\cZ(\wt{G})$ by gauging the normal subgroup $K\subset \widetilde{G}$. 

In $d$ spacetime dimensions, this means that $\cC_{\IR}$ will include a 0-form symmetry $G^{(0)}=\widetilde{G}/K$ as well as  a dual $\widehat{K}^{ (d-2)}$ $(d-2)$-form symmetry.
Notably, while $G$ is isomorphic to $G_\UV$ as a group, it has a trivial self 't Hooft anomaly inside $\cC_\IR$. On the other hand it has a mixed anomaly with $\widehat{K}^{(d-2)}$ determined by the extension \eqref{eq:group ext} \cite{Tachikawa:2017gyf}:
\be
S_{\text{inflow}} = 2 \pi i \int \beta\left(A_1^{(\IR)}\right) \cup C^{(\IR)}_{d-1} \, ,
\ee
where $A_1^{(\IR)}$ is the $G$-gauge field, $C^{(\IR)}_{d-1}$ the gauge field for the emergent $(d-2)$-form symmetry $\widehat{K}$ and $\beta$ the Bockstein map associated to the short exact sequence of groups.
We conclude that the IR symmetry then is
\be
\cC_{\IR} = G^{(0), \omega=0} \times^\nu \widehat{K}^{(d-2)} \,,
\ee
where $\times^\nu$ denotes the mixed 't Hooft anomaly \cite{Tachikawa:2017gyf}. In the SymTFT our discussion leads to the following setup: 
\be 
\label{RGQuicheWWWfull}
\begin{split}
\begin{tikzpicture} 
 \draw [\UVcolor,  fill=\UVcolor]
(0,0) -- (0,3) --(3,3) -- (3,0) -- (0,0) ; 
 \draw [\IRcolor,  fill=\IRcolor]
(3,0) -- (3,3) --(6,3) -- (6,0) -- (3,0) ; 
\draw [very thick] (3,0) -- (3,3)  ;
\draw [very thick] (0,0) -- (6,0);
\node at (1.3,1.5) {$\fZ(G^{(0),\omega}_\UV)$}; 
\node at (4.5,1.5) {$\fZ(\widetilde{G}) = \fZ (\cC_{\IR})$}; 
\node[above] at  (3,3) {$\cI_{\IFun}$};
\node[below] at (1.5,0) {$ G^{(0),\omega}_\UV$};
\node[below] at (4.5,0) {$G \times^\nu \widehat{K}^{(d-2)}$};
\end{tikzpicture}
\end{split}
\ee
The anomaly $\omega$ of the UV symmetry group $G_{\UV}^{(0),\omega}$ is non-trivially realized in the IR by $\cC_\IR$, which contains an anomaly-free 0-form symmetry $G$: while the topological defects of $G_\UV^{(0),\omega}$ are mapped into those of the anomaly-free IR subgroup $G \subset \cC_\IR$, the mapping of their morphisms is non-trivial. Specifically certain configuration of UV defects that intersect on a line, are mapped into configuration of IR symmetry defects where the line is dressed with a generator of the 2-form symmetry $\widehat{K}^{(d-2)}$. The UV anomaly is matched by a nontrivial symmetry fractionalization for the IR 0-form symmetry and encoded in this non-trivial modification of the junctions. We will show this very explicitly through examples.

\end{enumerate}

\subsubsection{\boldmath{Example: (3+1)d with $\bZ_N^{(0),k}$}}
Let us illustrate the previous discussion in $(3+1)$d space-time dimensions for $G_\UV=\bZ_N$. The group-cohomology anomaly $k$ takes value in $k\in H^5(B\bZ_N,U(1))=\bZ_N$, i.e. 
\be
\cC_{\UV} = \Z_N^{(0), \omega=k} \,,
\ee
and is encoded in the anomaly-inflow action
\begin{equation}\label{eq:UV anomaly Z3}
    S_{\text{inflow}}^{(\UV)}=\frac{2\pi i k}{N}\int_{X_5}A_1 ^{(\UV)}\cup \beta \left(A_1^{(\UV)}\right) \cup \beta \left(A_1^{(\UV)}\right)\, ,
\end{equation}
where $\beta$ is the Bockstein homomorphism. 
Such an anomaly can be trivialized by the extension 
\begin{equation}
    1\rightarrow \Z_N \rightarrow \Z_{N^2} \rightarrow \Z_N \rightarrow 1 \ ,
\end{equation}
from the previous discussion, the $\bZ_N$ anomaly can be matched by an IR symmetry $\bZ_N ^{(0)} \times \bZ_N ^{(2)} $ with a mixed anomaly
\begin{equation}
     S_{\text{inflow}}^{(\IR)}=\frac{2\pi i}{N}\int _{X_5} B_3 \cup \beta\left(A_1^{(\IR)}\right) \ .
\end{equation}
We now show how to obtain this result within the framework of the SymTFT.

The SymTFT for the UV symmetry is the (4+1)d twisted DW theory, with action (see \cite{Kaidi:2023maf} for a discussion of its salient properties)
\begin{equation}
    S_{\SymTFT} =\frac{2\pi i}{N}\int_{X_5}\Big(a \cup d b +k\, a \cup \beta (a) \cup \beta (a) \Big) \ .
\end{equation}
Here $a \in C^1(X_5,\bZ_N), b\in C^3(X_5,\bZ_N)$. The relevant topological operators are: 
\begin{enumerate}
    \item \textbf{Line operators} $\Q_1^{(n_e)}$ labeled by $n_e \in \bZ_N$
    \begin{equation}
        \Q_1^{(n_e)}(\gamma_1)=\exp{\left(\frac{2\pi i n_e}{N}\int _{\gamma_1} a\right)}  \ .
    \end{equation}
    \item \textbf{3-surface operators} $\Q_3^{(n_m, l)}$ labeled by $n_m, \,l \in \bZ_N$
    \begin{equation}
       \Q_3^{(n_m,l)}(\Sigma_3)=\exp{\left(\frac{2\pi i}{N}\int _{\Sigma_3}n_m\, b+l\, a\cup \beta(a) \right)} \ .
    \end{equation}
    The defects $\Q_3^{(0,l)}$ can be interpreted as theta defects \cite{Bhardwaj:2022kot, Bhardwaj:2022lsg} characterized by a gauged SPT $l\in H^3(B\bZ_N,U(1))\cong \bZ_N$ on their worldvolume.
\end{enumerate}
 Electric and magnetic defects, $\Q_1^{(n_e)}(\gamma_1)$ and $\Q_3^{(n_m,l)}(\Sigma_3)$, have canonical (ordinary) linking given by 
\begin{equation}
\langle \Q_3^{(n_m,l)}(\Sigma_3) \, \Q_1^{n_e}(\gamma_1) \rangle  =  \exp{\left(\frac{2\pi i}{N}n_e n_m \text{Lk}(\gamma_1,\Sigma_3)\right)} \ .
\end{equation}
Furthermore, there is a nontrivial (type 2) triple linking between $\Q_3^{(n_m^i,l^i)}(\Sigma^i_3)$ defects:\footnote{The geometric triple linking invariant $\text{Lk}(\Sigma_3^1,\Sigma_3^2,\Sigma_3^3)_2$ can be described as the ordinary linking number between $\Sigma_3^1$ and the line $\Sigma_3^2\cap \Sigma_3^3$.} 
\begin{equation} \label{eq:triple linking}
\ba
&\langle \Q_3^{(n_m^1,l^1)}(\Sigma_1) \Q_3^{(n_m^2,l^2)}(\Sigma_2) \Q_3^{(n_m^3,l^3)}(\Sigma_3) \rangle \cr 
&=    \displaystyle  \exp \left(\frac{4\pi i \text{Lk}(\Sigma_3^1,\Sigma_3^2,\Sigma_3^3)_2}{N^2}
    \Big(n_m^1n_m^2l^3+n_m^1l^2n_m^3+ \right.  \cr 
&  \qquad  \ \ \ \ \  \left.  + l^1 n_m^2 n_m^3 -\frac{3 k}{N} n_m^1n_m^2n_m^3\Big) \right)  \,.
\ea
\end{equation}
This is particularly relevant for us, as it is the simplest bulk observable detecting the anomaly. 
We give a derivation of this equation and more details which are needed to understand algebras in appendix \ref{app:link}.

The canonical Dirichlet boundary $\Bsym_\UV$ with $G_\UV$ symmetry corresponds to a Lagrangian algebra:
\begin{equation}
  \cL_{\UV}=\Big\langle \Q_1^{(1)}, \Q_3^{(0,1)} \Big\rangle=\left\{ \Q_1^{(n_e)}, \Q_3^{(0,l)},  \ n_e,l\in \bZ_N \right\} \ .  
\end{equation}
Notice that all of the theta-defects $\Q_3^{(0,l)}$ must be added to get a Lagrangian algebra, as they braid trivially with $\Q_1^{(n_e)}$. Once this is done, the set becomes maximal: all the other defects have non-trivial braiding with at least one element of $\cL_{\UV}$. 
The symmetry generators are the projection of $\Q_3^{(n_m,l)}$ onto the boundary: 
\begin{equation}
     \D_3^{(n_m)}=\pi_\UV \left(\Q_3^{(n_m,l)}\right) \ .
 \end{equation}

\vspace{2mm} \noindent \textbf{The RG-quiche.} We will now present an explicit derivation of the RG-quiche leading to the enhanced $\cC_\IR$ symmetry. For simplicity we will restrict our attention to the case of odd $N$ and $\gcd(N,k)=1$. The more general case is similar, but requires further sophistication.\footnote{The case of $N$ even is technically more subtle, as the interfaces are not obtained by simple condensation, but by condensation with discrete torsion valued in $H^5(B^2 \bZ_N,U(1))\cong\bZ_{\text{gcd}(N,2)}$.}
To satisfy \eqref{RGQuiche WWW0} we take $K=\bZ_N$ and $\widetilde{G}=\bZ_{N^2}$. We will denote topological defects in $\fZ(\wt{G})$ by $\wt{\Q}$.
The IR gapped boundary condition corresponding to 
\be
\cC_\IR= \widehat{\bZ}_N^{(2)} \times^\nu \bZ_N^{(0)}
\ee
is implemented by the Lagrangian algebra:
\be
\cL_\IR = \left\{ \wt{\Q_1}^{N r_e} \, , \wt{\Q_3}^{N r_m, N s}\ | \ r_e,r_m,s=0,...,N-1 \right\} \, ,
\ee
the algebra multiplication map is nontrivial and assigns to a codimension-two intersection $\gamma_{ij} = \Sigma_i \cap \Sigma_j$ a dressing by the line
\be \wt{\Q_1}^{m_e^{ij}} \, , \quad m_e^{ij} = -( r_m^i s^j + r_m^j s^i  ) + N \wt{k} r_m^i r_m^j \, , \ee 
where $\wt{k}$ will be related to the possible symmetry fractionalization classes.
This ensures that the triple linking between generators is always trivial. It can be checked that the triple linking of no other 3-surface operator  can be trivialized by the same token. The IR symmetry operators are:
\be 
\ba 
   &\widetilde{\D}_3^{(n_m)} =\pi _{\IR}\left(\widetilde{\Q}_3^{(n_m+N r_m, N s)}\right) \,, \ \   &&n_m=0,...,N-1 \cr 
    &\widetilde{\L}_1^{(n_e)} \ =\pi_{\IR}\left(\widetilde{\Q}_1^{(n_e+N r_e)}\right) \,, \ \   &&n_e=0,...,N-1 \ .
\ea 
\ee
together with the generalized theta-defects constructed from the 2-form symmetry:
\be
\widetilde{\Theta}_3^{(l)}=\pi_{\IR}\left( \widetilde{\Q}_3^{(0,l+Ns)}\right) \ , \ \  \ \ \ l=0,...,N-1 \ .
\ee
We now want to construct the interface algebra $\cA_k$ landing us on $\fZ(G^{(0), \omega=k})$. As per our previous discussion, such algebra should be \textbf{electric} in order to give rise to a tensor functor. 
The correct algebra is:
\be
\cA_k = \Big\langle \wt{\Q}_3^{(N r_m,0)}   \Big\rangle \, , \ \ \ \ m_e^{ij} = N \wt{k} r_m^i r_m^j \, ,
\ee
with $\wt{k} = - 2^{-1} k^{-1}$. Notice that this is an electric algebra, that can be completed into $\cL_\IR$.
The condensation map identifies:
\be \ba \label{eq:mapping bulk defects}
&\wt{\Q}_1^{(N n_e)} &&\mapsto \quad \Q_1^{(n_e)} \, , \\
&\wt{\Q}_3^{(n_m, - N k n_m)} &&\mapsto \quad \Q_3^{(n_m,0)} \, .
\ea \ee 
By computing the triple linking of $\wt{\Q}_3^{(n_m, - N k n_m)}$, it is straightforward to check that they reproduce the cubic anomaly $k \in H^5(\bZ_N, U(1))$.
This is compatible with the tensor functor:
\be
\IFun(\D_3^{(n_m)}) = \wt{\D}_3^{(n_m)} \, .
\ee
More interestingly, the symmetry fractionalization class in the IR can be detected via the dressing of certain junctions of 0-form symmetry generators. To see this, consider a boundary UV defect $\D_3^{(1)}(\Sigma_3)$ on a three-manifold $\Sigma_3$ for which ($\PD$ denotes the Poincare dual):
\be
\PD(\Sigma _3) \wedge d \PD(\Sigma_3) = N \PD(\gamma_1) \, .
\ee
Using \eqref{eq:mapping bulk defects}, we see that the tensor functor maps it into $\pi_\IR(\wt{\Q}_3^{(1,-Nk)})$ in the IR. The surface $\Sigma_3$ turns on the symmetry fractionalization background we want to detect.
Lifting the setup slightly into the bulk, this is encoded in the action:
\be \ba
S[\Sigma] &= \frac{2\pi i}{N^2}\int_{X_5} (da_\IR + d \PD(\Sigma) ) \, b_\IR  \\
&{} \quad \quad - \frac{2 \pi i k}{N^3} \int_{X_5} d\PD(\Sigma) a_\IR d a_\IR  \,,
\ea \ee
shifting $a_\IR \to a_\IR - \PD(\Sigma)$ we find:
\be
S[\Sigma] = S + \frac{2 \pi i}{N^2} \int_{\partial X_5} \PD(\gamma) \, a_\IR \, .
\ee
Which corresponds to the insertion of an emergent symmetry defect $\pi_{\IR}(\wt{\Q}_1^{k})=\wt{\L}_1^k$ along $\gamma$. 
This result is consistent with the symmetry fractionalization of backgrounds:
\be \label{eq:identification background}
A_1^{(\IR)} = A_1^{(\UV)} \, , \quad B_3^{(\IR)} = k A_1^{(\UV)} \cup \beta(A_1^{(\UV)}) \, ,
\ee
and we have recovered it geometrically in the SymTFT.

\vspace{2mm} \noindent \textbf{RG-Quiche from Gauge Fields.} We conclude by presenting an alternative perspective on the RG-quiche interface using the bulk action.
The SymTFT actions in the UV and IR are, respectively:
\be \ba
S_{\UV} &= \frac{2 \pi i}{N} \int a_\UV \cup d b_\UV + \frac{k}{N} a_\UV\cup \beta(a_\UV)^2 \, , \\
S_{\IR} &= \frac{2 \pi i}{N} \int a'\cup d \wt{b} + \wt{a}\cup d b' + \frac{2 \pi i}{N^2} \int a' \cup d b' \, ,
\ea \ee
where we have decomposed the IR gauge fields in $a_\IR = a' + N \wt{a}$ and $b_\IR =b' + N \wt{b}$, respectively. We now consider an interface $\cI_\IFun$ between the two theories. Gauge invariance requires:
\be \ba
0&=\frac{2 \pi i}{N} \int_{\cI_\IFun} \biggl( d \lambda' \cup \left[ \widetilde{b} + \frac{1}{N} b' \right] + d \wt{\lambda} \cup b' \biggr. \\
&\biggr. \qquad    - d \lambda_\UV \cup \left[ b_\UV + \frac{k}{N} a_\UV \cup \beta( a_\UV) \right]   \biggl) \, .
\ea \ee
To cancel the gauge variation we identify $a'=a_\UV$ at the interface, together with:
\be \label{eq: interfacecont}
\wt{b} = b_\UV \, , \quad b'  =k a_\UV \cup \beta( a_\UV) \, ,
\ee
furthermore $\wt{a}$ remains dynamical on the interface, and the gauge variation is compensated by an interface term $\wt{a} \cup b'$. This corresponds to a Neumann boundary condition for $\wt{a}$.
A UV symmetry defect:
\be
\Q_3^{(n_m,0)} = \exp\left( \frac{2 \pi i n_m}{N} \int_\Sigma b_\UV \right) \, ,
\ee
is mapped into:\footnote{We write $\frac{1}{N} \wt{b}$ as $\frac{1}{N^2}(b' + N\wt{b}) - \frac{1}{N^2} b'$ and use the gluing conditions.}
\be \ba
\phi\left(\Q_3^{(n_m,0)}\right) = &\exp\left( \frac{2 \pi i n_m}{N^2}\int b_\IR - k N a_\IR \cup \beta(a_\IR)  \right) \cr 
\equiv & \wt{\Q}_3^{(n_m, - k N n_m)} \, , 
\ea \ee
matching our previous discussion. Notice that, above, $\beta(a_\IR) = d a_\IR / N^2$ is the IR Bockstein map.
The IR boundary conditions are simply Dirichlet boundary conditions for $a'$ and $b'$. The matching equation \eqref{eq:masterequation} requires that we should impose the same gapped boundary condition by first acting with $\cI_\IFun$ and then giving Dirichlet boundary condition for $a_\UV$. This follows immediately from our discussion above.

The symmetry fractionalization can also be seen right away. Consider equation \eqref{eq: interfacecont} in the presence of the $\cL_\IR$ boundary conditions $a' = A_1^{(\UV)}$, $b'=B_3^{(\IR)}$, it reads:
\be
B_3^{(\IR)} = k A_1^{(\UV)}\cup \beta(A_1^{(\UV)}) \, ,
\ee
implementing anomaly matching via the fractionalization of the emergent two-form symmetry.

This example shows clearly how the related phenomena of symmetry transmutation and symmetry fractionalization are described in our general framework of SymTFT interfaces.

\subsection{SETs and Relation to Wang-Wen-Witten} \label{sec:SET}

In section \ref{sec:AnoEme} we considered an anomalous $\cC_\UV= G^{(0),\omega}$ and constructed an embedding into $\cC_{\IR}$, that contains a non-anomalous $G_\UV^{(0),\omega=0}$ sub-symmetry. 

In this section we will furthermore trivialize the symmetry $G^{(0),\omega=0}$. Specifically, we will  consider a symmetric phase for this non-anomalous sub-symmetry. In this phase, the higher-form symmetry will be broken, corresponding to a {\bf symmetry-enriched topological order (SETO)}.

\subsubsection{General Analysis Starting with $G_\UV^{(0),\omega}$}

We will now discuss IR gapped phases preserving the non-anomalous 0-form symmetry, in other words the Wang-Wen-Witten construction \cite{Wang:2017loc}. 
Building on the construction of the RG-quiche in the last subsection, we construct symmetry preserving gapped phases for $G$ by choosing suitable (gapped) physical boundary conditions in $\fZ(\cC_\IR)$. 
Requiring the 0-form symmetry $G^{(0)}$ to be unbroken in such a phase will give rise to a SETO -- namely a nontrivial topological order.

We start our analysis with the setup in section \ref{sec:AnoEme}, where $\cC_\IR$ includes 
\be
G^{(0), \omega=0} \times^\nu \widehat{K}^{(d-2)} \,,
\ee
as well as the theta defects and condensates constructed out of $\widehat{K}^{(d-2)}$.
The mixed anomaly $\nu$ follows from the extension  
\be
1\rightarrow K \rightarrow \widetilde{G}\xrightarrow{p} G\rightarrow 1\,.
\ee
Recall that  $K$ is abelian.
Notice, that according to our SymTFT criterion in section \ref{sec:normal subcat}, $G^{(0),\omega=0}\subset \cC_\IR$ is an anomaly-free but non-normal subcategory. 
Hence, a $G^{(0)}$-symmetric gapped phase necessarily requires the spontaneous breaking of further emergent symmetries.

To study the symmetric gapped phase, we close the SymTFT sandwich and insert a {gapped} physical boundary $\Bphys$: 
\be 
\label{RGsemisandiwch}
\begin{split}
\begin{tikzpicture} 
 \draw [\UVcolor,  fill=\UVcolor]
(-0.5,0) -- (-0.5,3) --(3,3) -- (3,0) -- (-0.5,0) ; 
 \draw [\IRcolor,  fill=\IRcolor]
(3,0) -- (3,3) --(6,3) -- (6,0) -- (3,0) ; 
\draw [very thick] (3,0) -- (3,3)  ;
\draw [very thick] (-0.5,0) -- (6,0)  ;
\draw [very thick] (3,3) -- (6,3)  ;
\node[below] at (1.5,0) {$\Bsym_{\UV}$}; 
\node[below] at (4.5,0) {$\Bsym_{\IR}$}; 
\node[above] at (4.5,3) {$\Bphys$}; 
\node at (1.3,1.5) {$\fZ(G_\UV^{(0),\omega})$}; 
\node at (4.5,1.5) {$\fZ(\widetilde{G}^{(0)})= \fZ (\cC_\IR)$}; 
\node[above] at  (3,3) {$\cI_{\cA}$};
\end{tikzpicture}
\end{split}
\ee
We would like to make the sub-symmetry $G^{(0),\omega=0} \subset \cC_\IR$ act trivially in this phase. To this end, we need $\Bphys$ to be a magnetic boundary condition with respect to $\widetilde{G}$, i.e. condensing the fluxes (labeled by conjugacy classes of $\wt{G}$)\footnote{For a non-abelian $G$, genuine elements of the center are labeled by conjugacy classes $\Q_3^{[g]}$ and irreducible representations $\Q_1^{\rho}$. The full categorical structure is encoded in $\cZ(n\Vec_{G}) = \displaystyle \boxplus_{[g] \in G} n\Rep(H_g)$, where $H_g$ is the centralizer of $[g]$. This structure generalizes the $H^3(G,U(1))$ labels $l$ we have encountered in the previous discussion. We will not need these subtleties in what follows.}: 
\be
\cL_{\text{phys}}=\cL_{\text{mag}, \wt{G}^{(0)}}  = \bigoplus_{\wt{g} \in \wt{G}}\wt\Q_{d-1}^{[\wt{g}]}   \,.
\ee
With this choice, the UV $G_\UV^{(0),\omega}$-symmetry generators $\Q_{d-1}^{[g]}$, $g\in G$, can cross the interface and end on $\Bphys$. On the other hand the defects $\wt\Q_{d-1}^{k}$, $k\in K$, can end both on the physical and $\Bsym_\IR$ boundaries:  
\be 
\label{RGClopenSandwich}
\begin{split}
\begin{tikzpicture} 
 \draw [\UVcolor,  fill=\UVcolor]
(-0.5,0) -- (-0.5,3) --(3,3) -- (3,0) -- (-0.5,0) ; 
 \draw [\IRcolor,  fill=\IRcolor]
(3,0) -- (3,3) --(7.5,3) -- (7.5,0) -- (3,0) ; 
\draw [ultra thick] (3,0) -- (3,3)--(7.5,3)  ;
\draw [ultra thick] (-0.5,0) -- (7.5,0)  ;
\node[below] at (1.5,0) {$\Bsym_{\UV}$}; 
\node[below] at (5.5,0) {$\Bsym_{\IR}$}; 
\node[above] at (5.5,3) {$\Bphys$}; 
\node at (1.3,1) {$\fZ(G_\UV^{(0),\omega})$}; 
\node at (4.4,1) {$\fZ(\widetilde{G}^{(0)})= \fZ (\cC_\IR)$}; 
\node[above] at  (3,3) {$\cI_{\cA}$};
\draw[thick, rounded corners](-0.5,2) -- (5.5,2) -- (5.5,3);
\draw [black,fill=black] (5.5, 3) ellipse (0.05 and 0.05);
\node[above] at (1.5, 2) {$\Q_{d-1}^{[g\in G]}$};
\node[above] at (4.2, 2) {$\wt{\Q}_{d-1}^{[\wt{g}: p(\wt{g})=g]}$};
\draw[thick] (5.95,0) -- (5.95,3);
\draw [black,fill=black] (5.95, 3) ellipse (0.05 and 0.05);
\draw [black,fill=black] (5.95, 0) ellipse (0.05 and 0.05);
\node[right] at (5.95, 1.5) {$\wt{\Q}_{d-1}^{k\in K}$};
\end{tikzpicture}
\end{split}
\ee
The boundary of $\widetilde{\Q}_{d-1}^{k}$, for $k\in K$, produces a $(d-2)$-dimensional topological defect 
\be
    \U_{d-2}^{k} = \partial \widetilde{\Q}_{d-1}^{k} \,, \quad k\in K\,,
\ee
which generates an {\bf emergent $1$-form symmetry} $K^{(1)}$ in the gapped phase. The full symmetry $\cD_\IR$ of this phase does not have 0-form symmetries acting on local operators (it is a $G^{(0)}$-preserving phase) but a 1-form symmetry and $(d-2)$-form symmetry 
\be 
\cD_\IR \supset     K^{(1)}\times \widehat{K}^{(d-2)}\,,
\ee
 with a mixed anomaly
\begin{equation}\label{mixedanomi}
    S^{(\IR)}_{\text{inflow}}=2\pi i \int _{X_{d+1}}C_{2}^{(\IR)}\cup B_{d-1}^{(\IR)}
\end{equation}
with $C_2^{(\IR)}\in H^2(X,\widehat{K})$, $B_{d-1}^{(\IR)}\in H^{d-1}(X,K)$. 
The IR phase is a {\bf Symmetry Enriched Topological Order (SETO)} .  
The effective field theory for this phase is the standard, untwisted, $(d+1)$-dimensional Dijkgraaf-Witten theory \cite{Dijkgraaf:1989pz, Kapustin:2014gua} with gauge group $K$. As we will see very explicitly in an example below, the UV symmetry enforces that this theory is enriched with a symmetry $G$, in the sense that a background field $A_1^{(\UV)}$ for $G$ activates backgrounds for $K^{(1)}$ and $\widehat{K}^{(d-2)}$. This matches the anomaly without breaking the symmetry, and reproduces the general mechanism due to Wang-Wen-Witten \cite{Wang:2017loc}.

\subsubsection{Example: $(3+1)$d with $G_\UV^{(0),\omega}=\bZ_N^{(0),k}$}
Let us concretely consider again $(3+1)$d, with $G_\UV^{(0),\omega}=\bZ_N^{(0),k}$ and the anomaly \eqref{eq:UV anomaly Z3}. 
The physical boundary condition that makes the $G^{(0),\omega=0} \subset \cC_\IR$ act trivially is the magnetic boundary condition with respect to $\widetilde{G} = \Z_{N^2}$, 
\be
    \cL_{\text{phys}}=\bigoplus_{\widetilde{n}_m\in \widetilde{G}= \Z_{N^2}} \widetilde{\Q}_3^{(\widetilde{n}_m,0)} \,.
\ee
The boundary of $\widetilde{\Q}_3^{(\widetilde{n}_m\in K,0)}$ produces a 2-dimensional topological defect 
\begin{equation}
    \U_2^{(\widetilde{n}_m)} = \partial \widetilde{\Q}_3^{(\widetilde{n}_m,0)} \ , \ \ \ \ \  \ \widetilde{n}_m\in K
\end{equation}
 living on the boundary, that generates the {\bf emergent 1-form symmetry} $K^{(1)}$ in the gapped phase. The symmetry of this phase is $\cD_\IR=K^{(1)}\times \widehat{K}^{(2)}$
 with a mixed anomaly
\begin{equation}\label{eq:4d DW anomaly}
    S^{(\IR)}_{\text{inflow}}=2\pi i \int _{X_5}C_2^{(\IR)} \cup B_3^{(\IR)}
\end{equation}
with $B_3^{(\IR)}\in H^3(X,\widehat{K})$, $C_2^{(\IR)}\in H^2(X,K)$.
The effective theory is again the untwisted, $(3+1)d$ Dijkgraaf-Witten theory with gauge group $K$.

To understand how the UV anomaly is reproduced in this gapped phase, we need again to inspect how the $G_\UV$-symmetry defects $\D_3^{(n_m)}$ are mapped in the IR, equivalently how the backgrounds $B_3^{(\IR)}, C_2^{(\IR)}$ are related with $A_1^{(\UV)}$. The second identification of \eqref{eq:identification background} remains valid. On the other hand, as $G$ is extended by $K$ in $\cZ(\widetilde{G})$, fusing $\D_3^{(n_m)}$ and $\D_3^{(n_m')}$ we get $\D_3^{(n_m+n_m')}$ dressed with $\widetilde{\Q}_3^{(\nu(n_m,n_m'),0)}$:
\be
\centering
\begin{tikzpicture}[scale=3]
    \filldraw[color=blue, fill=white!70!blue, opacity=0.5] (0.6,-0.2) --(0.6,0.8) -- (1.35,1.25) -- (1.35,0.25);
 \filldraw[color=red, fill=white!70!red, opacity=0.5] (0.5,0.3) --(0.5,1.3) -- (1.35,1.25) -- (1.35,0.25); 
 \filldraw[color=violet, fill=orange!50!violet, opacity=0.5] (1.35,0.25) -- (1.35,1.25) -- (2.15,1.35) -- (2.15,0.35); 
 \node [] at (0.3,1.3) {\color{red} $\D_3^{(n_m)}$};
 \node [] at (0.44,-0.26) {\color{blue} $\D_3^{(n_m')}$};
 \node [] at (2.35,0.15) {\color{violet} $\D_3^{(n_m+n_m')} \color{black} \times \color{orange} \widetilde{\Q}_3^{(\nu(n_m,n_m'),0)}$};
\end{tikzpicture}
\ee
with $\nu\in H^2(BG_\UV,K)$ the extension class. But $\widetilde{\Q}_3^{(\nu(n_m,n_m'),0)}$ becomes trivial and leaves the 1-form symmetry defect $\U_2^{(\nu(n_m,n_m'))}$ on its boundary, that lives at the junction:
\be
\centering
\begin{tikzpicture}[scale=3]
    \filldraw[color=blue, fill=white!70!blue, opacity=0.5] (0.6,-0.2) --(0.6,0.8) -- (1.35,1.25) -- (1.35,0.25);
 \filldraw[color=red, fill=white!70!red, opacity=0.5] (0.5,0.3) --(0.5,1.3) -- (1.35,1.25) -- (1.35,0.25); 
 \filldraw[color=violet, fill=white!25!violet, opacity=0.5] (1.35,0.25) -- (1.35,1.25) -- (2.15,1.35) -- (2.15,0.35); 
 \node [] at (0.3,1.3) {\color{red} $\D_3^{(n_m)}$};
 \node [] at (0.44,-0.26) {\color{blue} $\D_3^{(n_m')}$};
 \node [] at (2.45,0.4) {\color{violet} $\D_3^{(n_m+n_m')} \color{black}  $};
 \node [] at (1.6,0.13) {\color{orange} $\U_2^{(\nu(n_m,n_m'))}$};
 \draw[line width=2mm, color=orange] (1.35,1.25) -- (1.35, 0.25);
\end{tikzpicture}
\ee
This is the the statement that $G_\UV$, while does not act on local operators, has symmetry fractionalization with $K^{(1)}$ controlled by the class $\nu\in H^2(G_\UV,K)$. 
This implies that the background $A_1^{(\UV)}$ activates a nontrivial background field for the emergent one-form symmetry:
\begin{equation}
    C_2^{(\IR)}=A_1^{(\UV) \, *}(\nu)=\beta \left( A_1^{(\UV)}\right) \ .
\end{equation}
Plugging this back into \eqref{eq:4d DW anomaly} together with the second equation in \eqref{eq:identification background} we immediately reproduce the UV anomaly \eqref{eq:UV anomaly Z3}.

\subsection{Non-Minimal Interfaces}
\label{sec:nonmin}

As we have seen in the last section, the symmetry $\cC_\IR$, that is the target of the functor $\IFun : G_\UV^{(0),k}\rightarrow \cC_\IR$ can be realized in a $G$-preserving gapped phase which  includes an emergent one-form symmetry. If we denote the symmetry of this phase by $\cD_\IR$, this suggests that there must also exist a different functor
\begin{equation}
    \Fun:G_\UV^{(0),k} \rightarrow \cD_\IR \,.
\end{equation}
To come full circle, we show how to describe this functor directly, without invoking the presence of the IR physical (gapped) boundary condition or  the symmetry $\cC_\IR$. 
In other words, we will construct the corresponding SymTFT interface $\cI_\Fun$ between $\fZ(\cC_\UV)$ and $\fZ(\cD_\IR)$.
The construction is quite general, and makes use of recent results about \textbf{non-minimal} gapped boundary conditions for higher-dimensional topological orders \cite{Bhardwaj:2024qiv, Wen:2024qsg, Bhardwaj:2025piv}.

The first observation is that $\cD_\IR$ is modular: any faithfully acting topological defect braids non-trivially with at least one other such defect. 
In our case the defects $\U_2^{(\widetilde{n}_m)}, \L_1^{(n_e)} $ generate $K^{(1)}$ and $\widehat{K}^{(d-2)}$ respectively, and have non-trivial braiding. The linking corresponds to the canonical action of $\widehat{K}$ onto $K$.

If a symmetry $\cC$ is modular, then the category of genuine topological defects of its SymTFT $\cZ(\cC)$ is trivial \cite{Decoppet:2023uoy}. In our case:
\begin{equation}
    \cZ(\cD_\IR)=(d-1)\Vec  \,.
\end{equation}
This means that, as a TQFT, $\fZ(\cC_\IR)$ must be an invertible theory, see e.g. \cite{Argurio:2024oym}.

An interface $\cI_{\Fun}$ between $\cZ(\cC_\UV)$ and an invertible TQFT is a boundary condition for the former. 
This appears puzzling at first sight: to satisfy the Matching Equation \eqref{eq:masterequation}, the boundary condition $\cI_\Fun$ needs to be magnetic. However, the absence of a magnetic boundary condition is precisely the way in which a 't Hooft anomaly for the $\cC_\UV$ symmetry is detected via the SymTFT.
The key point is that, although $\fZ(\cD_\IR)$ is invertible, it may still correspond to a nontrivial SPT phase.

The boundary condition $\cI_\Fun$ can then be a \textbf{non-minimal} one \cite{Bhardwaj:2024qiv, Wen:2024qsg, Bhardwaj:2025piv}. 
To construct it we start from the canonical Dirichlet boundary condition $\Bsym_\UV$ of $\fZ(\cC_\UV)$ and stack it with a decoupled $d$-dimensional TQFT $\cT$. The theory $\cT$ has two important properties: 
\begin{enumerate} 
\item First, it admits a $\cC_\UV$ action. For $\cC_\UV = G^{(0),\, \omega}_\UV$ this means that it can be coupled to $G_\UV^{(0)}$ gauge fields consistently. 
\item Second, it is a gapped boundary condition for the invertible theory $\fZ(\cD_\IR)$. Thus $\cT$ is modular, and the nontrivial braiding between its objects is encoded in the SPT $\fZ(\cD_\IR)$.
\end{enumerate}
This setup describes a factorized interface $\Bsym_\UV \boxtimes \cT$ between $\cZ(\cC_\UV)$ and $\cZ(\cD_\IR)$. 

Precisely when the Wang-Wen-Witten construction applies, we can furthermore choose the $G_\UV^{(0), \, \omega}$ action on $\cT$ as to cancel the 't Hooft anomaly stemming from the Dirichlet boundary condition $\Bsym$. We denote this theory by $\cT^G$. In practice, constructing $\cT^G$ from $\cT$ corresponds to the choice of appropriate symmetry fractionalization classes.

This setup provides us with an anomaly-free diagonal $G_\UV^{(0)}$ symmetry action on $\Bphys \boxtimes \cT^G$, which we can then gauge. The interface $\cI_F = \left(\Bsym_\UV \boxtimes \cT^G\right)/G$ is precisely the magnetic boundary condition we are after:
\be 
\label{RGsemisandiwch_nonmin}
\begin{split}
\begin{tikzpicture}[scale=1.2] 
 \draw [\UVcolor,  fill=\UVcolor]
(-0.5,0) -- (-0.5,3) --(3,3) -- (3,0) -- (-0.5,0) ; 
 \draw [white,  fill=gray!20!white]
(3,0) -- (3,3) --(6,3) -- (6,0) -- (3,0) ; 
\draw [very thick] (3,0) -- (3,3)  ;
\draw [very thick] (-0.5,0) -- (6,0)  ;
\node[below] at (1.5,0) {$\Bsym_{\UV}$}; 
\node[below] at (4.5,0) {$\Bsym_{\IR}$}; 
\node at (1.3,1.5) {$\fZ(G_\UV^{(0),\omega})$}; 
\node at (4.5,1.5) {$\fZ(\cD_\IR)=\text{invertible}$}; 
\node[above] at  (3,3) {$\displaystyle \cI_F=\frac{\Bsym_\UV \boxtimes \cT ^G}{G}$};
\end{tikzpicture}
\end{split}
\ee
Notice that in this picture, the IR theory does not need to be gapped. In fact the IR physical boundary can very well be gapless, and the modular symmetry $\cD_\IR$ can be realized in a CFT. However we do not gain much out of this observation, as a modular symmetry can always be decoupled from the rest of the theory \cite{muger2003structure,muger2003subfactors,Hsin:2018vcg,Cordova:2025eim}.

The lesson that we learn from this, is that in dimension $d>2$ it becomes natural to consider topological interfaces that are non-minimal, in the sense explained above. Including them in the analysis of SymTFT interfaces satisfying the Matching Equation can considerably enlarge the landscape of possible UV/IR functors, making the study in $d>2$ much richer and complicated, but nevertheless under control.

\vspace{2mm}
\noindent
{\bf Example: $(3+1)$d with $G_\UV^{(0),\omega}=\bZ_N^{(0),k}$.}
For concreteness, consider the construction above in the case of an anomalous $\bZ_N$ symmetry in $(3+1)d$. We take the $(3+1)d$ TQFT $\cT$ as the $\bZ_N$ DW theory:
\begin{equation}
    \cT: \quad \frac{2\pi i}{N}\int _{X_4} x_2 \cup dy_1  .
\end{equation}
We denote by $C_2$ and $B_3$ the background field for $\bZ_N^{(1)}$ and $\bZ_N^{(2)}$ respectively, that act on $e^{\frac{2 \pi i}{N} n_e\int y_1}$ and $e^{\frac{2 \pi i}{N} n_m\int x_2}$. 
The corresponding center is an invertible theory, with action:
\be
S_{\text{SymTFT}} = \frac{2\pi i}{N} \int b_2 \cup d c_2 + c_1 \cup d b_3 + c_2 \cup b_3 \, .
\ee
The theory $\cT$ is simply its Dirichlet boundary condition $b_3 = B_3, \, c_2 = C_2$.

The enrichment $\cT^{\bZ_N}$ with the 0-form symmetry is achieved as follows. We declare that the theory has a $0$-form symmetry $\bZ_N$ that does not permute lines and surfaces, but such that a background field $A_1$ activates 
\begin{equation}
    C_2=\beta(A_1) \ , \ \ \ \ B_3=-k A_1\cup \beta(A_1) \ .
\end{equation}
As we discussed above, the $0$-form symmetry defined in this way is anomalous. Its 't Hooft anomaly precisely cancels the one arising from the Dirichlet boundary of $\fZ(\bZ_N^{(0),k})$. Thus the stacking 
\begin{equation}
    \Bsym_\UV \boxtimes \cT^{\bZ_N}
\end{equation}
has an anomaly free diagonal $\bZ_N$ symmetry, which we can gauge. This produces the non-minimal interface corresponding to the functor $F: \bZ_N^{(0),k}\rightarrow \cD_\IR$.

\subsection{Non-Invertible Generalizations}

\label{sec:noninv WWW}
Using the mechanism described in the previous sections it is possible to discuss generalizations of the Wang-Wen-Witten mechanism in which an anomalous non-invertible symmetry is preserved in gapped phase. This provides a non-invertible analogue of the Wang-Wen-Witten construction.

As an example, we consider triality defects in (3+1)d \cite{Choi:2022zal}. These arise when a theory $\cT$ with $\bZ_n^{(1)}$ 1-form symmetry is self-dual under the operation $ST$:
\begin{equation}
    Z_{\cT}[B_2] = \sum _{b_2'} \exp{\left(\frac{2\pi i}{n} \int _{X_4} B_2\cup b_2' +\frac{\mathfrak{P}(b_2')}{2}\right)}\, Z_{\cT}[b_2']
\end{equation}
corresponding to gauging with discrete torsion. Above $\mathfrak{P} : H^2(X_4,\bZ_n) \to H^4(X_4,\bZ_{\gcd(2,n)n})$ is the Pontryagin square operation.
The relevant data characterizing the triality symmetry have been discussed in the literature. On top of the 1-form symmetry group $\bZ_n$, they are \cite{Antinucci:2023ezl, Cordova:2023bja}:
\begin{enumerate}
    \item A non-degenerate symmetric bicharacter on the 1-form symmetry. In the $\bZ_n$ under consideration this is a number $p$ coprime with $n$. We will take $p=1$ without loss of generality.
    \item A generalized Frobenius-Schur indicator labeled by $\epsilon \in H^5(\bZ_3,U(1))$ $= \bZ_3$, describing a cubic anomaly for the triality defect.
\end{enumerate}
This category can be referred to as a generalized 3-Tambara-Yamagami category 
\be
\cC_\UV= 3\TY^{\Z_3} (\Z_n ^{(1)}, \chi, \epsilon) \,,
\ee
where the superscript $\Z_3$ indicates that we are considering triality $\Z_3$. 
We take this categorical symmetry as our $\cC_\UV$. The SymTFT $\fZ(\cC_\UV)$ has been constructed in \cite{Kaidi:2022cpf, Antinucci:2022vyk, Antinucci:2023ezl, Cordova:2023bja, Bhardwaj:2024xcx}. It can be described starting with the SymTFT for the 1-form symmetry $\bZ_n^{(1)}$ alone, namely the (4+1)d 1-form $\bZ_n$ Dijkgraaf-Witten theory
\begin{equation}\label{eq:1fs DW}
    S=\frac{2\pi i}{n}\int _{X_5} b_2 \cup dc_2 \,,
\end{equation}
and then gauging the $\bZ_3^{(0)}$ triality symmetry
\begin{equation}
    \mathsf{ST}: \ \ \  b_2 \mapsto -b_2-c_2 \ , \ \ \ \ \ c_2\mapsto b_2 \,,
\end{equation}
while adding a discrete torsion $\tau \in H^5(B\bZ_3,U(1))$ in the bulk, with $\tau =\epsilon$.

The symmetry $\cC_\UV$ can be anomalous (lacks a fiber functor), depending on $n$ and $\epsilon\in \bZ_3$. Again, as in the case of standard TY categories, the obstruction theory to a fiber functor stems follows a two-level structure \cite{Antinucci:2023ezl, Cordova:2023bja}:
\begin{enumerate}
    \item  \textbf{First obstruction:} If the original Dijkgraaf-Witten theory \eqref{eq:1fs DW} does not have any $\bZ_3^{(0)}$-invariant topological boundary condition $\fB_{\text{invariant}}$, then there is no {triality}-invariant SPT phase for the one-form symmetry. This obstruction cancels only if the equation $r^2+r+1\equiv0 \pmod{n}$ admits a solution $r \in \bZ_n$. A triality symmetry with a first obstruction anomaly is either spontaneously broken, or must be realized in a gapless phase \cite{Apte:2022xtu,Damia:2023ses}.
    \item \textbf{Second obstruction:} When the first obstruction vanishes, we still have a second obstruction anomaly controlled by $\epsilon$. In the case that we will discuss the second obstruction only vanishes if $\epsilon \equiv 0 \mod 3$. \footnote{More generally, symmetry fractionalization can cancel some anomalies. The condition then takes the form $\epsilon + Y\neq 0$ for all allowed $Y\in \bZ_3$ computed in \cite{Antinucci:2023ezl}.}
\end{enumerate}
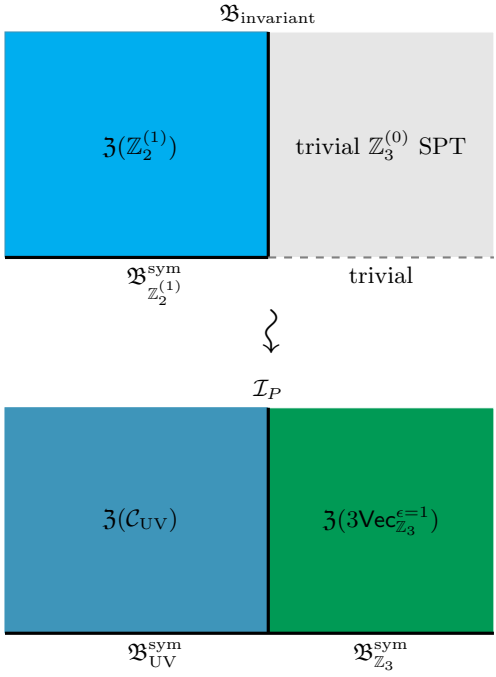
\begin{figure}[t]
\begin{tikzpicture}
\begin{scope}[shift= {(0,0)}]
 \draw [\UVcolor,  fill=\Ccolor]
(-0.5,0) -- (-0.5,3) --(3,3) -- (3,0) -- (-0.5,0) ; 
 \draw [white,  fill=gray!20!white]
(3,0) -- (3,3) --(6,3) -- (6,0) -- (3,0) ; 
\draw [very thick] (3,0) -- (3,3)  ;
\draw [very thick] (-0.5,0) -- (3,0)  ;
\draw [thick, dashed, color=gray] (3,0) -- (6,0)  ;
\node[below] at (1.5,0) {$\Bsym_{\bZ_2^{(1)}}$}; 
\node[below] at (4.5,0) {$\text{trivial}$}; 
\node at (1.3,1.5) {$\fZ(\bZ_2^{(1)})$}; 
\node at (4.5,1.5) {$\text{trivial  } \bZ_3^{(0)} \text{  SPT}$}; 
\node[above] at  (3,3) {$ \fB_{\text{invariant}}$};
 \node[rotate=-90] at (3,-1) {{\LARGE $\leadsto$}};
\end{scope}
\begin{scope}[shift= {(0,-5)}]
 \draw [\UVcolor,  fill=\UVcolor]
(-0.5,0) -- (-0.5,3) --(3,3) -- (3,0) -- (-0.5,0) ; 
 \draw [white,  fill=\Scolor]
(3,0) -- (3,3) --(6,3) -- (6,0) -- (3,0) ; 
\draw [very thick] (3,0) -- (3,3)  ;
\draw [very thick] (-0.5,0) -- (6,0)  ;
\node[below] at (1.5,0) {$\Bsym_{\UV}$}; 
\node[below] at (4.5,0) {$\Bsym_{\bZ_3}$}; 
\node at (1.3,1.5) {$\fZ(\cC_\UV)$}; 
\node at (4.5,1.5) {$\fZ(3\Vec_{\bZ_3}^{\epsilon=1})$}; 
\node[above] at  (3,3) {$\cI_\DFun$};
\end{scope}
\end{tikzpicture}
\caption{Top figure: the triality invariant boundary $\fB_{\text{invariant}}$ separates the 1-form SymTFT $\fZ(\bZ_2^{(1)})$, enriched with $\bZ_3^{(0)}$ from a trivial $\bZ_3$-SPT. Gauging $\bZ_3^{(0)}$ with discrete torsion $\epsilon=1$ produces an interface $\cI_\DFun$ between $\fZ(\cC_\UV)$ and $\fZ(3\Vec_{\bZ_3}^1)$.
\label{fig:fig:triality gauging}}
\end{figure}

We want to show that if the first obstruction anomaly vanishes but the second does not, the non-invertible triality symmetry can be realized in a SET phase that does not break it.

We focus on the simplest example $n=2$, $\epsilon =1$. The first obstruction anomaly vanishes for $r=1$, while we have a second obstruction anomaly. The key observation is that $\cC_\UV$ is not an ASCy: the 1-form symmetry $\bZ_2^{(1)}$ is a normal subcategory that is the kernel of a surjective functor 
\begin{equation}
    \DFun :\  \cC_\UV \ \rightarrow \ 3\Vec_{\bZ_3}^{(0),\epsilon=1}  \ .
\end{equation}
The SymTFT interface associated with $\DFun$ is built as follows. 
We start from the $\bZ_3^{(0)}$-invariant gapped boundary condition $\fB_{\text{invariant}}$ of \eqref{eq:1fs DW}, corresponding to $r=1$ and to the condensation of the $(n_e,n_m)=(1,1)$ dyonic surface.

This boundary condition can be viewed as an interface between a $\bZ_3$-enriched $\fZ(\bZ_2^{(1)})$ and an invertible $\bZ_3^{(0)}$-symmetric topological order. As the whole setup is triality-invariant, we then gauge $\bZ_3^{(0)}$ with discrete torsion $\epsilon=1$ throughout. On the left, this yields the SymTFT $\fZ(\cC_\UV)$ for triality symmetry; on the right, it gives the gauged $\bZ_3$ SPT with $\epsilon=1$, i.e., the twisted (4+1)d $\bZ_3$ DW theory discussed in section~\ref{sec:AnoEme}, which is the SymTFT for $\bZ_3$ with anomaly 1.
The interface between them is $\cI_\DFun$, as shown in figure~\ref{fig:fig:triality gauging}, while the boundary condition $\Bsym_{\bZ_3}$ of $\fZ(3\Vec_{\bZ_3})$ is the canonical Dirichlet boundary with $\bZ_3^{(0)}$ symmetry. 

The important fact is that $\fB_{\text{invariant}}$ is a magnetic boundary condition of $\fZ(\bZ_2^{(1)})$ with respect to $\Bsym_{\bZ_2^{(1)}}$, as it corresponds to the condensation of dyonic surfaces.
Therefore $\cI_\DFun$ 
defines a surjective functor $\DFun : \cC_\UV \rightarrow 3\Vec_{\bZ_3}^{\epsilon=1}$.

After we apply the functor $\DFun$ we are precisely in the same situation as at the beginning of section \ref{sec:AnoEme}: we have a $\bZ_3^{(0)}$ with anomaly $1$. Following the construction of section \ref{sec:SET}, and its reformulation in terms of a single functor as in section \ref{sec:nonmin}, we can further apply the invertible Wang-Wen-Witten construction:
\be
\begin{tikzpicture} 
\begin{scope}[scale=0.85] 
\draw [\UVcolor,  fill=\UVcolor]
(-3,0) -- (-3,3) --(0,3) -- (0,0) -- (-3,0) ; 
 \draw [\Scolor,  fill=\Scolor]
(0,0) -- (0,3) --(3,3) -- (3,0) -- (0,0) ; 
 \draw [white,  fill=gray!20!white]
(3,0) -- (3,3) --(6,3) -- (6,0) -- (3,0) ; 
\draw [very thick] (0,0) -- (0,3)  ;
\draw [very thick] (3,0) -- (3,3)  ;
\draw [very thick] (-3,0) -- (6,0)  ;
\node[below] at (-1.5,0) {$\Bsym_\UV$}; 
\node[below] at (1.5,0) {$\Bsym_{\bZ_3}$}; 
\node[below] at (4.5,0) {$\Bsym_{\cD_\IR}$}; 
\node at (-1.5,1.5) {$\fZ(\cC_\UV)$}; 
\node at (1.5,1.5) {$\fZ(3\Vec_{\bZ_3}^{\epsilon=1})$}; 
\node at (4.5,1.5) {$\fZ(\cD_\IR)$}; 
\node[above] at (0,3) {$\cI_{\DFun}$};
\node[above] at (3,3) {$\cI_{\Fun}$};
\end{scope}
\end{tikzpicture}
\ee
The composition 
\begin{equation}
    \DFun' =\Fun\circ \DFun : \quad \cC_\UV \rightarrow \cC_\IR ' \supset \left(\bZ_3^{(1)}\times \bZ_3^{(2)}\right)^{\text{anomaly}}
\end{equation}
describes the realization of the anomalous non-invertible triality symmetry in a symmetry preserving gapped phase, in the same spirit as Wang-Wen-Witten \cite{Wang:2017loc}.

\section{Symmetry Enriched SymTFT and LSM Anomaly-Matching}
\label{sec:LSM}
As a final application of our construction, let us discuss how discrete spacetime symmetries -- e.g. lattice translations -- can be matched in the continuum by emanant global symmetries. We focus on the IR matching of Lieb-Schultz-Mattis (LSM) anomalies \cite{Lieb:1961fr,Affleck:1986pq} in the continuum limit by 't Hooft anomalies for (emanant) global symmetries \cite{Cheng:2022sgb} and the constraints that the matching conditions impose on their structure.

In this section, we define an LSM anomaly as an obstruction to the realization of a symmetric, translationally invariant trivially-gapped phase on the lattice. 
The natural setup is that of a lattice system with a symmetry $\cC \rtimes \bZ_T$, where $\bZ_T$ are the lattice translations and $\cC$ is an anomaly-free symmetry, in the sense that it admits a fiber functor.\footnote{For simplicity, we will work in the infinite chain, so that the translation group is formally $\bZ$. This is inconsequential to our conclusions, as they concern the continuum limit.}
The notation $\cC \rtimes \cG$ stands for a $\cG$-crossed extension of $\cC$, called a crossed product (whose definition can be found in appendix~\ref{sec:GCross}), that is, the extension of a fusion category $\cC$ by the action of an outer automorphism group $\cG$. This encodes the fact that the translation symmetry acts nontrivially on the flavor symmetry, although the action is not necessarily a semidirect product of groups.
The LSM anomaly can then be interpreted as a sort of mixed 't Hooft anomaly between lattice translations $T$ and the global symmetry $\cC$ of the UV lattice system \cite{Cheng:2015kce,Seifnashri:2023dpa}.
It can be shown \cite{Seifnashri:2023dpa} in $(1+1)$d systems, that for $\cC= \Vec_G$ the LSM anomaly takes values in $H^2(G,U(1))$, which also describes bosonic SPT for the $G$ symmetry.
This resonates with the crystalline equivalence principle \cite{Thorngren:2016hdm}, according to which we would describe lattice translations as a $\bZ$ background gauge field $A_T$, and the LSM anomaly becomes:
\be
\omega_{\text{LSM}} = 2 \pi i \int A_T \cup \omega(A) \, , \ \ \ \omega \, \in  H^2(G,U(1)) \, .
\ee
In this setup, the twisting by lattice translations provides a map between $G$-SPTs resulting in the LSM anomaly. This perspective will be useful in generalizing our construction to non-invertible symmetries.

In the thermodynamic limit lattice translations act trivially on the low lying states and it is natural to expect that the LSM obstruction should be matched by an internal 't Hooft anomaly for an emergent symmetry.
In such a case, we say \cite{Cheng:2022sgb} that the global symmetry ``emanates" from lattice translations and is consistent with our proposed picture of 't Hooft anomaly matching between UV and IR (see also \cite{Seiberg:2025bqy}). 

In this section we use our SymTFT picture for UV/IR anomaly matching, together with the recent proposal \cite{Pace:2024acq,Pace:2025hpb} for the description of (1+1)d lattice translations (and other spacetime symmetries) via a {\bf Symmetry Enrichment} of the SymTFT construction (also known as a Symmetry Enriched Topological Order -- SETO -- in the Condensed Matter community), to derive an exact categorical sequence describing the LSM anomaly matching.
Our construction is conceptually simple, and gives a potent tool generalizing LSM anomaly matching to previously unknown setups involving non-invertible global symmetries, of which we provide the simplest example. We also expect many interesting examples to arise in higher dimensional setups, but do not analyze them in this work.

\subsection{LSM Anomaly from SESymTFT}
We will start by reviewing the construction of the Symmetry-Enriched SymTFT (SESymTFT) describing lattice spacetime symmetries proposed in \cite{Pace:2024acq,Pace:2025hpb}. 
We will focus for concreteness on enrichment by lattice translations, and encourage the reader to consult \cite{Pace:2024acq,Pace:2025hpb} for a more thorough presentation. Generalizations to orientation reversal symmetries should be possible.
We caution the reader that the construction of \cite{Pace:2024acq,Pace:2025hpb} differs in spirit from the standard SymTFT mantra: the lattice translation symmetry will not be gauged in the bulk, as this is not a natural operation to be performed on a lattice system, but it will rather be implemented as a background structure encoding its interplay with the remaining global symmetry.

As we will be dealing with $(2+1)$d non-isotropic -- in this case foliated -- systems, we will denote by $x,t$ the boundary coordinates and by $y$ the internal SymTFT direction. 
The enriched SymTFT for lattice translations is a foliated theory, which means it is non-trivially coupled to a foliation structure $\foli$ along the $x$ direction:
\be
\foli = a^{-1} dx \, ,
\ee
where $a$ is the lattice spacing. The foliation will act as a type of  background gauge field for the lattice translations, allowing the enrichment of the SymTFT at the cost of a mild loss of topological invariance.

The construction starts from an un-enriched theory $\fZ(\cC)$, describing the global (internal or flavor) symmetry of the lattice system. As remarked before, we will take $\cC$ to be anomaly-free. The $\cC$ symmetry thus admits several inequivalent fiber functors, $\FF_i$, describing the possible $\cC$-symmetric SPTs. As a matter of notation, we will take the trivially gapped phase to be one of these SPTs.

In order to have a non-trivial enrichment by lattice translations, we will further take $\cZ(\cC)$ to have nontrivial braided outer automorphisms, $\cU  \in \Aut\left( \cZ(\cC) \right)$. These describe the action of 0-form symmetries on the anyonic content of the SymTFT.
The insight of \cite{Pace:2024acq,Pace:2025hpb} is that the LSM anomaly can be described by appropriately coupling the foliation structure $\foli$ to an appropriately chosen $\cU$.
We will denote the associated enriched SymTFT by 
\be
\fZ(\cC)^{\cU T} \, ,
\ee
and will use the notation $\cU T$ to remind us of the choice of bulk automorphism which twists the translation symmetry.

Coupling the SymTFT to a nontrivial background foliation has the effect of enriching the SymTFT with an array of $\cU$ defects, stretching in the $(t,y)$ directions: 
\be
\begin{tikzpicture}
\begin{scope}[shift= {(0,-0.7)}]
\draw[->] (0,0) -- (0.5,0) node[anchor=north]{$(t,x)$};
    \draw[->] (0,0) -- (0,0.5) node[anchor= east]{$y$};
    \end{scope}
\begin{scope}
	\draw [\UVcolor,  fill=\UVcolor] 
	(0,0) -- (0,3) --(3,3) -- (3,0) -- (0,0) ; 
	\draw [white](0,0) -- (0,3) --(3,3) -- (3,0) -- (0,0) ;  ; 
	\draw [very thick] (0,0) -- (3,0)  ;
	\node at (1.5,1.5) {$\fZ(\cC)$} ;
	\node[below] at (1.5,0) {$\Bsym$}; 
\node at (4, 1.5) {$\text{\LARGE $\leadsto$}$}; 
\end{scope}
\begin{scope}[shift= {(5,0)}]
	\draw [\UVcolor,  fill=\UVcolor] 
	(0,0) -- (0,3) --(3,3) -- (3,0) -- (0,0) ; 
	\draw [white](0,0) -- (0,3) --(3,3) -- (3,0) -- (0,0) ;  ; 
	\draw [very thick] (0,0) -- (3,0)  ;
	\foreach \x in {1,...,3} {
		\draw[black!20!blue,opacity=0.5] (\x*0.75,0) -- (\x*0.75,3) node[above] {\footnotesize$\cU T$};
		\draw[fill=\Ncolor] (\x*0.75,0) circle (0.05);
	}
	\node at (1.5,1.5) {$\fZ(\cC)^{\cU T}$} ;
	\node[below] at (1.5,0) {$\Bsym$}; 
    \end{scope}
\end{tikzpicture}
\ee
Importantly, for this construction to be well-defined, we will need the symmetry boundary condition to be invariant under the action of $\cU$:
\be \label{eq: cUBsym}
\cU \times \Bsym = \Bsym \, .
\ee
In this case, $\cU$ descends to an automorphism of the global symmetry $\cC$ itself, and comes with a well-defined action on its structures, notably the space of $\cC$-symmetric gapped phases $\Mod_\cC$.

\vspace{2mm}
\noindent
{\bf LSM Anomaly.}
We are now ready to discuss how the LSM anomaly arises in this setup. As $\cC$ is anomaly-free, it can be trivially realized via (several) distinct SPTs, which we label by their fiber functors $\FF_i$, $i=1,...,M$.
These are in one-to-one correspondence with magnetic Lagrangian algebras, $\cL_i$ in $\cZ(\cC)$ with $\cL_i \cap \cL_\sym= \{1\} $. The SymTFT sandwich between $\Bphys$ and $\fB_{\cL_i}$ describes the $i$-th SPT phase.
The bulk automorphism $\cU$ acts on the $\cL_i$ by a permutation $\sigma$ as the number of ground states is preserved by the $\cU$ action using \eqref{eq: cUBsym}: 
\be
\cU\left(\cL_i\right) = \cL_{\sigma(i)} \, , \ \ \ \sigma \in S_M \, .
\ee
Consequently, in the enriched setup, a translationally-invariant SPT corresponds to a fixed point $i^*$ under $\sigma$: $\sigma(i^*)= i^*$. Thus we find:
\begin{center}
	\textit{The symmetry $\cC_{\UV} \rtimes \bZ_{\cU T}$ has an LSM anomaly if no magnetic Lagrangian algebra (i.e. SPT) is fixed under the $\cU$ action.}
\end{center}
Let us give a clarifying example.

\vspace{2mm}
\noindent \textbf{Example \boldmath$(\bZ_2 \times \bZ_2)\rtimes \bZ_T$.} To see the construction explicitly at work (and set the stage for our subsequent applications) let us consider $\cC=\Vec_{\bZ_2 \times \bZ_2}$. 
We will take $\Bsym$ to be described by electric condensation:
\be \label{eq: dualityinvZ2Z2}
\cL_{\text{sym}} = 1 \oplus e_1 \oplus e_2 \oplus e_1 e_2 \, .
\ee
The $\bZ_2 \times \bZ_2$ theory has two trivially gapped phases -- the trivial one and the $\bZ_2 \times \bZ_2$ SPT -- described by the Lagrangian algebras:
\be \ba
\cL_{\text{triv}} &= 1 \oplus m_1 \oplus m_2 \oplus m_1 m_2 \\
\cL_{\text{SPT}} &= 1 \oplus e_1 m_2 \oplus e_2 m_1 \oplus e_1 e_2 m_1 m_2 \, .
\ea \ee
The symmetry boundary condition $\Bsym$ is left invariant under the following automorphism $\cU$:
\be \ba
&\cU(e_1) = e_1 \, , \ \ &&\cU(e_2) = e_2 \, , \\
&\cU(m_1) = m_1 e_2 \, , \ \ &&\cU(m_2) = m_2 e_1 \, .
\ea \ee
As  $\cU(\cL_{\text{triv}})= \cL_{\text{SPT}}$, $\cU$ exchanges the trivial and SPT phases: 
\be \label{eq: LSMZ2Z2}
\begin{tikzpicture}[baseline={(0,1.5)}]
	\draw [\UVcolor,  fill=\UVcolor] 
	(0,0) -- (0,3) --(3,3) -- (3,0) -- (0,0) ; 
	\draw [white](0,0) -- (0,3) --(3,3) -- (3,0) -- (0,0) ;  ; 
	\draw [very thick] (0,0) -- (3,0)  ;
	\draw[very thick] (0,3) node[above] {$\fB^{\text{triv}}$}  -- (3,3) node[above] {$\fB^{\text{SPT}}$};
	\draw[black!20!blue,opacity=0.5] (1.5,0) -- (1.5,3) node[above] {$\cU T$};
	\draw[fill=\Ncolor] (1.5,0) circle (0.05); \draw[fill=\Ncolor] (1.5,3) circle (0.05);
	\node[below] at (3,0) {$\Bsym$}; 
	\node at (1.5,1.5) {$\fZ(\bZ_2\times \bZ_2)^{\cU T}$};
\end{tikzpicture}  = \  \ \ \ 
\begin{tikzpicture}[baseline={(0,0)},rotate=90]
	\draw[very thick] (0,0) node[above] {triv} -- (0,3) node[above] {SPT};  \draw[fill=\Ncolor] (0,1.5) node[below] {$T$} circle (0.05);
\end{tikzpicture}
\ee
This leads to an LSM anomaly for the combined symmetry, as no $\bZ_2 \times \bZ_2$ SPT is lattice-translation invariant. The $\cU$ automorphism of $\cZ(\Vec_{\bZ_2 \times \bZ_2})$ can also be described explicitly as:
\be
\cU = \exp\left( \pi i \int a_1 \, a_2 \right) \, ,
\ee
where $a_1, \, a_2$ are the electric gauge fields of the $\bZ_2 \times \bZ_2$ untwisted DW theory. Coupling $\cU$ to the foliation structure simply leads to the following action for the SymTFT:
\be
S = \pi i \int \left( \sum_{i=1,2} a_i d b_i  + \foli \cup a_1 \cup a_2 \right) \, .
\ee

\subsection{Categorical Exact Sequences and LSM Anomaly-Matching} 

We will now move on to the description of the exact sequence encoding the continuum LSM anomaly matching by an emanant symmetry using the SymTFT.
We will focus on examples in which $\cU$ is a cyclic automorphism of order $N$, namely $\cU^N = 1$ for some $N\in \mathbb{N}$.\footnote{The structure of the automorphisms of a given topological order has been studied in detail in several works \cite{Kapustin:2010if,Fuchs:2012dt,Barkeshli:2014cna,Roumpedakis:2022aik,Buican:2023bzl}} 
Our aim is to construct a categorical exact sequence of the form:
\be
N \bZ_T \overset{\IFun}{\longrightarrow} \cC_{\UV} \rtimes \bZ_{\cU T} \overset{\DFun}{\longrightarrow} \cC_{\IR} \, ,
\ee
where $\bZ_T$ stands for lattice translations and $\cC_{\IR}$ is the IR symmetry matching the LSM anomaly. The $\cC_\IR$ symmetry is generally larger than the $\cC_\UV$ symmetry, due to the presence of an emanant subsymmetry.
Graphically, this corresponds to an enriched RG-quiche:
\be 
\label{RGQuiche}
\begin{tikzpicture}
	\begin{scope}[scale=0.85] 
		\begin{scope}[shift={(-3,0)}]
			\draw [\Ncolor,  fill=\Ncolor]
			(0,0) -- (0,3) --(3,3) -- (3,0) -- cycle ; 
			\draw [very thick, dashed] (0,0) -- (3,0)  ;
			\foreach \x in {1,...,3} {
				\draw[dashed,opacity=0.5] (\x*0.75,0) -- (\x*0.75,3) node[above] {\footnotesize$T^N$};
				\draw[fill=\Ncolor] (\x*0.75,0) circle (0.05);
			}
		\end{scope}
		\draw [\Ccolor,  fill=\Ccolor]
		(0,0) -- (0,3) --(3,3) -- (3,0) -- (0,0) ; 
		\draw [very thick] (0,0) -- (3,0)  ;
		\foreach \y in {1,...,3} {
			\draw[black!20!blue, opacity=0.5] (\y*0.75,0) -- (\y*0.75,3) node[above] {\footnotesize $\cU T$};
			\draw[fill=\Ncolor] (\y*0.75,0) circle (0.05);
		}
		\draw [\Scolor,  fill=\Scolor]
		(3,0) -- (3,3) --(6,3) -- (6,0) -- (3,0) ; 
		\draw [very thick] (0,0) -- (0,3)  ;
		\draw [very thick] (3,0) -- (3,3)  ;
		\draw [very thick] (3,0) -- (6,0)  ;
		\node[below] at (-1.5,0) {$N\bZ_T$}; 
		\node[below] at (1.5,0) {$\cC_{\UV} \rtimes \bZ_T $}; 
		\node[below] at (4.5,0) {$\cC_{\IR}$}; 
		\node at (-1.5,1.5) {$\fZ(\Vec)^{T^N}$}; 
		\node at (1.5,1.5) {$\fZ(\cC_{\UV})^{\cU T}$}; 
		\node at (4.5,1.5) {$\fZ(\cC_{IR})$}; 
	\end{scope}
\end{tikzpicture}
\ee
The leftmost part is a trivial yet enriched topological order, it describes the subgroup of lattice translations which act trivially on the continuum Hilbert space, and thus are in the kernel of the UV/IR map.
The rightmost segment instead describes the IR continuum symmetry, where the LSM anomaly for $\cU T$ is matched by an emanant symmetry, so generically $\dim(\fZ(\cC_{\UV})) < \dim(\fZ(\cC_{\IR}))$. 
This is not in contradiction with the fact that the total quantum dimension should {\bf decrease} along a surjective RG-flow, as the UV dimension should be taken to be the one of the enriched SymTFT, which is formally infinite.
In order to describe the sequence we must provide the appropriate interfaces between the three topological orders.

We start with the description of the left interface. As the corresponding center is $\cZ(\Vec) = \Vec$ the interface must correspond to the condensation of a Lagrangian algebra in $\cZ(\cC_{\UV})$. Condition \eqref{eq:masterequation} forces this to be $\cL_{\UV}$, the Lagrangian algebra corresponding to the UV symmetry boundary:
\be
\begin{tikzpicture}[baseline={(0,1.5)}]
	\begin{scope}[shift={(-3,0)}]
		\draw [\Ncolor,  fill=\Ncolor]
		(0,0) -- (0,3) --(3,3) -- (3,0) -- cycle ; 
		\draw [very thick, dashed] (0,0) -- (3,0)  ;
		\foreach \x in {1,...,3} {
			\draw[dashed,opacity=0.5] (\x*0.75,0) -- (\x*0.75,3) node[above] {\footnotesize$T^N$};
			\draw[fill=\Ncolor] (\x*0.75,0) circle (0.05);
		}
	\end{scope}
	\draw [\Ccolor,  fill=\Ccolor]
	(0,0) -- (0,3) --(3,3) -- (3,0) -- (0,0) ; 
	\draw [very thick] (0,0) -- (3,0)  ;
	\foreach \y in {1,...,3} {
		\draw[black!20!blue, opacity=0.5] (\y*0.75,0) -- (\y*0.75,3) node[above] {\footnotesize $\cU T$};
		\draw[fill=\Ncolor] (\y*0.75,0) circle (0.05);
	}
	\node[below] at (1.5,0) {$\Bsym_{\UV}$}; 
	\node at (-1.5,1.5) {$\fZ(\Vec)^{T^N}$}; 
	\node at (1.5,1.5) {$\fZ(\cC_{\UV})^{\cU T}$}; 
	\draw[very thick] (0,0) -- (0,3) node[above] {$\Bsym_{\UV}$};
\end{tikzpicture}
\ee
Notice that this is compatible with the trivialization of the SymTFT enrichment by $\cU T$, as $\Bsym_{\UV}$ must be $\cU$-invariant by \eqref{eq: cUBsym}.

The rightmost interface instead implements a map between an enriched topological order and a standard one. The natural physical mechanism implementing for such a map corresponds to gauging the 0-form symmetry $\cG$ generated by $\cU$. 
In mathematical terms, the SESymTFT is described by a $\cG$-crossed braided tensor category \cite{Barkeshli:2014cna} and the gauging procedure corresponds to equivariantization. We refer the reader to \cite{Barkeshli:2014cna} for an in depth review. For our current purposes we will only require some of the results summarized in appendix \ref{sec:GCross}. 
As the $\cG$ automorphism group leaves $\Bsym$ invariant, it corresponds to an automorphism of $\cC_\UV$ itself. In these cases it is known, see \cite{2009arXiv0905.3117G}, that the center after gauging the 0-form symmetry is simply:
\be
\cZ(\cC_\IR) = \cZ(\cC_\UV \rtimes \cG) \, .
\ee
Alternatively, the interface $\cI_\cG$ can be described starting from $\fZ(\cC_{\UV} \rtimes \cG)$ and gauging the dual $\Rep(\cG)$ symmetry. 

As the starting symmetry boundary is $\cG$-invariant, it is naturally mapped to a corresponding boundary condition in $\fZ(\cC_{\UV}\rtimes \cG)$, described by enhancing the symmetric algebra $\cL_{\UV}$ by the generators of the dual symmetry $\Rep(\cG)$. This boundary condition encodes a $\cC_\UV \rtimes \cG$ symmetry, of which the $\cG$ subgroup is emanant.

Also in this case the Matching Equation is satisfied, as the action of the $\cG$-gauging interface on the invariant algebra $\cL_\UV$
provides the dressing by the $\Rep(\cG)$ generators.
The final quiche is thus:
\be
\begin{tikzpicture}[baseline={(0,1.5)}]
	\draw [\Ccolor,  fill=\Ccolor]
	(0,0) -- (0,3) --(3,3) -- (3,0) -- (0,0) ; 
	\draw [very thick] (0,0) -- (3,0)  ;
	\foreach \y in {1,...,3} {
		\draw[black!20!blue, opacity=0.5] (\y*0.75,0) -- (\y*0.75,3) node[above] {\footnotesize $\cU T$};
		\draw[fill=\Ncolor] (\y*0.75,0) circle (0.05);
	}
	\draw [\Scolor,  fill=\Scolor]
	(3,0) -- (3,3) --(6,3) -- (6,0) -- (3,0) ; 
	\draw [very thick] (3,0) -- (3,3) node[above] {$\cI_{\cG}$}  ;
	\draw [very thick] (3,0) -- (6,0)  ;
	\node[below] at (1.5,0) {$\cC_{\UV}\rtimes \bZ_T$}; 
	\node[below] at (4.5,0) {$\cC_{\UV} \rtimes \cG$}; 
	\node at (1.5,1.5) {$\fZ(\cC_{\UV})^{\cU T}$}; 
	\node at (4.5,1.5) {$\fZ(\cC_{UV} \rtimes \cG)$}; 
\end{tikzpicture}
\ee
In the UV, the enrichment by the lattice translation symmetry was implemented by the twist defect $Z^\cU$ of the $\cU$ symmetry. After gauging the $\cG$ symmetry the twist defects are genuine and become Gukov-Witten defects for the $\Rep(\cG)$ symmetry.
As they carry $\Rep(\cG)$ charge, they survive as nontrivial topological $\cG$ defects on the $\cC_\UV \rtimes \cG$ boundary condition, describing the action of the emanant symmetry $z \in \cG$ on the continuum degrees of freedom:
\be
\begin{tikzpicture}[baseline={(3,1.5)}]
\draw [\Ccolor,  fill=\Ccolor]
(1.5,0) -- (1.5,3) --(3,3) -- (3,0) -- (1.5,0) ; 
\draw [very thick] (1.5,0) -- (3,0)  ;
\draw[black!20!blue, opacity=0.5] (2.5,0) -- (2.5,3) node[above] {\footnotesize $\cU T$};
\draw[fill=\Ncolor] (2.5,0) node[below] {$Z^{\cU T}$} circle (0.05);
 \draw [\Scolor,  fill=\Scolor]
(3,0) -- (3,3) --(4.5,3) -- (4.5,0) -- (3,0) ; 
\draw [very thick] (3,0) -- (3,3) node[above] {$\cI_{\cG}$}  ;
\draw [very thick] (3,0) -- (4.5,0)  ;
\node[below] at (1.5,0) {$\Bsym_{\UV}$}; 
\node[below] at (4.5,0) {$\Bsym_{\IR}$}; 
\end{tikzpicture}  =  
\begin{tikzpicture}[baseline={(3,1.5)}]
\draw [\Ccolor,  fill=\Ccolor]
(1.5,0) -- (1.5,3) --(3,3) -- (3,0) -- (1.5,0) ; 
\draw [very thick] (1.5,0) -- (3,0)  ;
 \draw [\Scolor,  fill=\Scolor]
(3,0) -- (3,3) --(4.5,3) -- (4.5,0) -- (3,0) ; 
\draw [very thick] (3,0) -- (3,3) node[above] {$\cI_{\cG}$}  ;
\draw [very thick] (3,0) -- (4.5,0)  ;
\node[below] at (1.5,0) {$\Bsym_{\UV}$}; 
\node[below] at (4.5,0) {$\Bsym_\IR$}; 
\draw[fill=\Ncolor] (3.5,0) node[below] {$Z_\IR$} circle (0.05);
\end{tikzpicture}
\ee

\subsection{Invertible and Non-Invertible Examples}
The construction up to this point has been rather abstract. Let us conclude by providing two key examples: the simplest LSM anomaly matching involving $\cC_{\UV} = \Vec_{\bZ_2 \times \bZ_2}$ and the simplest non-invertible example: $\cC_{\UV}=\Rep(D_8)$ twisted by its triality automorphism.
As far as we are aware this is a novel result\footnote{The LSM anomaly of an $\Rep(G) \times Z(G)$ -- $Z$ here being the standard center of the group $G$ -- has instead been discussed in \cite{Pace:2024acq}.}.

\vspace{2mm}
\noindent \textbf{LSM Anomaly Matching: \boldmath $(\bZ_2 \times \bZ_2) \rtimes \bZ_T \to (\bZ_2)^3_{\text{III}}$.} 
We start with 
\be
\cC_{\UV} = \Vec_{\bZ_2 \times \bZ_2} \,.
\ee
We can now go back to our example of the LSM anomaly \eqref{eq: LSMZ2Z2} and our discussion of it.
The description of the normal subcategory is obvious and follows our prescription verbatim. 

To describe the target category of the surjective functor, {i.e. the IR symmetry}, we perform the condensation of the $\bZ_2$ automorphism $\cU$. 
The gauging of a bulk 0-form symmetry is known to lead to the center of a Tambara-Yamagami category \cite{Kaidi:2022cpf,Antinucci:2022vyk,Antinucci:2023ezl}, $\cZ(\TY(\bZ_2\times \bZ_2, \chi, \epsilon))$. The choice of $\chi$ can be read off from the action of the duality symmetry on electric and magnetic lines \cite{Antinucci:2023ezl}, while the FS indicator is described by gauging twisted by the discrete torsion $\epsilon \, \in \, H^3(\bZ_2, U(1))$.
We first perform the change of basis:
\be \ba
&e'_1 = e_1 m_2 \, , \ \ &&e'_2 = e_2 m_1 \, , \\ 
&m'_1 = m_1 \, , \ \ &&m'_2 = m_2  \, ,
\ea \ee
so that the duality action takes the form:
\be \ba
&\cU(e'_1) = m'_2 \, , \ \ &&\cU(e'_2) = m'_1 \, , \\
&\cU(m'_1) = e'_2 \, , \ \ &&\cU(m'_2) = e'_1 \, , \\
&\cU(e'_1 m'_2) = e'_1 m'_2 \, ,  \ \ &&\cU(e'_2 m'_1) = e'_2 m'_1 \, .
\ea \ee
This corresponds to the off-diagonal bicharacter $\chi_o$, leading to $\TY(\bZ_2\times \bZ_2, \chi_o,+) = \Rep(D_8)$. The automorphism action on the $\bZ_2 \times \bZ_2$ group is trivial on objects, but nontrivial on their three-valent junctions. Indeed the center of $\Rep(D_8)$ is the same as the one for the $\bZ_2^3$ group with a type-III anomaly:
\be
\omega(a_1,a_2,a_3) =  \exp\left( i \pi \int A_1 A_2 A_3 \right) \, . 
\ee
This anomaly correctly encodes the twisted action of $\bZ_T$ on $\bZ_2\times \bZ_2$, and we recover 
the $\bZ_2$-crossed braided extension of the center in the IR. 

We make this more precise by studying the induced boundary condition in $\cZ(\Rep(D_8))$. The anyon content is given in Table \ref{tab:spinsRepD8} \cite{Iqbal:2023wvm, Bhardwaj:2024qrf}. 
\begin{table}[t]
	\centering
	{
		\renewcommand{\arraystretch}{1.1}
		\begin{tabular}{|c|c|c|c|c|c|c|c|c|c|c|c|}
			\hline
			& 1 & $e_{\text{R}}$ & $e_{\text{G}}$ & $e_{\text{B}}$ & $e_{\text{RG}}$ & $e_{\text{GB}}$ & $e_{\text{RB}}$ & $e_{\text{RGB}}$ & $m_{\text{R}}$ & $m_{\text{G}}$ & $m_{\text{B}}$ \\
			
			\hline
			
			$\theta$ & $1$ & $1$ & $1$ & $1$ & $1$ & $1$ & $1$ & $1$ & $1$ & $1$ & $1$
			\\
			
			\hline

			$d$ & $1$ & $1$ & $1$ & $1$ & $1$ & $1$ & $1$ & $1$ & $2$ & $2$ & $2$ \\
			
			\hline \hline
			
			& $m_{\text{RG}}$ & $m_{\text{GB}}$ & $m_{\text{RB}}$ & $f_{\text{R}}$ & $f_{\text{G}}$ & $f_{\text{B}}$ & $f_{\text{RG}}$ & $f_{\text{GB}}$ & $f_{\text{RB}}$ & $s_{\text{RGB}}$ & $\bar{s}_{\text{RGB}}$ \\
			
			\hline
			
			$\theta$ & $1$ & $1$ & $1$ & $-1$ & $-1$ & $-1$ & $-1$ & $-1$ & $-1$ & $i$ & $-i$ \\
			\hline
			$d$ & $2$ & $2$ & $2$ & $2$ & $2$ & $2$ & $2$ & $2$ & $2$ & $2$ & $2$\\
			
			\hline 
	\end{tabular}}
	\caption{Spins and quantum dimensions of the simple lines of $\cZ(\Rep(D_8))$.}
	\label{tab:spinsRepD8}
\end{table}
The duality-invariant Lagrangian algebra \eqref{eq: dualityinvZ2Z2} is lifted into:
\be
\cL_{\text{sym}}^{\IR} =1 \oplus e_R \oplus e_G \oplus e_B \oplus e_{RG} \oplus e_{RB} \oplus e_{BG} \oplus e_{RGB} \, .
\ee
We can check explicitly that this is giving a tensor functor. The only lines which can traverse the interface are those that are uncharged under the dual $\Rep(\bZ_2)$, generated by $e_{RGB}$ and must be duality-invariant in order to end on $\Bsym_{\UV}$. These are exactly $1, e_R, e_G, e_B$ and their image under fusion with $e_{RGB}$:
\be
\begin{tikzpicture}[baseline={(0,1.5)}]
	\draw [\Ccolor,  fill=\Ccolor]
	(0,0) -- (0,3) --(3,3) -- (3,0) -- (0,0) ; 
	\draw [very thick] (0,0) -- (3,0)  ;
	\foreach \y in {1,...,3} {
		\draw[black!20!blue, opacity=0.5] (\y*0.75,0) -- (\y*0.75,3) node[above] {\footnotesize $\cU T$};
		\draw[fill=\Ncolor] (\y*0.75,0) circle (0.05);
	}
	\draw [\Scolor,  fill=\Scolor]
	(3,0) -- (3,3) --(6,3) -- (6,0) -- (3,0) ; 
	\draw [very thick] (3,0) -- (3,3) node[above] {$\cI_{\cG}$}  ;
	\draw [very thick] (3,0) -- (6,0)  ;
	\node[below] at (1.5,0) {$\Bsym_{\UV}$}; 
	\node[below] at (4.5,0) {$\Bsym_{\IR}$}; 
	\draw (3.5,0) arc (0:180:0.5 and 0.5);
	\draw (4.5,0) arc (0:180:1.5 and 1.5);
	\draw (5.5,0) arc (0:180:2.5 and 2.5);
	\node at (3.5,0.65) {$e_R$}; \node at (3.5,1.65) {$e_G$}; \node at (3.5,2.65) {$e_B$};
	\node at (2.5,0.65) {$e_1$}; \node at (2.5,1.65) {$e_2$}; \node at (2.5,2.65) {$e_1 e_2$};
\end{tikzpicture}
\ee
This boundary condition does not host a non-invertible symmetry, but rather the $(\bZ_2)^3$ symmetry with a type III anomaly, consistent with our previous remarks.

We are thus led to the enriched RG-quiche:
\be
\begin{tikzpicture} 
	\begin{scope}[scale=0.85]
		\begin{scope}[shift={(-3,0)}]
			\draw [\Ncolor,  fill=\Ncolor]
			(0,0) -- (0,3) --(3,3) -- (3,0) -- cycle ; 
			\draw [very thick, dashed] (0,0) -- (3,0)  ;
			\foreach \x in {1,...,3} {
				\draw[dashed,opacity=0.5] (\x*0.75,0) -- (\x*0.75,3) node[above] {\footnotesize$T^2$};
				\draw[fill=\Ncolor] (\x*0.75,0) circle (0.05);
			}
		\end{scope}
		\draw [\Ccolor,  fill=\Ccolor]
		(0,0) -- (0,3) --(3,3) -- (3,0) -- (0,0) ; 
		\draw [very thick] (0,0) -- (3,0)  ;
		\foreach \y in {1,...,3} {
			\draw[black!20!blue, opacity=0.5] (\y*0.75,0) -- (\y*0.75,3) node[above] {\footnotesize $\cU T$};
			\draw[fill=\Ncolor] (\y*0.75,0) circle (0.05);
		}
		\draw [\Scolor,  fill=\Scolor]
		(3,0) -- (3,3) --(6,3) -- (6,0) -- (3,0) ; 
		\draw [very thick] (0,0) -- (0,3)  ;
		\draw [very thick] (3,0) -- (3,3)  ;
		\draw [very thick] (3,0) -- (6,0)  ;
		\node[below] at (-1.5,0) {$N\bZ_T$}; 
		\node[below] at (1.5,0) {$[\bZ_2\times \bZ_2] \rtimes \bZ_T $}; 
		\node[below] at (4.5,0) {$(\bZ_2)^3_{\text{III}}$}; 
		\node at (-1.5,1.5) {$\fZ(\Vec)^{T^2}$}; 
		\node at (1.5,1.5) {$\fZ(\bZ_2 \times \bZ_2)^{\cU T}$}; 
		\node at (4.5,1.5) {$\fZ(\Rep(D_8))$}; 
	\end{scope}
\end{tikzpicture}
\ee
where 
\be
\cC_{\IR} = \Vec_{\Z_2 \times \Z_2\times \Z_2}^{\omega_{\text{III}}} \,.
\ee
We conclude this example by describing the UV/IR map between the generator of the translation symmetry $\cU T$ in the UV and the emanant $\bZ_2$ symmetry in the IR.

In the UV, the action of translation $\cU T$ is implemented by twist defects $Z_\rho$ of $\cU$, pushed on the $\Bsym_{\UV}$ boundary. There, the $Z_\rho$ are mapped to invertible objects:
\be
Z_{\rho_{(i,j)}} \mapsto 2 \eta^{\cU T} \eta_1^i \eta_2^j \, ,
\ee
where the superscript $\eta^{\cU T}$ enforces the fact that the (invertible) object is still non-local. Crossing the gauging interface, the tensor functor is simply:
\be
\DFun(\eta^{\cU T}) = \eta_3 \, ,
\ee
where $\eta_3$ generates the third (emanant) $\bZ_2$ symmetry of the IR description:
\be
\begin{tikzpicture}[baseline={(3,1.5)}]
	\draw [\Ccolor,  fill=\Ccolor]
	(1.5,0) -- (1.5,3) --(3,3) -- (3,0) -- (1.5,0) ; 
	\draw [very thick] (1.5,0) -- (3,0)  ;
	\draw[black!20!blue, opacity=0.5] (2.5,0) -- (2.5,3) node[above] {\footnotesize $\cU T$};
	\draw[fill=\Ncolor] (2.5,0) node[below] {$\eta^{\cU T}$} circle (0.05);
	\draw [\Scolor,  fill=\Scolor]
	(3,0) -- (3,3) --(4.5,3) -- (4.5,0) -- (3,0) ; 
	\draw [very thick] (3,0) -- (3,3) node[above] {$\cI_{\cG}$}  ;
	\draw [very thick] (3,0) -- (4.5,0)  ;
	\node[below] at (1.5,0) {$\Bsym_{\UV}$}; 
	\node[below] at (4.5,0) {$\Bsym_{\IR}$}; 
\end{tikzpicture} \ = \ 
\begin{tikzpicture}[baseline={(3,1.5)}]
	\draw [\Ccolor,  fill=\Ccolor]
	(1.5,0) -- (1.5,3) --(3,3) -- (3,0) -- (1.5,0) ; 
	\draw [very thick] (1.5,0) -- (3,0)  ;

	\draw [\Scolor,  fill=\Scolor]
	(3,0) -- (3,3) --(4.5,3) -- (4.5,0) -- (3,0) ; 
	\draw [very thick] (3,0) -- (3,3) node[above] {$\cI_{\cG}$}  ;
	\draw [very thick] (3,0) -- (4.5,0)  ;
	\node[below] at (1.5,0) {$\Bsym_{\UV}$}; 
	\node[below] at (4.5,0) {$\Bsym_{\IR}$}; 
	\draw[fill=\Ncolor] (3.5,0) node[below] {$\eta_3$} circle (0.05);
\end{tikzpicture}
\ee
As predicted by our formalism. 

\vspace{2mm}
\noindent \textbf{Non-invertible LSM Anomaly Matching.} Our SymTFT formulation gives a powerful handle on LSM anomaly-matching in cases where the underlying symmetry $\cC_{\UV}$ is non-invertible. The simplest example is given by the smallest fusion category which is not group-like but has several fiber functors: 
\be
\cC_{\UV}=\Rep(D_8)\,.
\ee
The $\Rep(D_8)$ fusion category has an $S_3$ outer-automorphism group permuting its three $\bZ_2$ subgroups. 
This can be extended to an automorphism of the center $\cZ(\Rep(D_8))$ which acts by permutation on the RGB labels in Table \ref{tab:spinsRepD8}. The $\Rep(D_8)$ symmetry boundary:
\be
\cL_{\Rep(D_8)} = 1 \oplus e_{RGB} \oplus m_{GB} \oplus m_{RB} \oplus m_{RG} \, ,
\ee
is invariant under this automorphism, while the three $\Rep(D_8)$ SPTs \cite{Bhardwaj:2024qrf, Warman:2024lir}:
\be \ba
&\cL_{\text{SPT}_R}= 1 \oplus e_B \oplus e_G \oplus e_{GB} \oplus 2 m_R      \, , \\
&\cL_{\text{SPT}_G}= 1 \oplus e_R \oplus e_B \oplus e_{RB} \oplus 2 m_G \, , \\
&\cL_{\text{SPT}_B}= 1 \oplus e_R \oplus e_G \oplus e_{RG} \oplus 2 m_B        \, ,
\ea \ee
are cyclically permuted:
\be
\begin{tikzcd}
	\text{SPT}_R \arrow[rr, "\cU"]  & & \text{SPT}_G \arrow[dl, "\cU"]\\
	& \text{SPT}_{B} \arrow[ul, "\cU"] &
\end{tikzcd}
\ee
{by $\cU$ which generates the $\bZ_3 \subset S_3$ automorphism subgroup. }

Decorating the translation generator by the triality automorphism $\cU$ and considering the enriched theory $\cZ(\Rep(D_8))^{\cU T}$ leads to an LSM anomaly between lattice translations and the $\Rep(D_8)$ symmetry.\footnote{
On the other hand, the three $\bZ_2^{R,G,B}$ subgroups of $S_3$ can give rise to no LSM anomaly, as they always leave the corresponding SPT fixed.}

Notice that the three $\Rep(D_8)$ SPTs are indistinguishable by the $\bZ_2 \times \bZ_2$ invertible subcategory of $\Rep(D_8)$ alone: the LSM anomaly truly involves the non-invertible symmetry generator.\footnote{More precisely, the three SPTs differ in their $\cD$-twisted sector \cite{Thorngren:2019iar}. The relevant datum is a group homomorphism $\nu_{R,G,B}$: $\bZ_2\times\bZ_2\to \bZ_2$ assigning a minus sign to two out of the three $\bZ_2$ subgroups. Clearly the permutation action cycles between $\nu_{R,G,B}$ providing the map on $\Rep(D_8)$ SPTs.}

The lattice translation symmetry acts by an invertible $\bZ_3$ symmetry $\eta^{\cU T}$ on $\bZ_2\times \bZ_2 \subset \Rep(D_8)$:
\be \ba
&\eta^{\cU T} \, \eta_R = \eta_G \, \eta^{\cU T} \, , \ \ &&\eta^{\cU T} \, \eta_G = \eta_B \, \eta^{\cU T} \, , \\
&\eta^{\cU T} \, \eta_B = \eta_R \, \eta^{\cU T} \, , \ \ &&\eta^{\cU T} \, \cD = \cD \, \eta^{\cU T} \, .
\ea \ee
In the continuum, the IR matching of the LSM anomalies gives rise to an emanant $\bZ_3$ symmetry. Indeed notice that, since the automorphism $\cG$ acts on $\Rep(D_8)$ by a simple permutation of the $\bZ_2$ subgroups, the gauged SymTFT is $\cZ(\Rep(D_8)\rtimes \bZ_3)$, where now $\rtimes$ really denotes the 
crossed product category, which is in the same spirit of semi-direct product, but defined for categories.
We discuss the details of this crossed extension $\cZ(\Rep(D_8)\rtimes \bZ_3)$ and crossed product $\rtimes$ in appendix \ref{sec:GCross}.
The UV boundary condition corresponds to the Lagrangian algebra:
\be
\cL_{\IR} = \left(\bigoplus_{a=0,1,2} 1^a \oplus e_{RGB}^a\right) \oplus \mu_{RG} \, ,
\ee
with $1^a$ the dual $\Rep(\bZ_3)$ representations and $\mu_{RG} = m_{RG} \oplus m_{RB} \oplus m_{GB}$. This is precisely the Lagrangian algebra which implements the canonical Dirichlet boundary condition in $\cZ(\Rep(D_8)\rtimes \Z_3)$.
The SymTFT description of the whole setup is: 
\be
\begin{tikzpicture} 
	\begin{scope}[scale=0.85]
		\begin{scope}[shift={(-3,0)}]
			\draw [\Ncolor,  fill=\Ncolor]
			(0,0) -- (0,3) --(3,3) -- (3,0) -- cycle ; 
			\draw [very thick, dashed] (0,0) -- (3,0)  ;
			\foreach \x in {1,...,3} {
				\draw[dashed,opacity=0.5] (\x*0.75,0) -- (\x*0.75,3) node[above] {\footnotesize$T^2$};
				\draw[fill=\Ncolor] (\x*0.75,0) circle (0.05);
			}
		\end{scope}
		\draw [\Ccolor,  fill=\Ccolor]
		(0,0) -- (0,3) --(3,3) -- (3,0) -- (0,0) ; 
		\draw [very thick] (0,0) -- (3,0)  ;
		\foreach \y in {1,...,3} {
			\draw[black!20!blue, opacity=0.5] (\y*0.75,0) -- (\y*0.75,3) node[above] {\footnotesize $\cU T$};
			\draw[fill=\Ncolor] (\y*0.75,0) circle (0.05);
		}
		\draw [\Scolor,  fill=\Scolor]
		(3,0) -- (3,3) --(6,3) -- (6,0) -- (3,0) ; 
		\draw [very thick] (0,0) -- (0,3)  ;
		\draw [very thick] (3,0) -- (3,3)  ;
		\draw [very thick] (3,0) -- (6,0)  ;
		\node[below] at (-1.5,0) {$N\bZ_T$}; 
		\node[below] at (1.5,0) {$\Rep(D_8) \rtimes \bZ_T $}; 
		\node[below] at (4.5,0) {$\Rep (D_8)\rtimes \Z_3$}; 
		\node at (-1.5,1.5) {$\fZ(\Vec)^{T^2}$}; 
		\node at (1.5,1.5) {$\fZ(\Rep(D_8))^{\cU T}$}; 
		\node at (4.5,1.5) {$\fZ (\Rep (D_8)\!\!\rtimes\! \Z_3)$}; 
	\end{scope}
\end{tikzpicture}
\ee
Thus, gauging the $\bZ_3$ symmetry generated by $\cU T$, the continuum IR symmetry is:
\be
\cC_{\IR} = \Rep(D_8) \rtimes \bZ_3 \, ,
\ee
with $\bZ_3$ generator $\eta_{\IR}$ emanating from the lattice translation symmetry. The UV/IR map reads:
\be
\DFun(\eta^{\cU T}) = \eta_{\IR} \, .
\ee
Due to the crossed product structure, the LSM anomaly is matched by a mixed 't Hooft anomaly of the $\Rep(D_8) \rtimes \bZ_3$ symmetry.

It would be interesting to extend our construction further to include non-invertible lattice translation symmetries \cite{Seiberg:2023cdc,Seiberg:2024gek} and various types of higher self-duality symmetries \cite{Antinucci:2022cdi,Diatlyk:2023fwf,Choi:2023vgk,Lu:2024lzf,Lu:2025gpt}.

\section{Conclusions and Outlook}
\label{sec:conclusion}
In this work, we have discussed how generalized symmetries can be matched along RG-flows via the mathematics of tensor functors. 
We have shown that tensor functors have in turn a simple SymTFT realization, which has allowed us to uncover several interesting aspects of this problem: Perhaps most striking of all the idea that 't Hooft anomalies for non-invertible symmetries can be explained and quantified using the language of short exact sequences of fusion categories, naturally giving rise to the notion of ASCies. 

While we have already given several applications of our construction, we expect that several new avenues are opened by our present work:
\begin{enumerate}
\item \textbf{A mathematical theory of ASCies.} The theory of short exact sequences of tensor categories -- and their possible higher-categorical generalizations -- remains subject of current mathematical development~\cite{bruguieres2011exact, bruguieres2014central, natale2020notionexactsequencehopf, ETINGOF20171187}. As a result, many structural properties of Anomalous Symmetry Categories (ASCies), as well as their interrelations, are still not fully understood.
As emphasized in this work, developing this mathematical framework offers a promising path toward a deeper understanding of 't~Hooft anomalies in both conventional and categorical settings.

    \item \textbf{Continuous symmetries.} The present work has focused exclusively on discrete finite symmetries. While it is widely believed that discrete symmetries can already capture the full anomaly structure of continuous ones, the language of tensor categories is inherently suited to describing symmetries with flat (i.e., topological) backgrounds.
    Recently, several works have developed a SymTFT-based approach to incorporate continuous symmetries~\cite{Antinucci:2024zjp, Brennan:2024fgj, Bonetti:2024cjk, Apruzzi:2024htg, Antinucci:2024bcm}. In this context, continuous spacetime symmetries and their anomalies naturally come into play. Our formalism may offer insight into how discrete spacetime symmetries on the lattice -- and their associated anomalies -- leave an imprint on the structure of their continuum counterparts.

\item \textbf{LSM anomaly matching for higher categories.} Our formulation of LSM anomaly matching via the SymTFT framework naturally extends to higher dimensions. These extensions yield simple and computable predictions for LSM anomalies and their matching in the presence of higher-form symmetries.    
Moreover, building on techniques from \cite{Aasen:2020jwb, Bhardwaj:2024kvy, Inamura:2025cum}, the SymTFT may provide a direct route to formulating UV lattice models that realize higher LSM anomalies -- and potentially offer insight into their IR dynamics.

\item \textbf{Anomalies of Weak and Strong Symmetries.} 
Recent developments have expended the study of symmetric phases for mixed states. In particular averaged or mixed state SPTs were discussed in \cite{degroot2022,ma2023average,zhang2022strange,sala2024,lee2025} for group-symmetries. 
Recently a general  SymTFT formulation of mixed state phases was put forward including mixed state SPTs in \cite{Schafer-Nameki:2025fiy, Luo:2025phx, Qi:2025tal}. In \cite{Schafer-Nameki:2025fiy} an extension to non-invertible weak and strong symmetries is developed, and it would be interesting to quantify anomalies for such mixed states combining the approaches of the present paper with that one.

\item \textbf{Constraints on RG-interfaces.} As our main motivating example, we have argued that tensor functors are realized by RG-interfaces between CFTs connected via symmetry-preserving relevant deformations. It is natural to expect that the mathematical structure of tensor functors imposes strong constraints on the physics of such interfaces.
For instance, the induced map on generalized charges can identify UV representations that become confined on the interface. It would also be interesting to explore whether phenomena such as symmetry enhancement on conformal interfaces~\cite{Gaiotto:2012np, Antinucci:2025uvj, Choi:2025ebk} can be more naturally understood within our framework.

\item \textbf{UV/IR map in the presence of defects.} A crucial ingredient in enriching the UV/IR correspondence is the presence of unscreened dynamical defects in the low-energy effective theory. This is a natural feature in, for example, the Higgs phase of a gauge theory, where Abrikosov–Nielsen–Olesen (ANO) or center vortices appear as massive excitations.  
The realization of generalized symmetries on extended dynamical defects has been studied from various perspectives in recent works~\cite{Copetti:2024onh, Choi:2024tri, Antinucci:2024izg, Brennan:2025acl, Copetti:2024rqj, Copetti:2024dcz, Cordova:2024iti, Gagliano:2025gwr}. These defects serve as prototypical examples of higher-charged objects or higher representations~\cite{Bhardwaj:2023ayw, Bartsch:2023wvv}. It would be interesting to synthesize these developments to further constrain both bulk and defect RG-flows.
    
\end{enumerate}
We hope to come back to these and other problems in the near future.

\vspace{2mm} \noindent {\bf Acknowledgments.}
We thank Fabio Apruzzi, Pieter Bomans, Ho Tat Lam, Sahand Seifnashri, Alison Warman, Matt Yu, Yunqin Zheng for discussions. 
We also thank the KITP for hospitality during the program GenSym25, during which this collaboration was initiated.
The work of AA, YG, and SSN is supported by the UKRI Frontier Research Grant, underwriting the ERC Advanced Grant ``Generalized Symmetries in Quantum Field Theory and Quantum Gravity”. The work of CC and in part of SSN is supported by the STFC grant ST/X000761/1.
This research was supported in part by grant NSF PHY-2309135 to the Kavli Institute for Theoretical Physics (KITP).


\bibliographystyle{ytphys}
{\baselineskip=0.85\baselineskip}
\bibliography{ref}


\appendix 

\section{Mathematical Background and Proofs}

In this appendix, we collect necessary background materials on tensor functors, their properties and their generalization to (2+1)d, along with a review of $G$-crossed product categories and equivariantizations. We also provide some proofs and computations, e.g., of the equivalence between the Matching Equation and the existence of a tensor functor between symmetry categories, and of the triple linking invariance in the (4+1)d twisted Dijkgraaf-Witten theory. 

\subsection{Mathematics of Tensor Functors}
\label{app:Math}

This section summarizes some basic notions from category theory, with a focus on tensor categories and short exact sequence of tensor categories, as needed for the main text. 

\vspace{2mm}
\noindent \textbf{General Category Theory Language.} We begin with the basic notions associated with functors in category theory. See \cite[Appendix A]{Weibel_1994} for further background on general category-theoretic concepts.

A \textbf{functor} $\Fun: \cC \rightarrow \cD$ from a category $\cC$ to a category $\cD$ is a rule that assigns: an object $\Fun(\D_1)$ of $\cD$ to every object $\D_1$ of $\cC$, and a morphism $\Fun(f):F(\D_1) \rightarrow F(\D_2)$ in $\cD$ to every morphism $f:\D_1 \rightarrow \D_2$ in $\cC$, such that $\Fun$ preserves the identity morphisms and compositions, i.e.
\begin{align}
    \Fun(\id_\D) & = \id_{\Fun(\D)}\,, \\
    \Fun(g \circ f) & = \Fun(g) \circ \Fun(f)\,.
\end{align}
A functor $\Fun: \cC \rightarrow \cD$ is faithful if, for each objects $\D_1,\D_2\in \cD$, the map between the hom-sets:
\begin{equation}
    \text{Hom}_{\cC}(\D_1,\D_2) \rightarrow \text{Hom}_{\cD}(\Fun(\D_1),\Fun(\D_2))\,,
\end{equation}
is injective. A functor is called an \textbf{embedding} if it is both injective on objects and faithful.

\vspace{2mm}
\noindent \textbf{Tensor Category Language.} We now focus on the relevant notions from tensor category theory. We refer to \cite{EGNO} for basic notions on tensor categories, and \cite{bruguieres2011exact,natale2020notionexactsequencehopf} for notions associated to exact sequences of tensor categories.

Given two tensor categories $\cC$ and $\cD$, with associators $\alpha$ and $\beta$, respectively, a \textbf{tensor functor} (or \textbf{monoidal functor}) 
\begin{equation}
    \Fun: \cC \rightarrow \cD
\end{equation}
is a functor equipped with a natural isomorphism $J$, called the \textbf{tensor structure} or \textbf{monoidal structure}: 
\begin{equation}
\label{eqn:monoidal_structure}
    J_{\D_1,\D_2}:\quad  \Fun(\D_1) \otimes \Fun(\D_2) \cong \Fun(\D_1 \otimes \D_2)
\end{equation}
for any objects $\D_1,\D_2$ in $\cC$, that guarantees the compatibility condition:
\begin{equation}
\label{eqn:monoidal_structure_compatibility}
\ba
&    J_{\D_1, \D_2 \otimes \D_3} \circ (\text{id}_{\Fun(\D_1)} \otimes J_{\D_2,\D_3}) \circ \beta_{\Fun(\D_1),\Fun(\D_2),\Fun(\D_3)} = \\
 &   \Fun(\alpha_{\D_1,\D_2,\D_3}) \circ J_{\D_1 \otimes \D_2, \D_3} \circ (J_{\D_1,\D_2} \otimes \text{id}_{\Fun(\D_3)})
\ea
\end{equation}
is satisfied for all objects $\D_1,\D_2,\D_3$ in $\cC$. This tensor structure $J$ is exactly the map between topological junctions of the UV and IR symmetry categories in figure~\ref{RGinterface_tensorproduct}. We say that a tensor functor $\Fun$, with tensor structure $J$, is compatible with the charge if
\begin{equation}
\label{eqn:compatibility_tensor_charge}
\ba
&    b_{n\,1, F(\D_1 \otimes \D_2)}^{\alpha} \circ (\id_{n\, 1} \otimes J_{\D_1,\D_2}) = \\
&   (J_{\D_1,\D_2} \otimes \id_{n\, 1}) \circ b_{n\,1, F(\D_1) \otimes F(\D_2)}^\alpha\,,
\ea
\end{equation}
for any objects $\D_1,\D_2$ in $\cC$ and for any half-braiding $b_{n\, 1}^\alpha$ on $n\,1$ in $\cD$.

We now turn to properties of tensor functors that are essential for defining a short exact sequence of tensor categories. Although many of these have already appeared in the main text, we restate them here for completeness.

A tensor functor $\Fun:\cC \rightarrow \cD$ is \textbf{normal} if for every object $\D$ of $\cC$, there is a subobject $\D_0 \subset \D$ such that $\Fun(\D_0)$ is the largest trivial subobject of $\Fun(\D)$. If both $\cC$ and $\cD$ are fusion categories, $\Fun$ is normal if and only if, for any simple object $\D$ in $\cC$, if $\Fun(\D)$ contains a copy of the tensor unit $1$, then $\Fun(\D)\cong n\, 1$ for some natural number $n$. We provide more details about the physical relevance of normal functors in Appendix~\ref{app:AllThingsShouldBeNormal}.

For a tensor functor 
\begin{equation}
    \Fun: \cC \rightarrow \cD\,,
\end{equation}
its \textbf{kernel} is the full tensor subcategory of $\cC$:
\begin{equation}
\label{eqn:kernal_tensor_functor}
    \ker(\Fun) = \Fun^{-1}(\langle 1 \rangle)\,,
\end{equation}
i.e., objects $\D$ of $\ker(\Fun)$ are such that $\Fun(\D)\cong 1^n$ for some natural number $n$.

A tensor functor $\DFun: \cC \rightarrow \cD$ is \textbf{dominant} (or \textbf{surjective}) if every object of $\cD$ is a subobject of $\DFun(\D)$ for some object $\D$ of $\cC$. This notion should not be confused with that of an essentially surjective functor: a functor $\Fun: \cC_1 \rightarrow \cC_2$ is essentially surjective if for every object $\D_2$ of $\cC_2$, there is object $\D_1$ of $\cC_1$ and isomorphism $\Fun(\D_1)\cong \D_2$ in $\cC_2$. Throughout, $\image(\Fun)$ denotes the essential image.

An \textbf{exact sequence of tensor categories} is a sequence 
\begin{equation}
    \cN \xrightarrow{\IFun} \cC \xrightarrow{\DFun} \cS\,,
\end{equation}
where $\cN$, $\cC$ and $\cS$ are tensor categories, $\DFun$ is a surjective tensor functor, and $\IFun$ is a full embedding whose essential image is equivalent to $\ker(\DFun)$. 

If such an exact sequence exist, we call $\cN$ a \textbf{normal} subcategory of $\cC$. If a tensor category $\cC$ does not have any normal subcategories other than $\Vec$, it is said to be \textbf{simple}. (It is also called simple with respect to rank-one module categories, if one considers exact sequences of tensor categories with respect to module categories, see \cite{ETINGOF20171187,natale2020notionexactsequencehopf}.)

It was shown in \cite{bruguieres2011exact} that a short exact sequence of fusion categories $\cN \rightarrow \cC \rightarrow \cS$ is of the form
\begin{equation}
    \langle A \rangle \rightarrow \cC \xrightarrow{-\otimes A} \Mod_{\cC}(A)\,,
\end{equation}
where $A$ is a connected (i.e., $\dim(\Hom(1, A))=1$) self-trivializing (i.e., for any $\D\subset A$, $\D\otimes A \cong nA$ for some $n$) semisimple commutative central (i.e., admits a half-braiding, making $A$ an algebra in $\cZ(\cC)$) algebra in $\cC$, and $\langle A \rangle$ denotes the full subcategory of $\cC$ generated by objects in $A$, closed under taking direct sums, subojects and quotients.

\subsection{Proof of the Matching Equation} 
\label{sec:proofOfME}

We now prove the equivalence between the condition that an interface satisfies the Matching Equation~\eqref{eq:masterequation} and the existence of a tensor functor $\Fun : \cC_\UV \rightarrow \cC_\IR$. This statement holds both when $\cC_\UV$ and $\cC_\IR$ are fusion 1-categories, where $\Fun$ is a tensor functor; and when $\cC_\UV$ and $\cC_\IR$ are fusion 2-categories, where $\Fun$ is a tensor 2-functor. In what follows, all constructions and arguments are phrased so as to apply uniformly in either setting.

Many of the steps in the proof rely on foundational results from \cite{EGNO} in the fusion 1-category case, and on the corresponding results from \cite{Decoppet:2023uoy} for fusion 2-categories. 

To begin, recall that the mathematical structure describing an interface between $\fZ(\cC_\UV)$ and $\fZ(\cC_\IR)$ is a $\cC_{\UV}$-$\cC_\IR$-bimodule (1- or 2-)category, or equivalently -- from the folded Lagrangian perspective -- a $(\cC_\UV \boxtimes \cC_\IR^{\text{op}})$-module (1- or 2-)category. Let $\cM$ denote this $\cC_{\UV}$-$\cC_\IR$-bimodule category. The Matching Equation~\eqref{eq:masterequation} translates to the equivalence of (right) $\cC_\IR$-module categories:
\begin{equation}
    \cC_\UV \boxtimes_{\cC_\UV} \cM \cong \cC_{\IR}\,.
\end{equation}

\vspace{2mm}
\noindent \textbf{From Tensor Functors to Interfaces.} 
Given a tensor (1- or 2-)functor 
\begin{equation}
    \Fun:\cC_\UV \rightarrow \cC_\IR\,,
\end{equation}
the (1- or 2-)category $\cM=\cC_\IR$ carries the structure of a $\cC_{\UV}$-$\cC_\IR$-bimodule category, with the left $\cC_\UV$ action given by the tensor structure in $\cC_\IR$:
\begin{align}
    \cC_\UV \times \cM\  & \rightarrow\  \cM\nn\\
   ( \D, m )\  & \mapsto \ \Fun(\D) \otimes m\,,
\end{align}
and module associator inherited from the associator of $\cC_\IR$ (similarly generalize to $\cC_{\UV}$-$\cC_\IR$-bimodule 2-categories). The universal property of the relative Deligne product guarantees
\begin{equation}
    \cC_\UV \boxtimes_{\cC_\UV} \cM \cong \cM = \cC_\IR
\end{equation}
as a $\cC_\IR$-module category. Hence the Matching Equation~\eqref{eq:masterequation} is satisfied, with the interface corresponding to the $\cC_{\UV}$-$\cC_\IR$-bimodule category $\cM=\cC_{\IR}$.

\vspace{2mm}
\noindent \textbf{From Interfaces to Tensor Functors.} 
Conversely, given a $\cC_{\UV}$-$\cC_\IR$-bimodule category $\cM$. In general, as a right $\cC_\IR$-module category, $\cC_\UV \boxtimes_{\cC_\UV} \cM$ decomposes into
\begin{equation}
    \cC_\UV \boxtimes_{\cC_\UV} \cM \cong \bigoplus_{i}\cN_i\,,
\end{equation}
where the (1- or 2-)categories $\cN_i$ are indecomposable $\cC_\IR$-module categories, i.e., there are equivalences of right $\cC_\IR$-module categories
\begin{align}
    T : \cC_\UV \boxtimes_{\cC_\UV} \cM  &\rightarrow \bigoplus_{i}\cN_i \cr 
    S : \bigoplus_{i}\cN_i& \rightarrow \cC_\UV \boxtimes_{\cC_\UV} \cM\,,
\end{align}
with natural isomorphisms to the identity functors
\begin{align}
    T\circ S & \cong \id_{\bigoplus_{i}\cN_i}\,,\\
    S\circ T & \cong \id_{\cC_\UV \boxtimes_{\cC_\UV} \cM}\,.
\end{align}
Using the equivalences of categories, one can define the tensor functor on the dual categories 
\begin{align}
    \Fun: \text{Fun}_{\cC_\UV}(\cC_\UV, \cC_\UV) & \rightarrow \text{Fun}_{\cC_\IR}\left( \oplus_{i}\cN_i, \oplus_{i}\cN_i \right)\nn\\
    \psi & \mapsto T \circ (\psi \boxtimes_{\cC_\UV}\id_{\cM}) \circ S\,.
\end{align}
The target category is a tensor (1- or 2-)category if and only if the right $\cC_\IR$-module category $\cC_\UV \boxtimes_{\cC_\UV} \cM$ is indecomposable (see section 7.12 of \cite{EGNO} for fusion 1-categories and \cite[Corollary 5.2.5]{Decoppet:2023uoy} for fusion 2-categories). 

Moreover, together with the tensor equivalence
\begin{equation}
    \text{Fun}_{\cC}(\cC,\cC) \cong \cC
\end{equation}
for any fusion category $\cC$ (valid in both the 1-categorical and 2-categorical settings), it is when the Matching Equation~\eqref{eq:masterequation} holds, i.e., 
\begin{equation}
    \cC_\UV \boxtimes_{\cC_\UV} \cM \cong \cC_\IR
\end{equation}
as right $\cC_\IR$-module categories, we have a tensor functor (or tensor 2-functor) between symmetry categories
\begin{equation}
    \Fun: \cC_\UV \rightarrow \cC_\IR\,.
\end{equation}

\subsection{Drinfeld Center of $\Rep(S_3)$} 
\label{sec:RepS3}

In this appendix we summarize some of the details of the SymTFT for $S_3$ finite non-abelian group, its condensable algebras and folded Lagrangians obtained in \cite{Bhardwaj:2023bbf, Bhardwaj:2024qrf}. This is used in the main text.
We present $S_3$ as 
\begin{equation}
    S_3=\langle a,b \ | \ a^3=b^2=1 \ , \ \ bab=a^2  \rangle \ .
\end{equation}
The Drinfeld center $\cZ(\Rep(S_3))= \cZ (\Vec_{S_3})$ is group-theoretical, and its anyons  are labeled by $([g], R)$, where $[g]$ is a conjugacy class of $S_3$ and $R$ an irreducible representation of its centralizer group. 
There are three conjugacy classes: $[1]$ with maximal centralizer, hence its representations are $\Rep(S_3)$ generated by $1,P,E$, which are the trivial, sign and 2d irreducible representations, where  $E\otimes P=E$, and
\be
E\otimes E=1\oplus P\oplus E \,.
\ee
The two non-trivial conjugacy classes are
\begin{itemize}
\item $[a]=\left\{a, a^2\right\}$ whose centralizer is $\bZ_3$, hence we label the corresponding lines by $a_{\chi=0,1,2}$.
\item $[b]=\left\{b,ab,a^2b\right\}$ whose centralizer is $\bZ_2$, hence we label these lines by $b_\pm$. 
\end{itemize}
The canonical $\Rep(S_3)$ symmetry Lagrangian algebra is 
\begin{equation}
    \cL_{\Rep(S_3)}=1\oplus a_0\oplus b_+ \ .
\end{equation}
The various condensable algebras (including the Lagrangian) assemble into the following Hasse diagram \cite{Bhardwaj:2024qrf}:
\be
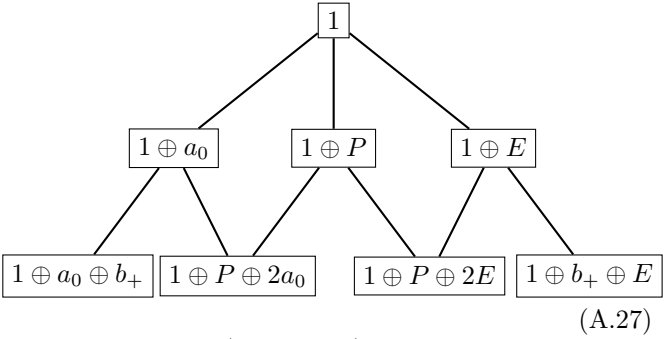
\begin{tikzpicture}[vertex/.style={draw}, scale=0.85]
\begin{scope}[shift={(0,0)}]
\node[vertex] (1)  at (0,0) {$1$};
\node[vertex] (2)  at (-2.5,-2) {$1\oplus a_0 $};
\node[vertex] (3) at (0, -2) {$1 \oplus P$} ;
\node[vertex] (4) at (2.5, -2) {$1 \oplus E$} ;
\node[vertex] (5) at (-4, -4) {$1 \oplus a_0\oplus b _+$} ;
\node[vertex] (6) at (-1.5, -4) {$1 \oplus P \oplus 2a_0 $} ;
\node[vertex] (7) at (1.5, -4) {$1 \oplus P \oplus 2E $};
\node[vertex] (8) at (4, -4) {$1 \oplus b_+ \oplus E$} ;
\draw (1) edge [thick] node[label=left:] {} (2);
\draw  (1) edge [thick] node[label=left:] {} (3);
\draw  (1) edge [thick] node[label=left:] {} (4);
\draw  (2) edge [thick] node[label=left:] {} (5);
\draw  (2) edge [thick] node[label=left:] {} (6);
\draw  (3) edge [thick] node[label=left:] {} (6);
\draw  (3) edge [thick] node[label=left:] {} (7);
\draw  (4) edge [thick] node[label=left:] {} (7);
\draw (4) edge [thick] node[label=left:] {} (8);
\end{scope}
\end{tikzpicture}
\ee
There are three (non-trivial) condensable but non-Lagrangian algebras
\begin{equation}
    \cA_{a_0}=1\oplus a_0 \ , \ \ \  \ \cA_P=1\oplus P \ ,\ \ \ \ \cA_E=1\oplus E \ .
\end{equation}
The last two, of dimensions $2$ and $3$ respectively, are magnetic. Their corresponding folded Lagrangian algebras have been computed in \cite{Bhardwaj:2023bbf}:
\begin{equation}
\ba
          \cL ^{(E)}_{\text{folded}}&=1\oplus E\oplus  P \, \overline{m} \oplus E\, \overline{m} \oplus b_+ \, \overline{e} \oplus b_- \, \overline{em} \cr 
    \cL^{(P)}_{\text{folded}} &=1\oplus P\oplus a_0 \overline{L}_{1,0} \oplus a_0 \overline{L}_{2,0}\oplus a_2\overline{L}_{1,1}\oplus a_2\overline{L}_{2,2}  \oplus \cr 
&    \quad \oplus E\overline{L}_{0,1} \oplus E\overline{L}_{0,2} \oplus a_1 \overline{L}_{1,2}\oplus a_1\overline{L}_{2,1}
\cr 
   \cL_{\text{folded}}^{(a_0)} &=1\oplus a_0\oplus P\, \overline{m} \oplus a_0\, \overline{m} \oplus b_+ \, \overline{e} \oplus b_- \, \overline{em}  \ .
\ea
\end{equation}

\subsection{Normal Functors and Subcategories}\label{app:AllThingsShouldBeNormal}

For non-invertible symmetries, a particularly important class of functors $\Fun: \cC_\UV \rightarrow \cC_\IR$ is given by normal functors \eqref{eq:normal fun}. For invertible symmetries, every tensor functor is automatically normal. Intuitively, normal functors avoid situations where a simple non-invertible defect $D$ in the UV flows to a sum of defects in the IR, where some of which act trivially and others non-trivially. Thus the kernel of a normal functor is sufficiently large to capture the entire portion of the UV symmetry that acts trivially in the IR.

A normal tensor functor has a simple characterization in terms of the interface $\cI$. We recall that by the folding trick the interface is equivalent to a boundary of $\fZ(\cC_\UV) \boxtimes \overline{\fZ(\cC_\IR)}$, described by the folded Lagrangian 
\begin{equation}
    \cL_\cI=\bigoplus _{i, j} n_{i, j} a^\UV_{i} \otimes \overline{b}^\IR_{j}\,.
\end{equation}
The coefficient $n_{i,j}$ is non-zero whenever the $a_i^\UV$ can transmute to $b_j^\IR$ through the interface. If $b_j^\IR \subset \cL_{\IR}$, a defect $a_i^\UV$ with $n_{i,j}\neq 0$ is the lift to the bulk of a symmetry defect that can be mapped to trivial by the tensor functor. Thus the condition for normality is 
\begin{equation} \label{eq:normal in folded}
    \forall b_i^\UV, n_{i,j}\neq 0, a_j^\IR \subset \cL_{\IR},\Longrightarrow n_{i,k}=0\  \forall a_k^\IR \not\subset \cL_{\IR} \ .
\end{equation}

As we will see shortly, normal functors are particularly interesting if they are also surjective, as their kernel define normal subcategories.\footnote{One can always restrict the target category to the image of a normal functor, hence making the functor surjective.} In this case the condition for normality can be rephrased into a simpler one in terms of the condensable magnetic algebra $\cA$ of $\cZ(\cC_\UV)$ that defines the functor. The surjective functor is normal if and only if there is a subalgebra $\cA_\IFun\subset \cL_{\UV}$ such that $\cA_\IFun \otimes \cA$ is a condensable algebra of $\cZ(\cC_\UV)$, and it is Lagrangian.

\vspace{0.2cm}

\noindent \textbf{An Example: $\Rep(S_3)$.} An  illustrative example is given by a $\Rep(S_3)$ symmetry, whose simple objects are $1,P,E$:
\begin{equation}
 E\otimes P=E  ,  \ \ P\otimes P=1 ,  \ \ E\otimes E=1\oplus P \oplus E \ .
\end{equation} 
$\Rep(S_3)$ has two natural surjective tensor functors given by the restriction to its $\bZ_2$ and $\bZ_3$ subgroups:
\begin{align}\label{eq:functors F23}
    F_2 : \Rep(S_3) & \rightarrow \Rep(\bZ_2) &  F_3 : \Rep(S_3) & \rightarrow \Rep(\bZ_3)\nn\\
    P & \mapsto \eta & P & \mapsto 1\nn\\
    E & \mapsto 1 \oplus \eta\,, & E & \mapsto \omega \oplus \omega^2\,.
\end{align}
Here $\eta$ and $\omega$ are the generators of $\Rep(\bZ_2)=\bZ_2$ and $\Rep(\bZ_3)=\bZ_3$ respectively. $\Fun_3$ is a normal functor, while $\Fun_2$ is not. 
 
Let us derive these facts from the SymTFT. From some known facts on $\cZ(\Rep(S_3))$ in Appendix \ref{sec:RepS3}, there are two (non-Lagrangian) magnetic algebras $\cA_P,\cA_E$ of dimensions $2$ and $3$ respectively. 

We start with $\cA_E = 1 \oplus E$. Its condensation produces $\fZ(\bZ_2)$, whose lines we label by $1,e,m,em$. The folded Lagrangian associated with the interface $\cI_{\cA_E}$ between $\cZ(\Rep(S_3))$ and $\cZ(\Vec_{\bZ_2})$ is \cite{Bhardwaj:2023bbf}:
\begin{equation}\label{eq:folded 1+E}
    \cL ^{(E)}_{\text{folded}}=1\oplus E\oplus  P \, \overline{m} \oplus E\, \overline{m} \oplus b_+ \, \overline{e} \oplus b_- \, \overline{em} \ .
\end{equation}
This gives a map of bulk lines across the interface, from which we read that $\cL_{\Rep(S_3)}=1 \oplus a_0\oplus b _+$ is mapped into $1\oplus e$, thus $\cI_{\cA_E}$ satisfies the Matching Equation \eqref{eq:masterequation}. From \eqref{eq:folded 1+E} we can read that the interface maps $P\mapsto m$ and $E\mapsto 1\oplus m$, and thus defines a functor that coincides with $\Fun_2$ of \eqref{eq:functors F23} ($m$ becomes $\eta$ on $\Bsym_{\bZ_2}$ via the forgetful functor). 

Notice that \eqref{eq:folded 1+E} does not satisfy \eqref{eq:normal in folded}, as expected, given that $\Fun_2$ is not a normal functor. Moreover, the only non-trivial subalgebra of $\cL_{\Rep(S_3)}$ is $\cA_{a_0}=1 \oplus a_0$, but $\cA_{a_0}\otimes \cA_E$ is not a condensable algebra, so the criterion for surjective normal functor is also not satisfied.

On the other hand, condensing $\cA_{P} = 1 \oplus P$ produces $\cZ(\Vec_{\bZ_3})=\DW(\bZ_3)$, whose lines we label by $L_{n,m}$, $n,m\in \{0,1,2\}$, and the folded Lagrangian is 
\begin{equation}
 \begin{array}{r}
    \cL^{(P)}_{\text{folded}}=1\oplus P\oplus a_0 \overline{L}_{1,0} \oplus a_0 \overline{L}_{2,0}\oplus a_2\overline{L}_{1,1}\oplus a_2\overline{L}_{2,2}  \oplus \vspace{0.2cm} \\
    \oplus E\overline{L}_{0,1} \oplus E\overline{L}_{0,2} \oplus a_1 \overline{L}_{1,2}\oplus a_1\overline{L}_{2,1}
 \end{array}   
\end{equation}
The Matching Equation is satisfied with $\cL_{\bZ_3}=1 \oplus L_{1,0} \oplus L_{2,0}$, and the action on $P$ and $E$ reproduces $\Fun_3$ from \eqref{eq:functors F23}. Now the condition \eqref{eq:normal in folded} is satisfied. The same is true for the condition about surjective normal functors, as 
\begin{equation}
    \cA_{a_0}\otimes \cA_P= 1\oplus P \oplus 2a_0
\end{equation}
is a Lagrangian algebra.

Now we return to the embedding of the anomaly-free subcategory of $\Rep(S_3)$, and see how it results in a short exact sequence of fusion categories.
Condensing $\cA_\IFun=\cA_{a_0}=1 \oplus a_0$ produces $\cZ(\Vec_{\bZ_2})$, and its corresponding folded Lagrangian in $\cZ (\Rep({S_3})) \boxtimes \ol{\cZ (\Vec_{\Z_2})}$ is \cite{Bhardwaj:2023bbf}:
\begin{equation}
    \cL_{\IFun}^{a_0}=1\oplus a_0\oplus P\, \overline{m} \oplus a_0\, \overline{m} \oplus b_+ \, \overline{e} \oplus b_- \, \overline{em}  \ .
\end{equation}
We see that \eqref{eq:masterequation} is satisfied if $\Bsym_\UV$ corresponds to $\cL_{\bZ_2}=1 \oplus e$. Thus we get a tensor functor $\IFun : \Vec_{\bZ_2} \rightarrow \Rep(S_3)$ that maps the $\bZ_2$ generator $\eta=\pi_\UV(m)$, where $\pi_\UV:\cZ(\Vec_{\bZ_2})\rightarrow \Vec_{\bZ_2}$ denotes the forgetful functor, into $\IFun(\eta)=P$. We see explicitly that $\Fun_3\circ \IFun$ is a fiber functor, and we get an exact sequence of categories
\begin{equation}
    \Vec_{\bZ_2} \rightarrow \Rep(S_3) \rightarrow \Vec_{\bZ_3} \ .
\end{equation}
This is a particular case of a general result: for any finite group $G$ and $N\triangleleft G$ normal, there is an exact sequence of categories \cite{bruguieres2011exact}
\begin{equation}
    \Rep(G/N) \rightarrow \Rep(G) \rightarrow \Rep(N) \ .
\end{equation}

\vspace{0.2cm}
\noindent 
\textbf{Example: Fibonacci Categories.}
We provide another example of a non-normal functor. 
Consider the Fibonacci category $\Fib$ with simple objects 1 and $\tau$, and the fusion rule:
\begin{equation}
    \tau \otimes \tau = 1 \oplus \tau\,.
\end{equation}
There is a surjective tensor functor
\begin{align}
    \DFun: \Fib\boxtimes \Fib & \rightarrow \Fib\nn\\
    \tau \boxtimes 1 & \mapsto \tau\,,\nn\\
    1 \boxtimes \tau & \mapsto \tau\,,\nn\\
    \tau \boxtimes \tau & \mapsto 1 \oplus \tau\,.
\end{align}
Note that this functor is not normal and has trivial kernel 
\begin{equation}
    \ker(\DFun)\cong \Vec\,.
\end{equation}
Still, one can view the target category $\Fib$ as the category of modules over the algebra $A = 1 \oplus (\tau \boxtimes \tau)$ in $\Fib\boxtimes \Fib$:
\begin{equation}
    \Fib \cong (\Fib\boxtimes \Fib)_{A}\,.
\end{equation}
This algebra $A$ fails to define a normal functor as it is not self-trivializing: 
\begin{equation}
    (\tau \boxtimes \tau) \otimes A \not \cong n A
\end{equation}
for any $n$. In this case, the kernel of $\DFun$ fails to agree with the full subcategory generated by $A$. 
This aligns with the observation that there is no non-trivial condensable algebras in $\Fib^{\boxtimes n}$ \cite{BOOKER2012176, Kong:2013aya, Neupert:2016pjk}.

\subsection{$G$-Crossed Categories and Equivariantizations}
\label{sec:GCross}

This appendix consists of the notion of a crossed product category, and its related short exact sequence of tensor categories, which are important to the study of symmetry enrichment and LSM anomaly matching. We also review a mathematical theorem and see how it computes the $G$-equivariantization of a $G$-crossed braided extension of the Drinfeld center. 

\vspace{0.2cm}
\noindent \textbf{Crossed Product Categories.}
We first review the construction of a crossed product category following \cite{tambara2001invariants}. In what follows we only present the categorical structure on objects, and we refer to \cite{tambara2001invariants} for the definition on morphisms, associator and unitors.

Given any (not necessarily finite) group $G$ acting on a tensor category $\cC$, 
\begin{align}
    G & \rightarrow \text{Aut}^{\ot}(\cC)\nn\\
    g & \mapsto \rho_g\,.
\end{align}
One can define the crossed product category $\cC \rtimes G$. As categories, 
\begin{equation}
    \cC \rtimes G = \bigoplus_{g\in G} \cC\boxtimes g \,,
\end{equation}
where the notation $\cC\boxtimes g \cong \cC$ for every $g\in G$. An object in $\cC \rtimes G$ is denoted by
\begin{equation}
    \bigoplus_{g\in G} \D \boxtimes g
\end{equation}
with $\D \in \cC$, and fusion rules can be computed from
\begin{equation}
    (\D_1 \boxtimes g_1) \ot (\D_2 \boxtimes g_2) = (\D_1 \ot \rho_{g_1}(\D_2)) \boxtimes g_1 g_2\,.
\end{equation}

Note that for any subgroup $H \subset G$, $\cC\rtimes G$ has a full fusion subcategory equivalent to $\Vec_H$ generated by objects of the form
\begin{equation}
    1 \boxtimes h\,,
\end{equation}
for $h\in H$.

\vspace{0.2cm}
\noindent \textbf{Short Exact Sequences from Crossed Product Categories.}
Given any $G$-action on a tensor category $\cC$ as before, if a normal subgroup $N\triangleleft G$ acts trivially on $\cC$, i.e., the $G$-action on $\cC$ factors through a $G/N$-action
\begin{equation}
    G/N \rightarrow \text{Aut}^{\ot}(\cC)\,,
\end{equation}
then there is a short exact sequence of tensor categories
\begin{equation}
    \Vec_{N} \xrightarrow{\IFun} \cC\rtimes G \xrightarrow{\DFun} \cC\rtimes G/N\,,
\end{equation}
where $\IFun$ is the embedding of the full subcategory $\Vec_N$ as discussed above, and the surjective functor $\DFun$ has trivial tensor structure, compatible with the associators as defined in \cite{tambara2001invariants}, and maps object $\D\boxtimes g$ in $\cC\rtimes G$ to
\begin{equation}
    \DFun(\D\boxtimes g) = \D \boxtimes gN \in \cC\rtimes G/N\,.
\end{equation}
Similarly, one can define its map on hom-sets, which we do not elaborate here. 
Note that $\image(\IFun) = \ker(\DFun)$, hence $\DFun$ is a normal functor.

\vspace{0.2cm}
\noindent \textbf{Equivariantization of Crossed Extensions.}
We now assume $G$ to be a finite group and turn to the $G$-equivariantization of a $G$-crossed braided extension of $\cZ(\cC)$\footnote{In general, there can be multiple $G$-crossed (braided) extensions of a given (braided) category.}, 
which shows up in the description of the IR TO in section~\ref{sec:LSM}. 

It is shown in \cite[Theorem 3.5]{2009arXiv0905.3117G} that there is an equivalence of braided fusion categories 
\begin{equation}
    \cZ_{\cC}(\cC\rtimes G)^{G} \cong \cZ(\cC\rtimes G)\,,
\end{equation}
where $\cZ_{\cC}(\cC\rtimes G)^{G}$ denotes the $G$-equivariantization of the relative center $\cZ_{\cC}(\cC\rtimes G)$, which is a braided $G$-crossed fusion category 
\begin{equation}
    \cZ_{\cC}(\cC\rtimes G) = \bigoplus_{g\in G} \cZ_\cC (\cC\boxtimes g)\,,
\end{equation}
with objects of $\cZ_\cC (\cC\boxtimes g)$ being pairs $(\D \boxtimes g, \gamma)$, where $\D \in \cC$ and $\gamma$ the natural isomorphisms
\begin{equation}
    \gamma_{\D'}: \D' \otimes \D \xrightarrow{\cong} \D \otimes g(\D')\,,
\end{equation}
for each $\D'\in \cC$, satisfying natural compatibility conditions. Note that the trivially-graded component $\cZ_\cC(\cC\boxtimes e) = \cZ(\cC)$, where $e\in G$ is the identity element. One can conclude that $\cZ_{\cC}(\cC\rtimes G)$ is indeed a $G$-crossed braided extension of $\cZ(\cC)$. 

Applying the above result with $\cC=\Rep(D_8)$ and $G=\bZ_3$, we conclude the $\bZ_3$-equivariantization of the $\bZ_3$-crossed extension of $\cZ(\Rep(D_8))$ is
\begin{equation}
    \cZ_{\Rep(D_8)}(\Rep(D_8) \rtimes \bZ_3)^{\bZ_3} \cong \cZ(\Rep(D_8)\rtimes \bZ_3)\,.
\end{equation}
Explicitly, the trivially-graded component of the $\bZ_3$-crossed extension $\cZ_{\Rep(D_8)}(\Rep(D_8) \rtimes \bZ_3)$ is simply $\cZ(\Rep(D_8))$ itself, and for $a=1,2$, the $a$-graded component contains objects
\begin{equation}
\begin{split}
    & 1_a\,, \, (e_{\text{RGB}})_a \,, \,  (s_{\text{RGB}})_a\,, \, (\bar{s}_{\text{RGB}})_a\,, \\
    & (e_{\text{R}}\oplus e_{\text{G}} \oplus e_{\text{B}})_a \,,\, (e_{\text{RG}}\oplus e_{\text{GB}} \oplus e_{\text{RB}})_a\,, \\
    & (m_{\text{R}}\oplus m_{\text{G}} \oplus m_{\text{B}})_a \,,\, (m_{\text{RG}}\oplus m_{\text{GB}} \oplus m_{\text{RB}})_a\,,\\
    & (f_{\text{R}}\oplus f_{\text{G}} \oplus f_{\text{B}})_a \,, (f_{\text{RG}}\oplus f_{\text{GB}} \oplus f_{\text{RB}})_a\,.
\end{split}
\end{equation}
The fusion rules are induced from the fusion rules of $\cZ(\Rep(D_8))$ and the group law of the grading.

\subsection{Tensor Functors and ASCies in (2+1)d}
\label{app:F2C}

As suggested earlier, we would like to use the SymTFT approach to study tensor functors and ASCies, in order to formulate a higer-dimensional version of these notions -- without having to revert to higher tensor functors. This is physically extremely well-motivated. In the case of (2+1)d, i.e. fusion 2-categories, we can make some of this very precise as the SymTFT for such symmetries is extremely well-developed. 

The first important point is that all so-called bosonic fusion 2-categories have been classified \cite{decoppet2022drinfeld} and their Drinfeld center is given by $\fZ (\TwoVec_{G}^\omega)$ for a finite 0-form symmetry group $G$ with anomaly $\omega\in H^4 (G, U(1))$. 

First we would like to generalize the notion of tensor functors to fusion 2-categories, i.e. define a functor  
\be\label{blabla2}
\Fun: \  \cC_\UV \to \cC_\IR \,,
\ee
where the symmetry categories in the UV and IR are now 2-categories, describes symmetries of $(2+1)$d systems. Examples are 0-form and 1-form symmetries, as well as 2-group symmetries.  
Although tensor 2-functors have been defined in \cite{baez1996higher}, there is not too much detailed study of properties of 2-functors in the literature. Instead, we use the alternative definition in terms of SymTFT interfaces, developed in section~\ref{sec: SymTFTTF}. 
Gapped boundary conditions and interfaces for the SymTFTs of any bosonic fusion 2-category have been classified in \cite{Bhardwaj:2024qiv, Bhardwaj:2025piv,  Bhardwaj:2025jtf, Wen:2025thg}.

\vspace{2mm}
\noindent
{\bf Tensor 2-Functors and SymTFT Interfaces.}
A tensor 2-functor (\ref{blabla2}) 
is equivalent to an RG-interface $\cI_\Fun$ that satisfies the Matching Equation (\ref{eq:masterequation}), i.e., an interface that maps the $\cC_\UV$ symmetry boundary Lagrangian algebra $\cL_\UV$ to the $\cC_\IR$ symmetry boundary Lagrangian algebra $\cL_\IR$. This is proven in appendix \ref{sec:proofOfME} using general results from \cite{Decoppet:2023uoy}. Thus, although the properties of tensor 2-functor may be hard to check, the SymTFT formulation is extremely useful.

In order to quantify anomalies, we also need to define the short exact sequence of fusion 2-categories 
\begin{equation}\label{lala2}
    \cN \xrightarrow{\IFun} \cC \xrightarrow{\DFun} \cS\,.
\end{equation}
Neither injective nor surjective 2-functors are defined in the literature. We use our SymTFT setup as a starting point: 
instead of defining 2-functor $\IFun: \cN \rightarrow \cC$ to be an embedding, we consider an interface $\cI_\IFun$ such that any topological defect that ends on the symmetry boundary $\cC$ has to end on the interface $\cI_\IFun$, i.e., the interface corresponds to condensation of a subalgebra of the symmetry Lagrangian algebra, similar to~\eqref{RGQuicheInj}.

Likewise, the placeholder for a  2-functor $\DFun:\cC \rightarrow \cS$ to be surjective (dominant), is an interface where  only topological defect that can end on both the $\cC$-symmetry boundary and the interface $\cI_\DFun$ to be the trivial defects, analogous to \eqref{RGQuicheSurj}.

The exactness of (\ref{lala2})
can again be formulated as follows: any topological defect in $\cZ(\cN)$ gets mapped, after passing through the interfaces $\cI_\IFun$ and $\cI_\DFun$, to a topological defect in $\cZ(\cS)$ that ends on the symmetry boundary; and conversely, any topological defect in $\cZ(\cS)$ that can end on the symmetry boundary arises from a topological defect in $\cZ(\cN)$. This is similar to figure \ref{fig:greybluegreen}.

\vspace{2mm}
\noindent
{\bf UV-IR Symmetries and ASCies.}
In $(2+1)$d, fixing symmetry 2-categories $\cC_\UV$ and $\cC_\IR$ for the UV and IR theories, an interface between the corresponding SymTFT $\fZ(\cC_\UV)$ and $\fZ(\cC_\IR)$ is described by a Lagrangian algebra in the folded theory 
\begin{equation}
    \fZ(\cC_\UV) \boxtimes \overline{\fZ(\cC_\IR)}\,.
\end{equation}
As the center of all (bosonic) fusion 2-categories are group-theoretical, the  SymTFTs for the UV and IR symmetries are described by 
\begin{equation}
    \fZ(\TwoVec_{G_\UV}^{\omega_\UV})\, , \quad \fZ(\TwoVec_{G_\IR}^{\omega_\IR})\, ,
\end{equation}
for some finite groups $G_\UV$, $G_\IR$ and cocycles $\omega_\UV \in H^4(G_\UV, U(1))$, $\omega_\IR\in H^4(G_\IR, U(1))$. Furthermore, a Lagrangian algebra in 
\begin{equation}
    \cZ(\TwoVec_{G_\UV \times G_\IR}^{\omega_{\UV} \overline{\omega_\IR}})\,,
\end{equation}
corresponds to an interface between the two topological orders.

\vspace{2mm}
\noindent
{\bf Examples.}
As an example, consider a short exact sequence of finite groups
\begin{equation}
    1 \rightarrow N \rightarrow G \xrightarrow{p} G/N \rightarrow 1\,.
\end{equation}
Associated to this, there is a short exact sequence of fusion 2-categories
\begin{equation}
    \TwoVec_{N} \rightarrow \TwoVec_{G}^{p^*\omega} \xrightarrow{\DFun} \TwoVec_{G/N}^\omega\,,
\end{equation}
for $\omega \in H^4(G/N, U(1))$, where $\DFun$ is induced from the projection $p:G \rightarrow G/N$. The interface corresponding to $\DFun$ is the condensation interface labeled by $\cI_{(G, N, \omega, \Vec_{G^{\text{diag}}})}$ following the notation in \cite{Bhardwaj:2025jtf}. 

Likewise, for any short exact sequence of 2-groups (in terms of fiber sequences, or crossed-modules)
\begin{equation}
    1 \rightarrow \cH \rightarrow \cG \xrightarrow{p} \cK \rightarrow 1\,,
\end{equation}
and anomaly $\omega \in H^4(B\cK, U(1))$, we propose a short exact sequence of fusion 2-categories
\begin{equation}
    \TwoVec_\cH \rightarrow \TwoVec_\cG^{p^*\omega} \rightarrow \TwoVec_\cK^\omega\,.
\end{equation}
The bosonic fusion 2-categories are classified \cite{Decoppet:2024htz} in terms of group-theoretical data and fusion 1-categories. This would open up a way to classification of ASCies for symmetry 2-categories to be possible by combining short exact sequences of groups and fusion 1-categories, and it would be very interesting to develop this direction in the future. 

What this appendix shows is that although not all the higher categorical structures may be known by mathematicians, the physical approach using the SymTFT reformulation is powerful and provides a useful way to extend our results to higher dimensions. We have sees this at work in section~\ref{sec:WWW}.

\subsection{Link Invariants and Algebras in (4+1)d} 
\label{app:link}

In this appendix we briefly detail the computation of (triple) link invariants for the (4+1) dimensional twisted Dijkgraaf-Witten theory. Our methods follow closely \cite{Putrov:2016qdo} and are a slight generalization of \cite{Kaidi:2023maf}. These are used in section \ref{sec:WWW}.

\vspace{2mm} \noindent \textbf{Link invariants.} First we write down the action in the presence of multiple $\Q_3^{(n_m,l)}$ defects:
\be \ba
S_{\text{SymTFT}}[\Sigma_i, (n_m^i,l^i)] & = \frac{2 \pi i}{N} \int \biggl[ (da + \sum_i \, n_m^i \PD(\Sigma_i)) \, b \\
&\!\!\!\!\! + \left( \sum_i l^i \PD(\Sigma_i) + k \beta(a)\right) a \, \beta(a) \biggr] \, ,
\ea \ee

where $\PD$ denotes the Poincare dual. Let $\widehat{\Sigma}_i$ be a surface filling $\Sigma_i$, then $d \PD(\widehat{\Sigma_i}) = \PD(\Sigma_i)$. After the change variables $a \to a - \sum_i \, n_m^i \PD(\widehat{\Sigma}_i)$, the action picks up a phase (we assume that all the surfaces have trivial self-intersection):
\be \ba
& S_{\text{SymTFT}} [\Sigma_i, (n_m^i,l^i)] = S_{\text{SymTFT}} \\
&\ + \frac{2 \pi i}{N^2} \sum_{i > j,k} \int 2 n_m^i n_m^j l^k \int \PD(\widehat{\Sigma}_i)  d \PD(\widehat{\Sigma}_j) d \PD(\widehat{\Sigma}_k) \\
&\ - \frac{2 \pi i}{N^3} \sum_{i>j>k} 6 k n_m^i n_m^j n_m^k \int \PD(\widehat{\Sigma}_i)  d \PD(\widehat{\Sigma}_j) d \PD(\widehat{\Sigma}_k) \, .
\ea \ee 
The term inside the integral is the type-2 linking number:
\be
\text{Lk}(\Sigma_1, \Sigma_2, \Sigma_3)_2 = \int \PD(\widehat{\Sigma}_1)  d \PD(\widehat{\Sigma}_2) d \PD(\widehat{\Sigma}_3) \, .
\ee
We conclude that the triple linking between the magnetic operators $\Q_3^{(n_m, l)}$ is 
\be \ba
& \langle \Q_3^{(n_m^1,l^1)} \Q_3^{(n_m^2,l^2)} \Q_3^{(n_m^3,l^3)} \rangle = \\
& \ \exp \biggl[ \frac{4 \pi i \text{Lk}(\Sigma_1,\Sigma_2,\Sigma_3)_2}{N^2}\biggl(  n_m^1 n_m^2 l^3 +  n_m^1 n_m^3 l^2   \biggr. \\
& \qquad \qquad \qquad \quad +  n_m^2 n_m^2 l^1 - \frac{3 k}{N} n_m^1 n_m^2 n_m^3 \biggr) \biggr]\,.
\ea \ee
Furthermore, if we shift $n_m \to n_m + N$, changing variables to $a - N \, \PD(\widehat{\Sigma})$ imposes the identification:
\be
\Q_3^{N, 0} \sim \Q_3^{0, 3 k} \, ,
\ee
under which the triple linking is left invariant. 
In order to discuss the triple linking more carefully, we can also consider decoration of one-dimensional intersections $\gamma_{ij} = \Sigma_i \cap \Sigma_j$ by electric lines $\Q_1^{n_e^{ij}}$. The decorated triple linking is:
\be \ba
& \langle \Q_3^{(n_m^1,l^1)} \Q_3^{(n_m^2,l^2)} \Q_3^{(n_m^3,l^3)} \rangle_{n_e^{ij}} = \\
& \ \exp \biggl[ \frac{4 \pi i \text{Lk}(\Sigma_1,\Sigma_2,\Sigma_3)_2}{N^2}\biggl(  n_m^1 n_m^2 l^3 +  n_m^1 n_m^3 l^2  \\
& \qquad \qquad \quad +  n_m^2 n_m^2 l^1 - \frac{3 k}{N} n_m^1 n_m^2 n_m^3 \biggr) \\
& \qquad \qquad \quad + \frac{2 \pi i}{N}(n_m^1 n_e^{23} + n_m^2 n_e^{13} + n_m^{3} n_e^{12}) \biggr]\,.
\ea \ee
The decorated linking is invariant under the redefinitions:
\be \ba
n_m^1 &\to n_m^1 + N \, , \\ 
l^1 &\to l^1 + 3 k \, , \\ 
n_e^{13} &\to n_e^{13} - 2 l^3 \, , \\
n_e^{12} &\to n_e^{12} - 2 l^2 \, ,
\ea \ee
and similarly for the other labels. 

\vspace{2mm} \noindent \textbf{Condensable algebras.} 
We now discuss condensable algebras in the (4+1)d Dijkgraaf-Witten theory with trivial twist $k=0$. These are described by a set of objects, together with the assignment of electric lines $\wt{\Q}_1^{m_e^{ij}}$ on codimension-2 intersections of the algebra generators. These lines are needed to ensure mutual locality.

Apart from electric and magnetic Lagrangian algebras:
\be
\cL_e = \Big\langle \wt{\Q}_1^{(n_e)} \Big\rangle \, , \quad \cL_m = \Big\langle \wt{\Q}_3^{(n_m,0)} \Big\rangle \, ,
\ee
with $m_e^{ij} = 0$, dyonic algebras are also present. 
The algebra describing the $\bZ_N^{(2)} \times^\nu \bZ_n^{(0)}$ boundary condition is:
\be
\cL_\IR =  \Big\langle  \wt{\Q}_1^{(N r_e)} \, , \ \wt{\Q}_3^{(N r_m, N s)}  \Big\rangle \, ,
\ee
together with the assignment of lines at codimension-2 junctions:
\be
m_e^{ij} = - (r_m^i s^j + r_m^j s^i) + N \eta^{ij}_e \, .
\ee
The triple linking is:
\be \ba
&\Big\langle \wt{\Q}_3^{(N r_m^1, N s^1)} \wt{\Q}_3^{(N r_m^2, N s^2)} \wt{\Q}_3^{(N r_m^3, N s^3)}   \Big\rangle = \\
&\exp \biggl[ \frac{4 \pi i}{N}\left[ r_m^1 r_m^2 s^3 + r_m^1 r_m^3 s^2 + r_m^3 r_m^2 s^1   \right] \biggr]\,,
\ea \ee
and we can immediately verify that the choice of dressing is the correct one.

The non-maximal condensable algebras which are relevant for our discussions are variations of $\cL_m$:
\be
\cA_k = \Big\langle \wt{Q}_3^{(N r_m,0)} \Big\rangle \, , \quad m_e^{ij} = N \wt{k} r_m^i r_m^j \, ,
\ee
where $\wt{k} = - 2^{-1} k^{-1}$. Let us consider in detail the reduced TO $\fZ(\bZ_N^{(0)})/\cA_k$. Clearly we have electric lines:
\be
\Q_1^{n_e} \equiv \wt{\Q}_1^{N n_e} \, .
\ee
We now study the spectrum of surface operators. Consider the triple linking:
\be \ba
&\Big\langle \wt{\Q}_3^{(N r_m, 0)} \wt{\Q}_3^{(N r_m', 0)} \wt{\Q}_3^{(n,l)}   \Big\rangle = \\
&\exp \biggl[ \frac{2 \pi i}{N^2} \Bigl( r_m r_m' (2 l + N \wt{k} n) + N (r_m \mu' + r_m' \mu) \Bigr) \biggl]\,,
\ea \ee
where $\mu$ denotes the dressing of the junction between $\wt{\Q}_3^{(N r_m, 0)}$ and $\wt{\Q}_3^{(n,l)} $. 
The equation is solved by $l = N s$, $\mu = r_m \theta$ and $2 s + \wt{k} n + 2 \theta \equiv 0$ (mod $N$). Next we must consider a triple linking process involving only a single algebra object:
\be \ba
&\Big\langle \wt{\Q}_3^{(N r_m, 0)} \wt{\Q}_3^{(n, N s)} \wt{\Q}_3^{(n',N s')}   \Big\rangle = \\
&\exp \biggl[ \frac{2 \pi i}{N^2} \bigl( 2 r_m (s n' + n s') + r_m (n \theta' + n' \theta) \bigr) \biggl] \, ,
\ea \ee
eliminating $\theta$ we find:
\be
n s' + n' s - \wt{k}^{-1} n n' \equiv 0 \mod N^2 \, ,
\ee
which implies that $s = - k n$. Thus the remaining surface operators are:
\be
\Q_3^{(n_m,0)} = \wt{\Q}_3^{(n_m, - k N n)} \, .
\ee
Their triple linking is consistent with $k$ units of $\bZ_n$ anomaly.


\end{document}